\documentclass[12pt]{article}

\AtBeginDocument{
	\addtocontents{toc}{\normalsize}
}
\pdfoutput=1

\usepackage{amsmath,amssymb,amsfonts,amsthm,amscd,mathrsfs}
\usepackage{comment}
\usepackage{esvect}
\usepackage{xcolor}
\definecolor{darkblue}{rgb}{0.1,0.1,.7}
\usepackage[]{version}
\usepackage[]{graphicx}
\usepackage[]{latexsym}
\usepackage{geometry}
\usepackage[all,cmtip]{xy}
\usepackage{dsfont}

\usepackage[margin=10pt,font=small,labelfont=bf]{caption}
\usepackage{ifthen}
\usepackage{soul}
\usepackage{tikz}
\usepackage{array,mathrsfs,amsfonts,yfonts,dsfont,bbm,colonequals}
\usepackage[nodisplayskipstretch]{setspace}
\usepackage{dsfont}
\usepackage{cite}
\usepackage{xspace}
\usepackage{empheq}
\usepackage{extarrows}

\usepackage{xcolor}
\usepackage[colorlinks, linkcolor=darkblue, citecolor=darkblue, urlcolor=darkblue, linktocpage]{hyperref} 

\usepackage{subcaption}
\usepackage[utf8]{inputenc}
\usepackage{setspace}
\usepackage{float, graphicx}
\usepackage{bbm}

\numberwithin{equation}{section}

\newcommand{\be}{\begin{equation}}
\newcommand{\ee}{\end{equation}}
\newcommand{\bea}{\begin{eqnarray}}
\newcommand{\eea}{\end{eqnarray}}
\newcommand{\ba}{\begin{equation}\begin{equationed}}
\newcommand{\ea}{\end{equationed}\end{equation}}

\newcommand{\ud}{\mathrm d}

\def\g{\gamma}

\def\s{\sigma}

\def\a{\alpha}

\def\b{\beta}

\def\D{\Delta}
\def\G{\Gamma}
\def\l{\lambda}

\def\la{\langle}

\def\G{\Gamma}

\def\la{\label}
\def\rf{\eqref}
\usepackage{cancel}
\def\p{\phi}

\newcommand{\ddt}{\frac{\ud}{\ud \text{t}}}
\newcommand{\del}{\partial} 
\def\cG{c_{G}}

\newcommand{\ha}{\tfrac{1}{2}}

\newcommand{\no}{\nonumber}

\newcommand{\Tr}{{\rm Tr}}
\newcommand{\mc}{\mathcal}
\newcommand{\tn}[1]{\underline{\mathbf{#1}}}

\def\res{\text{res}}

\def\dd{\text{d}}
\newcommand{\td}[1]{\widetilde{#1}}
\newcommand{\kf}{\mathfrak{b}}
\newcommand{\df}{\mathfrak{d}}
\newcommand{\gf}{\mathfrak{g}}
\newcommand{\Z}{\mathbb{Z}}
\newcommand{\R}{\mathbb{R}}
\newcommand{\Zh}{\widehat{\mathbb{Z}}}
\newcommand{\Ph}{\widehat{\mathbb{P}}}
\newcommand{\gh}{\widehat{g}}
\newcommand{\vp}{\varphi}
\newcommand{\CP}{\mathbb{CP}^1}
\newcommand{\Nr}{N_{\mathrm{r}}}
\newcommand{\Nc}{N_{\mathrm{c}}}
\newcommand{\C}{\mathbb{C}}
\newcommand{\Xd}{\mathbb{X}}
\newcommand{\Yd}{\mathbb{Y}}
\newcommand{\Ad}{\mathbb{A}}
\newcommand{\Bd}{\mathbb{B}}
\newcommand{\Ld}{\mathbb{L}}
\newcommand{\ad}{\text{ad}}
\newcommand{\dlangle}{\left\langle\!\left\langle}
\newcommand{\drangle}{\right\rangle\!\right\rangle}
\newcommand{\Tg}[1]{\mathsf{T}^{#1}\mathfrak{g}}
\newcommand{\Tgc}[1]{\mathsf{T}^{#1}\mathfrak{g}^\C}
\newcommand{\TG}[1]{\mathsf{T}^{#1}G}
\newcommand{\TGc}[1]{\mathsf{T}^{#1}G^\C}
\newcommand{\zb}{\overline{z}}

\newcommand{\gj}{\mathbbm{g}}
\newcommand{\gb}{\overline{\mathbbm{g}}}

\newcommand{\uj}{\mathbbm{u}}

\newcommand{\Jc}{\mathcal{J}}
\newcommand{\ellb}{\bm{\ell}}
\newcommand{\DT}{C_{\text{defect}}}
\newcommand{\Cdist}{C_{\text{dist}}}
\newcommand{\Cvar}{C_{\text{var}}}
\newcommand{\dtom}{\frac{\text{d}\omega}{\text{dt}}\Bigr|_{\text{meas}}}

\usepackage{bm}

\usepackage{enumerate} 

\DeclareFontEncoding{LS1}{}{}
\DeclareFontSubstitution{LS1}{stix}{m}{n}
\DeclareSymbolFont{stixsymbols}{LS1}{stixscr}{m}{n}
\SetSymbolFont{stixsymbols}{bold}{LS1}{stixscr}{b}{n}
\DeclareMathSymbol{\kay}{\mathalpha}{stixsymbols}{"6B}

\begin{document}
	
	\vspace*{-.6in} \thispagestyle{empty}
	\begin{flushright}
		{\ttfamily CERN-TH-2025-107}
	\end{flushright}
	\vspace{0.2cm}
    {\Large
		\begin{center}
			{\bf 1-loop renormalisability of integrable sigma-models \\
            \vspace{0.1cm}from 4d Chern-Simons theory}\\
	\end{center}}
	\vspace{0.8cm}
	\begin{center}
		{\bf Sylvain Lacroix$^a$, \  \ Nat Levine$^{b}$  \ and  \   Anders Wallberg$^{c,d}$}\\[1.4cm] 
		{
			\small

\textit{${}^a$Sorbonne Université, CNRS, Laboratoire de Physique Théorique et Hautes \'Energies, LPTHE, F-75005 Paris, France} \\
            
\vspace{0.4cm}\textit{${}^b$Institute  for  Theoretical  Physics,  University  of  Amsterdam,  \\
             PO  Box  94485,  1090  GL Amsterdam, The Netherlands}\\

      \vspace{0.4cm}\textit{${}^c$Department of Theoretical Physics, CERN\\ 1211 Meyrin, Switzerland}
      \\
      
    \vspace{0.4cm}\textit{${}^d$Laboratory for Theoretical Fundamental Physics, EPFL\\1015 Lausanne, Switzerland}

			\normalsize
		}
		
	\end{center}

    \vspace{3mm}
	\begin{center}
		{ \small \texttt{ sylvain.lacroix@sorbonne-universite.fr ,\\  n.j.levine@uva.nl  \ \  , \ \   anders.heide.wallberg@cern.ch
  } 
		}
		\\
	\end{center}
	
	\vspace{3mm}
	
	\begin{abstract}
 \vspace{2mm}
 Large families of integrable 2d $\sigma$-models have been constructed at the classical level, partly motivated by the utility of integrability on the string worldsheet. It is natural to ask whether these theories are renormalisable at the quantum level, and whether they define quantum integrable field theories. By considering examples, a folk theorem has emerged: the classically integrable $\s$-models always turn out to be renormalisable, at least at 1-loop order. We prove this theorem for a large class of models engineered on surface defects in the 4d Chern-Simons theory by Costello and Yamazaki.
 We derive the flow of the `twist 1-form' (a 4d coupling constant that distinguishes different 2d models), proving earlier conjectures and extending previous results. Our approach is general, using the `universal' form of 2d integrable models' UV divergences in terms of their Lax connection and reinterpreting the result in the language of 4d Chern-Simons. These results apply equally to rational, trigonometric and elliptic models.
	\end{abstract}
	\vspace{2in}

	\newpage
	
	{
		\setlength{\parskip}{0.05in}
		\tableofcontents
		\renewcommand{\baselinestretch}{1.0}\normalsize
	}
	
	
	\setlength{\parskip}{0.1in}
 \setlength{\abovedisplayskip}{15pt}
 \setlength{\belowdisplayskip}{15pt}
 \setlength{\belowdisplayshortskip}{15pt}
 \setlength{\abovedisplayshortskip}{15pt}

\normalsize

\section{Introduction}
Integrable 2d $\sigma$-models play an important role in various domains of physics. For instance, in string theory, they have been crucial in our understanding of the AdS/CFT correspondence---see the reviews \cite{Beisert,AF,Bombardelli:2016rwb,vanTongeren:2013gva}. In that context, one is typically interested in conformally invariant $\s$-models, but it is also important and instructive to consider RG flows between or away from these fixed points. Various non-conformal integrable $\sigma$-models with either massive or massless infrared phases have found interesting applications in condensed matter physics~\cite{Haldane:1982rj,tsvelik2007quantum} and as toy models to explore non-perturbative aspects of QFT~\cite{Polyakov:1975rr,Hasenfratz:1990zz,Marino:2019eym}. It is therefore natural to try to explore and understand the space of all integrable $\sigma$-models, their RG flows and their conformal fixed-points. 

For the purposes of this paper, we will restrict to \textbf{bosonic 2d $\sigma$-models}, defined by the action\footnote{Here $x^a=(t,x)$ are the 2d coordinates, $\eta={\rm diag}(-,+)$ is the 2d Minkowski metric, $\epsilon_{ab}$ is the 2d Levi-Civita tensor with $\epsilon_{01}=1$, and $x^\pm = \ha(t\pm x)$ are the 2d lightcone directions.} 
\begin{align}
S = \frac{1}{4\pi \a'} \int \ud^2 x\, \mathcal L \ , \quad \qquad \mathcal L &= - (G_{mn}(X) \eta^{ab} + B_{mn}(X) \epsilon^{ab}) \, \del_a X^m \, \del_b X^n  \\
&= (G_{mn}(X) + B_{mn}(X)) \, \del_+ X^m \, \del_- X^n \no\,.
\end{align}
Here, the 2d scalar fields $X^m$ are the coordinates on a target space, which is equipped with a metric $G_{mn} = G_{(mn)}$ and a 2-form B-field $B_{mn}=B_{[mn]}$. We will be particularly interested in the \textbf{classically integrable} $\sigma$-models, corresponding to specific choices of metric and B-fields for which the theory admits an infinite number of symmetries.

These 2d $\sigma$-models generally have UV divergences that trigger an RG flow \cite{Friedan1,Friedan2,Ecker:1971xko,extra1, extra2}:
\be
\ddt (G_{mn} + B_{mn}) = R_{mn}-\tfrac{1}{4}(H^2)_{mn} - \ha (\nabla H)_{mn} + \ldots \ . \la{ric}
\ee
Here we indicate only the 1-loop flow, which is a generalisation of the famous Ricci flow to include the antisymmetric B-field (with field strength $H=\ud B$). This should be understood as a flow in a space of infinitely many couplings, the space of metrics and B-fields. 

The special $\sigma$-models that are \textbf{integrable} are expected to define consistent, self-contained quantum theories. In particular, they should represent consistent truncations of this flow, and thus should be renormalisable. One may hope to understand this property
as resulting from their `hidden' integrable symmetries protecting their structure under renormalisation. 
In the semiclassical limit, this yields the following conjectural statement~\cite{Fateev:1992tk,Fateev:1996ea,Lukyanov:2012zt}:

\vspace{-0.1 cm}
\hspace{0.05 cm}
\boxed{\hspace{0.05 cm}
\begin{minipage}[t]{3cm}
\vspace{0.1 cm}
\textbf{Folk theorem:} 
\vspace{0.1 cm}
\end{minipage}
\begin{minipage}[t]{\textwidth-4.9cm}
\vspace{0.1 cm}
    \textit{The class of classically integrable $\sigma$-models is stable under the 1-loop RG flow \rf{ric}.}
    \vspace{0.1 cm}
\end{minipage}\hspace{0.1 cm}}
\vspace{0.2cm}

\noindent Mathematically speaking, this seems rather non-trivial: it implies the existence of a non-trivial subspace --- among the infinite-dimensional space of geometries --- that is closed under the (generalised) Ricci flow. In recent years, several authors have studied the renormalisation of integrable $\sigma$-models, observing this conjecture to be true without exception \cite{extra3,Squellari:2014jfa,Itsios:2014lca,extra4, extra5, Demulder:2017zhz,Georgiou:2018gpe,Klimcik:2019kkf,Delduc:2020vxy,Hassler,LT,Hassler:2023xwn,LW2,Kotousov:2024lzg}.\footnote{See also the following studies extending these statements to higher-loop orders \cite{HLT1,HLT2,Georgiou:2019nbz,Schubring:2020uzq,Hassler:2020wnp,Hassler,LT,Alfimov:2021sir,Alfimov:2023evq,Bykov:2023klm}, as well as related earlier works~\cite{Kutasov:1989dt,Gerganov:2000mt}.} As well as our direct interest in integrable models, they are therefore a particularly simple class of $\sigma$-models, often with finitely many couplings, in which to study the RG flow and its fixed points.

We will prove this conjecture for a large class of integrable $\sigma$-models. First, however, we need to define exactly what is meant by `classically integrable'. A typical criterion for classical integrability is that the equations of motion can be written in \textbf{zero-curvature} form,
\begin{equation}
[\del_a + L_a, \del_b + L_b] = 0 \ , \la{ZCE}
\end{equation}
in terms of some \textbf{Lax connection} $L_a(X,z)$, which depends on the fields $X$ and meromorphically on an auxiliary `spectral' parameter $z$. This ensures the existence of infinitely many conserved charges, extracted from the monodromy of $L(X,z)$. In addition, one asks that these charges pairwise Poisson-commute, which is ensured by requiring that the Poisson brackets of the spatial component $L_x(X,z)$ are given by a so-called Maillet bracket. For the purposes of this article, we will take the existence of such a Lax connection as the definition of classical integrability.

In recent years, two \textbf{unifying frameworks} have been proposed to study the space of integrable $\s$-models at the classical level. They encompass all examples known to us and have allowed for the systematic construction of many new ones. The first, phrased in the Hamiltonian language, is the formalism of affine Gaudin models~\cite{Levin:2001nm,Feigin:2007mr,Vicedo:2017cge}; the second, defined in a Lagrangian setup, is the 4d Chern-Simons theory~\cite{CWY1,CY3,unif}. In fact, these two frameworks are closely related~\cite{Vicedo} but, in this work, we will focus on the \textbf{4d Chern-Simons theory}. This unifying 4d gauge theory, 
\be \label{4dIntro}
S_{4d}[A] = \frac{1}{16\pi^2 i} \int \omega \wedge  {\rm CS}[A] \ , \qquad \omega = \varphi(z) \, \ud z \ , 
\ee
is designed to engineer integrable 2d models on certain surface defects in 4d. The 4d space consists of the 2d space, plus the complex spectral parameter. The choice of integrable 2d model is then predominantly specified by the meromorphic 1-form $\omega$, which is like the `coupling constant' of the 4d theory. This object, called the \textbf{twist 1-form}, is allowed to have poles and zeros in the $z$-direction, which distinguishes those special points as 2d defects, where one needs to impose certain boundary conditions on the gauge field. The choice of those boundary conditions is then the other main ingredient defining the theory. The fundamental fields $X^m$ of the 2d $\sigma$-model are extracted from the defect degrees of freedom located at the poles of $\omega$. Similarly, the zeros of $\omega$, known as disorder defects, become the poles of the 2d Lax connection $L_a(X,z)$ and thus directly appear in the integrable structure of the model. In this construction, the complex spectral parameter can be valued either in the Riemann sphere, yielding rational integrable $\sigma$-models; on the torus, producing elliptic theories; or on higher-genus surfaces. The detailed classical construction of these models with `disorder' defects is owing to \cite{CY3,unif,Vicedo,Benini,Lacroix:2020flf,Liniado:2023uoo,LW1,Derryberry:2021rne} (see also~\cite{Lacroix:2021iit} and references therein for a review).

We will prove the folk theorem above for a large class of integrable $\sigma$-models derivable from 4d Chern-Simons theory with disorder defects. Specifically, we will argue that this class of theories is 1-loop renormalisable with the twist 1-form running as
\be
\boxed{\ddt \omega = \del_z \Psi(z) \, \ud z \ .} \la{tf}
\ee
Here $\Psi$ is a particular meromorphic function, determined by $\omega$ as the solution of an interpolation problem. Specifically, $\Psi$ is required to have the same pole structure as $\omega$ and to take specific values at its zeros (see equation \rf{psi} for details). The result \eqref{tf} then defines a consistent 1-parameter flow on the moduli space of meromorphic 1-forms with prescribed number and orders of poles and zeros. Let us briefly review its history. This formula was first conjectured to be the RG flow of rational integrable $\sigma$-models in \cite{Delduc:2020vxy}, based on various examples. This was later proven in~\cite{Hassler,Hassler:2023xwn} for theories with simple poles in the Lax connection and no gauge-symmetries, using the language of $\mathcal{E}$-models~\cite{Klimcik:1995dy,Lacroix:2020flf}. The proof for more general rational models has not appeared in the literature so far. Another related conjecture, proposed by Costello, was reported in the work~\cite{Derryberry:2021rne} of Derryberry. It expressed the RG flow as a remarkably simple flow of the periods of $\omega$, \textit{i.e.}\ its integrals over specific contours in the underlying Riemann surface. This proposal, based on more geometric considerations (and related to the so-called rel-flow~\cite{zorich2006flat,Bainbridge:2016hor}), applied also to higher-genus Riemann surfaces, therefore including elliptic models. More recently, it was proven~\cite{LW2} to be equivalent to the flow \eqref{tf}, with the function $\Psi$ introduced for rational models in~\cite{Delduc:2020vxy} and with an appropriate new definition of $\Psi$ for elliptic ones. In the latter case, the conjecture was then checked in a specific example~\cite{LW2} but remains open in general.

In this work, we will prove the formula \eqref{tf}, and thus the conjectures of~\cite{Delduc:2020vxy,Derryberry:2021rne}, in a very general case. Our result applies to both rational and elliptic models, with arbitrary pole structures for $\omega$ and the Lax connection---covering a large class of integrable $\s$-models. As mentioned above, these 2d theories are defined not just by the twist 1-form $\omega$ but also several other pieces of data, which essentially amount to a choice of boundary conditions for the 4d gauge field at the defects. A part of our work will also be to derive how these other data run under the RG flow. There are some known examples outside the reach of our analysis, for instance $\mathbb{Z}_T$-cosets~\cite{Eichenherr:1979ci,Young:2005jv}, higher-genus theories~\cite{CY3,Derryberry:2021rne} and models on flag manifolds~\cite{Bykov:2016rdv,Affleck:2021vzo}. While the latter two are beyond the scope of the present paper, we expect our approach to be applicable to $\mathbb{Z}_T$-cosets and their generalisations with only minor technical modifications. We refer to the Discussion for more details on these prospects. 

Our insight in proving the claim \eqref{tf} comes from a general approach to studying these integrable RG flows, independent of their origin from 4d Chern-Simons theory: the \textbf{`universal' 1-loop divergences} recently put forward in \cite{Levine,Levine4d}. The basic observation is that 1-loop divergences are determined by the classical equations of motion. In integrable 2d models, the equations of motion take a canonical zero-curvature form \rf{ZCE} in terms of the Lax connection: hence the claim is that the 1-loop divergences always take the same form when written in terms of the Lax connection. Concretely, under general assumptions, they are given by a universal path integral in terms of the zero-curvature equations following from the Lax connection.

This path integral is universal in the sense that it only depends on the positions and multiplicities of the poles of the Lax connection---not the particular form of the currents appearing as these poles' residues. Here, we will compute it in a very general case, finding that it takes the elegant form
\be
\ddt \widehat S^{(1)} = \frac{\cG}{2\pi} \, \int \ud^2 x \,  \oint_{{}_- \G_{+}} \frac{\ud z}{2\pi i} \, \langle L_+, \del_z L_-\rangle \ , \la{redi}
\ee
where ${}_- {} \G_{+}$ is a contour encircling only the poles of $L_+$ (and not those of $L_-$).
The contour integral over the spectral parameter gives a sum of residues, which sums up particular combinations of the currents in the Lax connection $L_\pm$ to produce a local 2d counterterm. Our general approach allows us to derive this result not only for rational models, but also for the elliptic ones. It is based on the observation that the pole structure of the Lax connection can be written in both cases in terms of an $\mathcal R$-matrix solving the classical Yang-Baxter equation. Formulated in this language---presumably related to the underlying integrable structure---our argument applies immediately to both the rational and elliptic cases, with the different $\mathcal R$-matrices appearing in each case. 

The universal counterterm \eqref{redi} is derived only from the existence of the zero curvature equation and minor technical assumptions on the Lax connection, independently of the details of the model. 
To prove renormalisability, we must show that this universal counterterm can be absorbed into a flow of the theory's classical couplings.
To do this, we will further leverage the models' origin from 4d Chern-Simons theory, by showing that \eqref{redi} can be rewritten as a local 4d counterterm
\be
\ddt \widehat S^{(1)} = -\frac{1}{16\pi^2i} \int  \Psi(z) \, \langle \text{F}[A], \text{F}[A] \rangle = -\frac{1}{16\pi^2i} \int  \Psi(z) \, \ud {\rm CS}[A] \ . \la{red2}
\ee
This expression is close to proving 1-loop renormalisability: after integrating by parts, one gets a term $\del_z \Psi(z) \wedge {\rm CS}[A]$ that can be absorbed as a flow of $\omega$ of the expected form \eqref{tf}, upon comparing with the classical action \rf{4dIntro}. However, making this statement precise requires additional care, as one needs to deal with various defect-localised terms arising from the flow of the poles of $\omega$ and of the boundary conditions imposed there. Similarly, in the elliptic case, one needs to take into account additional contributions stemming from the flow of the torus moduli. The careful treatment of these aspects is the last technical step required to prove \eqref{tf} and will be detailed in the main text. Although we rewrote the counterterm as a 4d integral, it is important to emphasise that, in this paper, we are studying the renormalisation of a 2d theory, and the theory that we quantise is the 2d one. The relation between this analysis and the quantisation of the 4d gauge theory has been discussed in~\cite{Levine4d}, where the equivalence between the classical 2d and 4d actions has been extended to the 1-loop effective actions. We will comment more in the Discussion section on the future prospects of quantisation directly in 4d.

\bigskip

 The paper is structured as follows. In Section \ref{sec:review}, we review the classical construction of rational integrable $\s$-models from 4d Chern-Simons theory with disorder defects. In Section \ref{sec:rational}, we use the `universal' computation of integrable models' 1-loop divergences to obtain a general expression for the log-divergent terms in the 1-loop effective action, written elegantly as a contour integral of the Lax connection. This computation is mostly independent of Section~\ref{sec:review}. We then recast this as a 4d counterterm, and show that it can be absorbed as a flow of the twist 1-form $\omega$ and the other data defining the theory. In Section \ref{sec:elliptic}, we review the construction of (equivariant) elliptic integrable field theories from 4d Chern-Simons. We then repeat and adapt the above analysis for the elliptic case, again proving 1-loop renormalisability. We discuss the results and future directions in Section \ref{sec:disc}. 

Various technical issues are confined to the appendices. In the main text, we focus on the case where the twist 1-form $\omega$ has simple, real poles; the general case is addressed in Appendix \ref{app:GenPoles} (and we stress that our results apply equally to that case). Appendix \ref{app:Integrals} explains the origin of certain distributional terms that arise when taking RG-derivatives of integrals of meromorphic quantities. As a supplement to Section \ref{sec:elliptic}, Appendix \ref{sec:RmatAppendix} contains a general version of the `universal' computation of 1-loop divergences starting from a Lax connection written in a general form in terms of a classical $\mathcal R$-matrix (which includes both the rational and elliptic cases). Finally, in Appendix \ref{app:PCM}, we exemplify our results in the familiar case of the Principal Chiral Model.

\section{Rational integrable \texorpdfstring{$\sigma$}{sigma}-models from 4d-CS theory}\label{sec:review}

Let us begin by reviewing the construction of a large class of rational integrable $\s$-models from 4d Chern-Simons theory. This construction was initiated in the important work~\cite{CY3} of Costello and Yamazaki and has been actively extended since then, leading to a rich variety of models. One of the main advantages of our approach to renormalisation is that it will apply uniformly to all of these theories: we will thus consider a quite general framework, following the recent works~\cite{Benini,Lacroix:2020flf,Liniado:2023uoo}. However, to make the presentation lighter, we will reserve the most general case for Appendix~\ref{app:GenPoles}, restricting to a simpler setup in the main text. We also refer the reader to the lecture notes~\cite{Lacroix:2021iit} for a more introductory review and further references.

\subsection{The 4d Chern-Simons theory}\label{subsec:4DCS}

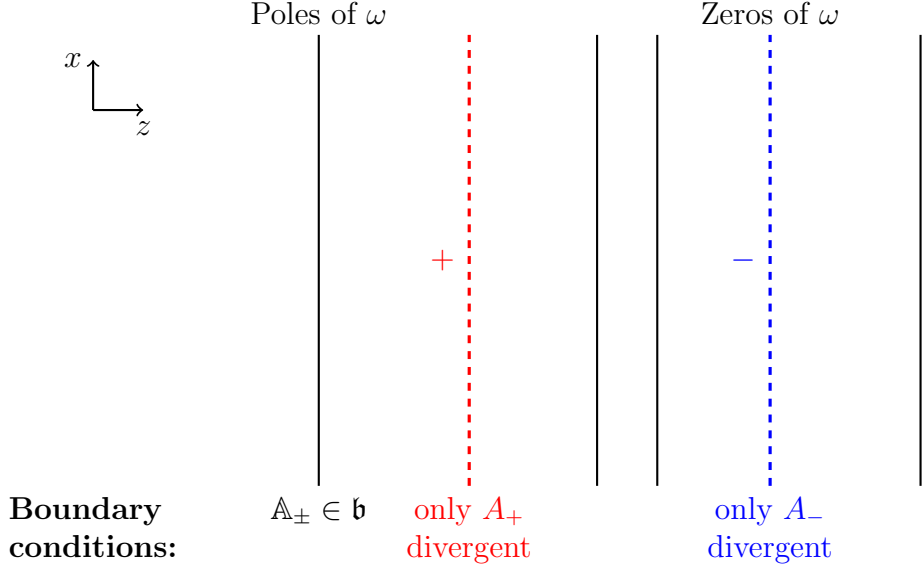
\begin{figure}[hb!]
\centering
\begin{tikzpicture}

  \def\ymin{0}
  \def\ymax{6}

  \draw[solid,thick]  (0,\ymin) -- (0,\ymax);
  \draw[dashed,very thick,red] (2,\ymin) -- (2,\ymax);
  \draw[solid,thick]  (3.7,\ymin) -- (3.7,\ymax);
  \draw[solid,thick]  (4.5,\ymin) -- (4.5,\ymax);
  \draw[dashed,very thick,blue] (6,\ymin) -- (6,\ymax);
  \draw[solid,thick]  (8,\ymin) -- (8,\ymax);

  \draw[->, thick] (-3,\ymax-1) -- ++(2/3,0) node[anchor=north] {$z$};
  \draw[->, thick] (-3,\ymax-1) -- ++(0,2/3) node[anchor=east]  {$x$};

  \node[anchor=south] at (0,\ymax) {Poles of $\omega$};
  \node[anchor=south] at (6,\ymax) {Zeros of $\omega$};

  \node[anchor=north, align=left] at (-3,\ymin) {\bf Boundary \\ \bf conditions:};
  \node[anchor=north] at (0,\ymin) {$\mathbb{A}_\pm\in\kf$};
  \node[anchor=north, align=center,red] at (2,\ymin) {only $A_+$\\divergent};
  \node[anchor=north, align=center,blue] at (6,\ymin) {only $A_-$\\divergent};

  \node at (1.95,\ymin+3) [anchor=east,red] {$+$};   
  \node at (5.95,\ymin+3) [anchor=east,blue] {$-$};   

\end{tikzpicture}

\caption{A cartoon of the defining data of the 4d Chern-Simons theory on $\Sigma\times \mathbb{CP}^1$. The vertical ($x$) direction depicts the 2d manifold $\Sigma$, while the horizontal ($z$) direction is the spectral $\mathbb{CP}^1$. The meromorphic twist 1-form $\omega$ is defined by its poles (solid lines) and zeros (dashed lines). These marked points in the spectral variable define 2d defects where certain boundary conditions must be imposed. The zeros of $\omega$ are so-called `disorder defects', where the 4d gauge field may become singular; they split into two subsets $\Zh^+$ (in red) and $\Zh^-$ (in blue) where the light-cone components $A_+$ and $A_-$ may diverge. At the poles of $\omega$, where the 2d theory lives, the boundary condition is a choice of maximally isotropic subalgebra $\kf$ of the defect algebra $\df$, where the jet of the 4d gauge field must lie.\label{fig:cartoon}}
\end{figure}

\paragraph{General setup.} Our aim in this section is to engineer integrable $\s$-models in 2 dimensions from a 4-dimensional gauge theory. The 2d worldsheet of these $\s$-models will be denoted by $\Sigma$, with light-cone coordinates $(x^+,x^-)$ and a locally flat metric $\ud s^2 = \ud x^+ \ud x^-$. The integrability of these theories will be encoded in the flatness of a Lax connection $L_\pm(z)$, valued in a complexified Lie algebra $\gf^\C$ and depending rationally on a complex spectral parameter $z\in\CP$ (as well as smoothly on the 2d fields). The central gauge theory will be defined on the 4d manifold $M=\Sigma\times \CP$, with coordinates $(x^+,x^-,z,\zb)$ that incorporate the spectral parameter as part of the geometry. As we will explain, each of the resulting integrable $\s$-models will be characterised by a choice $(\omega,\Zh^\pm,G,\kf)$ of the following ingredients:\vspace{-7pt}
\begin{itemize}\setlength\itemsep{1pt}
    \item a meromorphic 1-form $\omega=\vp(z)\,\dd z$ on $\CP$ called the \textit{twist 1-form} ;
    \item a partition of the zeros of $\omega$ into two subsets $\Zh^+$ and $\Zh^-$ ;
    \item a simple, real Lie group $G$ with Lie algebra $\gf$ ;
    \item a maximally isotropic subalgebra $\kf$ of the `defect algebra' associated to $(\omega,G)$.\vspace{-4pt}
\end{itemize}
Surface defects will be located in the 4d space at the poles and zeros of the twist 1-form $\omega$, and their positions will be the main continuous parameters of the resulting integrable 2d $\s$-models. In particular, the zeros $\Zh^\pm$ will correspond to the poles of the light-cone components $L_\pm(z)$ of the Lax connection and will thus play an important role in the integrable structure. The Lie group $G$ will be the gauge group of the 4d theory and will determine the Lie algebra $\gf^\C$ in which the 2d Lax connection is valued. The final ingredient is more subtle and is related to the choice of boundary conditions imposed at the surface defects at the poles of $\omega$. See Figure \ref{fig:cartoon} for a schematic depiction of these defining data.

\paragraph{The twist 1-form.} The \textit{twist 1-form} $\omega$ is the most important ingredient, so we shall describe it in some detail. Locally, it can be written as $\omega = \vp(z)\,\dd z$, where $\vp(z)$ is a rational function of the spectral parameter $z$, often called the `twist function'. We will denote the set of \textit{poles} of $\omega$ as
\begin{equation}
    \Ph = \lbrace p_1, \dots, p_N \rbrace \ .
\end{equation}
For simplicity in the main text, we will restrict ourselves to the case of simple poles on the real axis: the general case, with higher-order and complex poles, will be discussed briefly at the end of this subsection and in detail in Appendix~\ref{app:GenPoles}. This restriction is purely for pedagogical reasons: we insist that all the results of this paper also apply to an arbitrary pole structure, as explained in the appendix.

We now consider the \textit{zeros} of $\omega$. We will assume them to be real and, for reasons that will become clear later in this section, separate them into two subsets
\begin{equation}\label{eq:Zeros}
    \Zh^+ = \bigl\lbrace z_1^+, \dots, z_{N_+}^+ \bigr\rbrace\,, \qquad  \Zh^- = \bigl\lbrace z_1^-, \dots, z_{N_-}^- \bigr\rbrace\,.
\end{equation}
Furthermore, we denote by $m^\pm_i \in \Z_{\geq 1}$ the multiplicity of the zero $z_i^\pm$, which is allowed to be arbitrary. We will demand that there are as many zeros in $\Zh^+$ as in $\Zh^-$ when counted with multiplicities, \textit{i.e.}\ that
\begin{equation}\label{eq:M1}
    \sum_{i=1}^{N_+} m_i^+ = \sum_{i=1}^{N_-} m_i^- \equiv M\,, 
\end{equation}
with the last equality defining $M$. By the Riemann-Hurwitz theorem, we then have
\begin{equation}\label{eq:M2}
    N = 2M+2\,,
\end{equation}
where $N$ is the number of poles. In particular, note that this forces $\omega$ to have an even number of poles and of zeros (counted with multiplicities).\\

We can write $\omega$ in a factorised form, in terms of its poles $p_r$, its zeros $z_i^\pm$ and a real proportionality factor $K$,\footnote{Here, we assumed that the points $p_r$ and $z_i^\pm$ are all in the finite complex plane $\C \subset \CP$, which can always be ensured by a Möbius transformation of $z$.}
\begin{equation}\label{eq:OmegaFac}
    \omega = K \frac{\prod_{i=1}^{N_+} (z-z_i^+)^{m_i^+} \prod_{i=1}^{N_-} (z-z_i^-)^{m_i^-}}{\prod_{r=1}^{N} (z-p_r)}\, \dd z \, ,
\end{equation}
so that it is parametrised by the numbers $(K,p_r,z_i^\pm)$. Note that we can eliminate $3$ of these parameters by Möbius transformations of $z$. For fixed numbers and multiplicities of the poles and zeros, and taking into account this redundancy,  $\omega$ has $N+N_++N_--2$ continuous, real parameters. We finally define the \textit{levels} of the model as the residues\footnote{\label{foot:WZ}These levels will eventually be related to coefficients of Wess-Zumino terms in the resulting 2d $\s$-models.}
\begin{equation}\label{eq:Levels}
    \ell_r = \res_{z=p_r} \,\omega\,,
\end{equation}
which are, by construction, rational functions of the parameters $(K,p_r,z_i^\pm)$.

\paragraph{The 4d gauge theory.} As mentioned at the beginning of this subsection, part of the defining data of the 4d Chern-Simons theory is the choice of a simple, connected gauge group $G$, with Lie algebra $\gf$. The fundamental field of this theory is then a gauge field
\begin{equation}
    A = A_+\,\dd x^+ +  A_-\,\dd x^- +  A_z\,\dd z +  A_{\zb}\,\dd \zb \,,
\end{equation}
defined as a 1-form on $M=\Sigma\times\CP$ valued in the (complexified) gauge algebra $\gf^\C$. As usual, the Chern-Simons 3-form of $A$ is defined as
\begin{equation}
    \label{eq:CSA}\text{CS}[A] = \left\langle A, \dd A + \frac{2}{3} A \wedge A \right\rangle\,.
\end{equation}
Here, $\langle\cdot,\cdot\rangle$ is the invariant bilinear form\footnote{\label{fn:Form}Since $G$ is simple, the invariant bilinear form $\langle\cdot,\cdot\rangle$ coincides with the Killing form up to a proportionality factor. Throughout this paper, we will normalise it so that
\begin{equation}
    \langle X,Y \rangle = -\frac{1}{2\cG} \Tr_{\gf}\bigl[\ad_{X}\, \ad_Y \bigr]\,, \qquad \forall \, X,Y\in\gf^\C\,, \no
\end{equation}
where $\cG \in \Z_{\geq 1}$ is the dual Coxeter number of $G$. The minus sign ensures that $\langle \cdot,\cdot \rangle$ is positive-definite if $G$ is compact.} on $\gf^\C$ and an exterior product $\wedge$ is implicit when taking pairings of $\gf^\C$-valued forms. The 4d Chern-Simons theory is then defined by the action~\cite{Nekrasov,Costello1,CY3}
\begin{equation}\label{eq:Action4d}
    S_{4d}[A] = \frac{1}{16 \pi^2 i} \int_{M} \omega \wedge \text{CS}[A]\,.
\end{equation}
We refer to the appendix \ref{app:Int} for a more detailed discussion of how such integrals against meromorphic 1-forms are defined and computed. We note that the presence of $\omega = \vp(z)\,\dd z$ means that the only component of the 3-form $\text{CS}[A]$ contributing to the action is the one parallel to $\dd \zb \, \wedge \, \dd x^+ \wedge \dd x^-$. As a consequence, the $A_z$ component of the gauge field completely decouples from the theory: it is therefore unphysical and can be set to any value. The relevant dynamical degrees of freedom (subject to further gauge symmetry) are thus the remaining components $(A_+,A_-,A_{\zb})$.

To ensure the reality of the action \eqref{eq:Action4d}, we impose that these components are equivariant with respect to complex conjugation on $\CP$ and $\gf^\C$, \textit{i.e.} that
    \begin{equation}\label{eq:RealityA}
    A_\mu(x^\pm,\zb,z) = \tau\bigl( A_\mu(x^\pm,z,\zb) \bigr) \ , \qquad\mu=+,-,\bar z\  .
\end{equation}
Here, $\tau : \gf^\C \to \gf^\C$ is the antilinear involutive automorphism of $\gf^\C$ induced by complex conjugation, whose fixed-point subalgebra is the real form $\gf = \lbrace X\in \gf^\C \; | \; \tau(X) = X \bigr\rbrace$.

\paragraph{Disorder defects.} Since the 1-form $\omega$ has zeros at $\Zh^\pm = \lbrace z_i^\pm \rbrace_{i=1}^{N_\pm}$, then the action can remain finite even if the gauge field diverges at these points. For the action to be finite, the different components of the gauge field should not blow up too much at the same time, so we demand the following pattern of singularities:
    \begin{equation}\label{eq:Disorder}
        A_\pm = O\left( \frac{1}{(z-z_i^\pm)^{m_i^\pm}} \right)\,.
    \end{equation}
In other words, the two sets of zeros $\Zh^+$ and $\Zh^-$ are distinguished by the respective light-cone components $A_+$ and $A_-$ blowing up.\footnote{Note that the reason why we distributed these singularities along the light-cone components of $A$ is because we are eventually interested in relativistic $\s$-models. We expect that similar requirements can also be found that lead to Euclidean theories, using the components of $A$ along complex coordinates on $\Sigma$. More generally, one could consider an even larger class of theories, neither relativistic or Euclidean, corresponding to singularities distributed along arbitrary directions in $\Sigma$. We believe that the methods developed in this paper to study the 1-loop RG flow of the relativistic models could also apply to this more general class, but leave a thorough analysis of this question for the future.} Otherwise, we ask that the gauge field be smooth and regular on $M=\Sigma\times\CP$.
The singularities \eqref{eq:Disorder} will play an important role in the resulting 2d integrable model and are called \textit{disorder defects}. 

\paragraph{Variation and equations of motion.} Let us now consider the variational problem associated with the action \eqref{eq:Action4d}. Under infinitesimal transformations $\delta A$ of the gauge field, one finds
\begin{equation}\label{eq:varA}
    \delta S_{4d} = \frac{1}{8\pi^2 i} \int_M \langle \omega \wedge \text{F}[A], \delta A \rangle + \delta S^{\text{defect}}_{4d}+\delta S^{\text{bdry}}_{4d}\,,
\end{equation}
where $\text{F}[A]=\dd A + A\wedge A$ is the curvature of $A$, and the last two terms are given by
\begin{subequations}
\begin{equation}\label{eq:DeltaDefect}
    \delta S^{\text{defect}}_{4d} = - \frac{1}{16\pi^2 i} \int_M \dd\omega \wedge \langle A,\delta A \rangle\,,
\end{equation}
\begin{equation}
    \delta S^{\text{bdry}}_{4d}= \frac{1}{16\pi^2i}\int_M\dd \left(\omega\wedge\left<A,\delta A\right>\right)\,.\label{eq:DeltaBoundary}
\end{equation}
    \end{subequations}
Since $\omega = \varphi(z) \, \ud z$ is a meromorphic 1-form, its exterior derivative $\dd\omega= \del_{\bar z} \varphi(z) \, \ud\bar z \wedge \ud z$ is a distribution on $\CP$ supported on the poles $\Ph$ (see appendix \ref{app:Int} for more details). The 4d integral in equation \eqref{eq:DeltaDefect} thus localises to a surface integral on the poles defect $\Sigma \times \Ph$. This defect term requires a careful treatment and will be the subject of its own paragraph later. Similarly, the integral in \eqref{eq:DeltaBoundary} localises to the boundary of $M=\Sigma\times\mathbb{CP}^1$: since the boundary $\partial\mathbb{CP}^1$ is empty and we will assume that $A$ decreases sufficiently fast towards the boundary of $\Sigma$, this term can be discarded.
In contrast, the first term in the variation \eqref{eq:varA} is a proper 4d integral and thus corresponds to a bulk contribution.
For the action to be extremised, this term must vanish for all $\delta A$, giving the bulk equation of motion
\begin{equation}
    \omega \wedge \text{F}[A] = 0\,.\label{eq:4DCSEOM}
\end{equation}
This is almost a zero curvature equation for the gauge field $A$, as in standard Chern-Simons theory. The difference here is the presence of an additional fourth direction, tempered by the exterior product with $\omega=\vp(z)\,\dd z$, which removes the $A_z$-component of the gauge field and all $\partial_z$-derivatives from the equation. We thus see that---away from zeros of $\omega$---this equation amounts to the flatness of the gauge field $\text{F}_{\mu\nu}[A]=0$ along the directions $\mu,\nu=(x^+,x^-,\zb)$  (but not along the $z$-direction).

\paragraph{Defect term from jets.} Let us now discuss the defect term \eqref{eq:DeltaDefect} in detail. As mentioned earlier, the derivative $\dd\omega$ is a distribution, more precisely a sum of Dirac distributions localised at the poles $\Ph=\lbrace p_r \rbrace_{r=1}^N$ of $\omega$ (see appendix \ref{app:Int} and equation \eqref{eq:dOmDist} for more details). The integration over $z\in\CP$ in \eqref{eq:DeltaDefect} can thus be performed explicitly, yielding a surface integral along $\Sigma$. One then finds
\begin{equation}\label{eq:DeltaDefect2}
    \delta S^{\text{defect}}_{4d} = -\frac{1}{8\pi}\sum_{r=1}^N\; \ell_r\, \int_\Sigma \,\langle A,\delta A \rangle |_{z=p_r}\,,
\end{equation}
where $\{\ell_r\}_{r=1}^N$ are the levels \eqref{eq:Levels}. Note that the 2-form $\langle A,\delta A \rangle$ is integrated along the surface $\Sigma \times \lbrace p_r\rbrace$, so only the spatial components $A_\pm$ of the gauge field contribute to the above expression. This formula naturally leads us to consider the so-called \textit{jet} of $A_\pm$ along the pole defect $\Sigma\times \Ph $, defined as
\begin{equation}\label{eq:Jet}
 \Ad_\pm = \bigl( A_\pm \bigl|_{z=p_r}\bigr)_{r=1,\dots,N} \qquad \in \qquad \df = \gf^N\,.
\end{equation}
This object is just the set of evaluations of $A_\pm$ at the poles $\Ph$. Due to the reality condition \eqref{eq:RealityA} and our assumption that the poles $p_r$ are real, each of these evaluations is valued in the real Lie algebra $\gf$. The jet $\Ad_\pm$ is thus naturally interpreted as an element of the direct product $\df = \gf^N$, which we call the \textit{defect Lie algebra}. The latter will play a crucial role in the rest of this section. It has dimension
\begin{equation}\label{eq:DimD}
    \dim\df = 2(M+1)\,\dim\gf\,,
\end{equation}
with $M$ as in equations \eqref{eq:M1} and \eqref{eq:M2}. Moreover, it is naturally equipped with a Lie bracket
\begin{equation}\label{eq:LieDefectSimple}
     [ \Xd, \Yd ]_{\df} = \bigl(  [X_r, Y_r] \bigr)_{r=1,\dots,N}\,,
\end{equation}
where $\Xd = (X_r)_{r=1,\dots,N}$ and $\Yd = (Y_r)_{r=1,\dots,N}$ are generic elements of $\df=\gf^N$. The relevance of this Lie algebraic structure will be made clear later when discussing gauge symmetries.

We also introduce the notation $\Ad = \Ad_+ \, \dd x^+ + \Ad_- \, \dd x^-$, which is a $\df$-valued 1-form on $\Sigma$. The defect term \eqref{eq:DeltaDefect2} can then be compactly rewritten as
\begin{equation}\label{eq:DefectJet}
    \delta S^{\text{defect}}_{4d} = -\frac{1}{8\pi} \int_\Sigma \,\dlangle \Ad,\delta \Ad \drangle_{\df}\,,
\end{equation}
where we have defined a symmetric bilinear form $\dlangle \cdot, \cdot \drangle_{\df}$ on the defect algebra by
\begin{equation}\label{eq:FormDefectSimple}
    \dlangle \Xd, \Yd \drangle_{\df} = \sum_{r=1}^N \ell_{r} \, \langle X_r, Y_r \rangle\,.
\end{equation}
One easily checks that $\dlangle\cdot,\cdot\drangle_{\df}$ is non-degenerate and ad-invariant with respect to the Lie bracket \eqref{eq:LieDefectSimple}.

\paragraph{Boundary condition and the maximally isotropic subalgebra.} Having rewritten the defect term in the compact form \eqref{eq:DefectJet}, we can return to the analysis of the variation of the action \eqref{eq:varA}. For the variational problem to be well defined, this defect term needs to vanish. This will be arranged by imposing a \textit{boundary condition} on the gauge field's jet $\Ad_\pm$. Recall that the latter was defined as a field in the defect algebra $\df$. We will demand that it takes values in a \textit{maximally isotropic subalgebra} $\kf$ of $\df$:
\begin{equation}\label{eq:BC}
        \Ad_\pm  \; \in \; \kf \,.
\end{equation}
Different choices of $\kf$ correspond to different boundary conditions, and such a choice is the last ingredient defining the 4d Chern-Simons theory.

More concretely, an isotropic subalgebra of the defect algebra is a subspace $\kf \subset \df$ satisfying the following conditions:
\begin{equation}
    [\Xd,\Yd]_{\df} \in \kf \qquad \text{ and } \qquad \dlangle \Xd, \Yd \drangle_{\df} = 0\,, \qquad \text{ for all }\,\Xd,\Yd \in \kf\,.
\end{equation}
Clearly the second property, \textit{i.e.}\ the isotropy of $\kf$, combined with the boundary condition \eqref{eq:BC}, implies the vanishing of the defect term \eqref{eq:DefectJet}, as desired. The first property---that $\kf$ should also close under the Lie bracket---will be justified in the next paragraph and has to do with gauge symmetries. We finally ask that $\kf$ is of maximal dimension for an isotropic subspace of $\df$, \textit{i.e.} that it is half-dimensional:
\begin{equation}\label{eq:dimK}
     \dim\kf = \frac{1}{2} \dim\df = (M+1)\dim\gf\,.
\end{equation}
This assumption will be important later on when we will use the boundary condition to relate the 4d gauge field to 2d fields localised on the defects.

\paragraph{Gauge symmetry.} The theory is gauge-invariant under
local transformations by a parameter $u(x^+,x^-,z,\, \zb)$, which is a smooth function on $M$ valued in the gauge group $G^\C$. The gauge field transforms as a connection, \textit{i.e.}
\begin{equation}\label{eq:GaugeTransf}
    A \longmapsto A^u = u^{-1}A\,  u +  u^{-1}\dd u\,.
\end{equation}
In order to preserve the reality condition \eqref{eq:RealityA} imposed on $A$, we ask that the gauge parameter satisfies a similar condition $u(x^\pm,\zb,z) =\tau\bigl(  u(x^\pm,z,\zb) \bigr)$, where $\tau$ is the involutive automorphism of $G^\C$ characterising its real form $G$.

It is well-known that the curvature of $A$ is covariant under gauge transformations: $\text{F}[A^u]=u^{-1}\text{F}[A]u$. The equation of motion \eqref{eq:4DCSEOM} is thus clearly gauge-invariant. To be true gauge symmetries of the model, they also need to preserve its action \eqref{eq:Action4d}. This is more subtle: we summarise the main ideas here and refer to~\cite{Benini} for a detailed analysis. A direct computation shows that the gauge-transformed Chern-Simons form $\text{CS}[A^u]$ is equal to $\text{CS}[A]$ plus a closed 3-form. While this yields the same bulk equation of motion for $A$, it implies that $S_{4d}[A^u]$ is equal to $S_{4d}[A]$ plus a defect term located at the poles defect $\Sigma \times \Ph$, which is non-vanishing in general. This is in fact to be expected: recall that we imposed the boundary condition \eqref{eq:BC} on the configurations of the gauge field at these poles. The gauge symmetries therefore need to preserve this boundary condition. Crucially, this will also ensure that no defect terms arise in the gauge transformation of the action.

Recall that the boundary condition \eqref{eq:BC} was phrased in terms of the jet $\Ad_\pm$ of the gauge field at the poles $\Ph$, defined in equation \eqref{eq:Jet}. The gauge transformation $A \mapsto A^u$ acts on this jet as
\begin{equation}\label{eq:GaugeJet}
 \Ad_\pm \longmapsto \Ad_\pm^{\uj} = \uj^{-1}\Ad_\pm\, \uj + \uj^{-1}\partial_\pm \uj\,,
\end{equation}
where $\uj = ( u|_{z=p_r} )_{r=1}^N$ is the jet of the gauge parameter $u$. It is valued in the direct product $D=G^N$ of $N$ copies of the gauge group, which we interpret as the Lie group of the defect algebra $\df=\gf^N$. The equation \eqref{eq:GaugeJet} thus takes the form of a 2d gauge transformation of the jet $\Ad_\pm : \Sigma \to \df$ by the function $\uj : \Sigma \to D$. Note that, for this to make sense, it was necessary to establish a Lie group structure on $D$ and thus a Lie bracket \eqref{eq:LieDefectSimple} on the defect algebra $\df$. The boundary condition \eqref{eq:BC} that $\Ad_\pm$ be valued in the maximal isotropic subalgebra $\kf$ of $\df$ is clealy preserved under the gauge transformation \eqref{eq:GaugeJet} if we impose the similar boundary condition
\be
\uj \in B \ , \la{gbc}
\ee
on the gauge parameter (where $B$ is the unique closed subgroup of $D$ such that $\text{Lie}(B)=\kf$).\footnote{By a slight abuse of terminology, we will sometimes refer to $B$ as a maximal isotropic subgroup of $D$.} This is why we asked earlier for $\kf$ to be a subalgebra of $\df$.\footnote{If $\kf$ was only an isotropic subspace but not a subalgebra, the preservation of the boundary condition $\Ad_\pm\in\kf$ would impose stronger conditions on $\uj$, as for instance being valued in the trivial subgroup of $D$ rather than a half-dimensional one $B$. This would lead to more complicated 2-dimensional $\s$-models, with less `transparent' integrable structures. We refer to the recent work~\cite{Cole:2023umd} for more details on these cases.} From now on, we will reserve the term \textit{gauge symmetry} for a transformation \eqref{eq:GaugeTransf} whose gauge parameter satisfies the boundary condition \eqref{gbc}. As shown in~\cite[Theorem 4.2]{Benini}, the action \eqref{eq:Action4d} is indeed invariant under this class of transformations. In contrast, we shall refer to transformations \eqref{eq:GaugeTransf} by a parameter $u$ not respecting the boundary condition as \textit{formal gauge transformations}: these generally do not preserve the action and are not physical gauge symmetries.

\paragraph{General pole structure.} In the main text, we are making the assumption that the poles $\Ph = \lbrace p_r \rbrace_{r=1}^N$ of $\omega$ are real and simple. We end this subsection with a brief glimpse of the case of general pole structures, which is discussed in detail in Appendix \ref{app:GenPoles}. If $p_r$ is a pole of order $m_r\in \Z_{\geq 1}$, the defect term \eqref{eq:DeltaDefect} involves not only the evaluation $A_\pm|_{z=p_r}$ of the gauge field 
at $z=p_r$, but also 
its $z$-derivatives $\partial_z^k A_\pm|_{z=p_r}$ of order $k\leq m_r-1$. This naturally leads to a generalised definition for the jet \eqref{eq:Jet} of $A$ as $\Ad_\pm = \bigl( \partial_z^k A_\pm|_{z=p_r} \bigr)_{r=1,\dots,N}^{k=0,\dots,m_r-1}$. The latter then has to be interpreted as an element of a more complicated defect algebra $\df$ than the one $\gf^N$ considered in the case of simple real poles. In particular, the Lie bracket and invariant pairing of this defect algebra then take the form \eqref{eq:LieDefect} and \eqref{eq:FormDefect}, which are intricate generalisations of \eqref{eq:LieDefectSimple} and \eqref{eq:FormDefectSimple}. In this setup, poles of higher order are associated with so-called \textit{Takiff algebras} and complex poles always come in pairs of conjugates, associated with complexified algebras rather than real ones. The crucial property of this construction is that with the defect algebra generalised in this way, the rest of this section stays exactly the same. We insist that all the results of this paper also apply to this general case and refer to the Appendix \ref{app:GenPoles} for detailed explanations.

\subsection{Extracting the 2d integrable \texorpdfstring{$\sigma$}{sigma}-model}
\label{subsec:int}

In the previous subsection, we defined the 4d Chern-Simons theory associated with the data $(\omega,\Zh^\pm,G,\kf)$. We now explain how this theory localises classically to a 2d integrable $\sigma$-model. While the derivation is classical, and involves solving some of the 4d theory's equations of motion, note that the relation between the 2d and 4d theories was also argued to persist to the level of 1-loop effective actions \cite{Levine4d}.

\paragraph{Fields $\bm{\gh}$ and $\bm{L}$.} We start by re-parametrising the gauge field as
\begin{equation}\label{eq:AgL}
    A = \gh\, L\,\gh^{\,-1} - \dd\, \gh\,\gh^{\,-1}\,,
\end{equation}
in terms of a new group-valued field $\gh : M \to G^\C$ and a $\gf^\C$-connection
\begin{equation}
    L = L_+\,\dd x^+ + L_-\,\dd x^-
\end{equation}
with components only along the spatial directions $\Sigma$. We note that equation \eqref{eq:AgL} can be rephrased as $L=A^{\gh}$, using the notation \eqref{eq:GaugeTransf} for formal gauge transformations. In this sense, $L$ can be thought of as a `formal axial gauge' in which the $\dd z$- and $\dd\zb$-components are set to zero. We recall that the $\dd z$-component $A_z$ completely decouples from the theory and thus can be fixed to any given value: this freedom can always be used to ensure that $L=A^{\gh}$ has no $\dd z$-component. The part of \rf{eq:AgL} that defines $\gh$ is the $\dd\zb$-component, $\partial_{\zb}
\gh\,\gh^{\,-1}=-A_{\zb}$, which admits a solution for $\gh$.\footnote{Note however that $\gh$ is not uniquely determined. This will be discussed at the end of the subsection.} We note that the passage from $A$ to $L=A^{\gh}$ is only a formal gauge transformation: indeed, the function $\gh$ defined through the above differential equation will generally not satisfy the boundary condition \eqref{gbc} imposed on gauge parameters $u$ to define physical gauge symmetries. Thus, $A$ and $L=A^{\gh}$ are not physically equivalent configurations of the gauge field.

\paragraph{$\bm{L}$ is for Lax connection.} To interpret $L$, let us 
consider how it appears in the equations of motion. Since the bulk equation of motion \eqref{eq:4DCSEOM} is invariant under formal gauge transformations and $L=A^{\gh}$, it reduces to $\omega \wedge \text{F}[L] = 0$. Projecting this equation along $\dd z \wedge \dd x^+ \wedge \dd x^-$ and $\dd z \wedge \dd \zb \wedge \dd x^\pm$ yields
\begin{equation}\label{eq:EoML}
    \omega\bigl( \partial_+ L_- - \partial_- L_+ + \bigl[ L_+, L_- ] \bigr) = 0 \qquad \text{ and } \qquad \omega\,\partial_{\zb} L_\pm = 0\,.
\end{equation}
The simple form of the second equation relies on the defining property of $L$, namely that it has no $\zb$-component. This equation tells us that, away from the zeros $\Zh$ of $\omega$, $L_\pm$ is a holomorphic function of $z$. More precisely, as we will explain below, $L_\pm$ is a meromorphic function of $z\in\CP$ with poles at the points $\Zh$. On the other hand, the first equation in \eqref{eq:EoML} is the flatness of $L$ seen as a connection on the 2d-manifold $\Sigma$, for all $z\in\CP\setminus\Zh$. We thus see that $L$ can be interpreted as a connection on the 2d-manifold $\Sigma$, which is flat on-shell and depends meromorphically on an auxiliary complex parameter $z$. These are exactly the properties of a \textit{Lax connection} for an integrable 2d field theory on $\Sigma$, with $z \in \CP$ playing the role of the \textit{spectral parameter}. In 4d Chern-Simons theory, the two key features of this Lax connection---meromorphicity and flatness---thus arise from the 4d bulk dynamics. In the rest of this subsection, we will describe the resulting integrable 2d models in more detail.

\paragraph{Analytic structure of $\bm L$.} Let us describe more precisely how $L$ depends on $z$. As explained above, the second equation of motion in \eqref{eq:EoML} implies that $L$ is holomorphic in $z$ away from the zeros of $\omega$. In fact, by Liouville's theorem, any holomorphic function on all of $\CP$ is constant: so for $L_\pm$ to depend non-trivially on $z$, it must have poles at the zeros of $\omega$. Recall from equations \eqref{eq:Zeros} and \eqref{eq:Disorder} that we separated these zeros, the locations of the `disorder' defects, into two disjoint subsets $\Zh^\pm = \bigl\lbrace z_i^\pm \bigr\rbrace_{i=1}^{N_\pm}$, where the light-cone components $A_+$ and $A_-$ are respectively allowed to be singular. Since $A_\pm$ are related to $L_\pm$ by a formal gauge transformation by $\gh$, which is regular everywhere, $L_\pm$ can have poles of order $m_i^\pm$ at the points $z_i^\pm$, but not at the points $z_i^\mp$. The Lax connection $L_\pm$ thus takes the following meromorphic form:
\begin{equation}\label{eq:Lz}
    L_\pm(z) = \sum_{i=1}^{N_\pm} \sum_{k=0}^{m_i^\pm-1} \frac{\Jc_\pm^{(i,k)}}   {(z-z_i^\pm)^{k+1}} + \Jc_\pm^{(0,0)}\,,
\end{equation}
where $\Jc_\pm^{(i,k)} : \Sigma \to \gf$ are Lie algebra-valued\footnote{The fact that they are valued in the real form $\gf$ and not the complexified algebra $\gf^\C$ follows from the reality condition \eqref{eq:RealityA} on $A$ and our assumption that the zeros $z_i^\pm$ are real.} 
2d currents on $\Sigma$. This is the most general solution of the second equation of motion in \eqref{eq:EoML} compatible with the condition \eqref{eq:Disorder}. In particular, the zeros of $\omega$ and their splitting into the subsets $\Zh^\pm = \bigl\lbrace z_i^\pm \bigr\rbrace_{i=1}^{N_\pm}$ exactly control the respective poles of $L_+$ and $L_-$. This highlights their important role in the integrable structure of the model.

\paragraph{The fundamental 2d fields.} Let us now describe how the fundamental fields of the 2d model are extracted from the 4d setup. Recall that the 4d theory is invariant under gauge symmetries $A\mapsto A^u$, with $u$ satisfying appropriate boundary conditions \rf{gbc} at the poles of $\omega$. In terms of the new variables $(\gh,L_\pm)$, these gauge symmetries act as
\begin{equation}\label{eq:gLu}
    \gh \longmapsto u^{-1} \gh\, \qquad \text{ and } \qquad L_\pm \longmapsto L_\pm\,.
\end{equation}
In particular, the Lax connection $L$ is left invariant. In contrast, these transformations allow us to `gauge away' most of the degrees of freedom contained in $\gh$. In fact, one could naively think that choosing $u=\gh$ would eliminate all of these degrees of freedom; however, $\gh$ does not satisfy the boundary condition \rf{gbc} imposed on gauge parameters, so it does not define a proper gauge symmetry.
In contrast, gauge parameters $u$ with trivial evaluations at the poles $\Ph=\lbrace p_r \rbrace_{r=1}^N$ of $\omega$ correspond to true symmetries. We can thus use those to eliminate all the degrees of freedom in $\gh$ except for its jet along the poles defect:
\begin{equation}\label{eq:gJet}
    \gj = \bigl( \gh|_{z=p_r} \bigr)\null_{r=1}^N \;\, : \;\, \Sigma \; \longrightarrow \; D \,.
\end{equation}
The latter is naturally interpreted as a 2d field on $\Sigma$, valued in the defect group $D=G^N$.  It could equivalently be treated as an edge mode on the defect $\Sigma \times \Ph$ (see~\cite{Benini} for more details). 

The field $\gj$ is not yet completely gauge invariant, since the gauge freedom $\gh \mapsto u^{-1}\gh$ 
has a residual part 
$\gj \mapsto \uj^{-1} \gj$, where $\uj = (u|_{z=p_r})_{r=1}^N \in D$ is the jet of $u$ (and multiplication is that of the defect group $D$). 
The jet $\uj$ was constrained to be valued in the maximally isotropic subgroup $B$ of $D$, by the boundary condition \eqref{gbc} imposed on the gauge parameter $u$.
Hence the gauge-invariant field is the image of $\gj$ under the canonical projection $D \to B \hspace{-1pt}\setminus\hspace{-1pt} D$,
\begin{equation}\label{eq:gQuot}
    \gb  \;\, : \;\, \Sigma \; \longrightarrow \;  B \hspace{-1pt}\setminus\hspace{-1pt} D\,.
\end{equation}
We finally note that $\gb$ is not yet completely physical: it will be subject to one more 2d local symmetry, arising from a different mechanism, and which will be discussed at the end of this subsection.

\paragraph{Lax connection in terms of $\bm{\gb}$.} Having extracted the gauge-invariant 2d degrees of freedom contained in $\gh$, let us now return to the Lax connection $L_\pm$. As seen in equation~\eqref{eq:Lz}, it is described by the $\gf$-valued currents $\lbrace \Jc_\pm^{(i,k)} \rbrace_{i=0,\dots,N_\pm}^{k=0,\dots,m_i^\pm-1}$, where we define $m_0^\pm=1$. In total, each of $L_+$ and $L_-$ therefore contains the following number of components, by \eqref{eq:M1}:
\begin{equation}\label{eq:CompCurr}
    \left( \sum_{i=0}^{N_\pm} m_i^\pm \right) \dim\,\gf = (M+1)\dim\,\gf \ .
\end{equation}
 Importantly, these components are not additional physical degrees of freedom: rather, the boundary condition \eqref{eq:BC} will relate them to the field $\gj$ introduced above in equation \eqref{eq:gJet}. 
 This boundary condition involves the jet of $A_\pm$ at the poles $\Ph=\lbrace p_r \rbrace_{r=1}^N$, which reads
\begin{equation}\label{eq:ConstBC}
    \Ad_\pm = \gj\,\Ld_\pm\,\gj^{-1} - \partial_\pm\gj\,\gj^{-1} \quad \text{ with } \quad \Ld_\pm = \left( \sum_{i=1}^{N_\pm} \sum_{k=0}^{m_i^\pm} \frac{\Jc_\pm^{(i,k)}}{(p_r - z_i^\pm)^{k+1}} + \Jc_\pm^{(0,0)} \right)_{r=1,\dots,N}\,,
\end{equation}
where we have written $A$ in terms of $(\gh,L)$ using \eqref{eq:AgL} and used the expression \eqref{eq:Lz} for $L_\pm$.
The boundary condition imposes that $\Ad_\pm$ is valued in the subalgebra $\kf\subset\df$. This translates to a system of $\dim\,\df - \dim\,\kf = (M+1)\dim\,\gf$ linear equations on the currents $\lbrace \Jc_\pm^{(i,k)} \rbrace_{i}^{k}$ (by \eqref{eq:dimK}). Being the same in number as the components of the currents \eqref{eq:CompCurr}, the boundary conditions thus completely fix the currents in terms of $\gj$ and its derivatives $\partial_\pm\gj$. We will not need their explicit expression here and refer to~\cite{Lacroix:2020flf,Liniado:2023uoo} for more details. 

We can now think of the currents $\Jc_\pm^{(i,k)}$ as functionals of the $D$-valued field $\gj$. One further checks that they are invariant under the residual gauge transformations $\gj \mapsto \uj^{-1} \gj$ with $\uj\in B$.\footnote{This is in agreement with the second equation in \eqref{eq:gLu}, which states the invariance of the Lax connection under all 4d gauge transformations.} Therefore, we can treat the currents as functionals $\Jc_\pm^{(i,k)}[\gb]$ of the gauge-invariant field $\gb$, obtained by projecting $\gj$ onto the quotient $B \hspace{-1pt}\setminus\hspace{-1pt} D$. In the end, the Lax connection \eqref{eq:Lz} then becomes a local functional $L_\pm[\gb]$ of $\gb$, linear in the derivative $\del_\pm \gb$ and depending meromorphically on $z$, as is standard for the Lax connection of an integrable $\s$-model. We note that there are as many independent current components in each of $L_+[\gb]$ and $L_-[\gb]$ as scalar fields encoded in $\gb$, since $\dim B \hspace{-1pt}\setminus\hspace{-1pt} D$ coincides with the number \eqref{eq:CompCurr}.

\paragraph{The 2d action.} The field $\gh$ may be decomposed into its gauge-invariant part $\gb$ (built from the values $(\gh|_{z=p_r})_{r=1}^N$ at the defects $\Ph$) and a pure-gauge part $\gh_{\rm bulk}$ (essentially its behaviour away from the defects $\Ph$). By solving two of the equations of motion, we expressed $L=L[\gb]$ in terms of only $\gb$. However, by equation \eqref{eq:AgL}, the 4d gauge field depends on both $\gb$ and $\gh_{\rm bulk}$. Nevertheless, since the 4d action is gauge invariant, the pure-gauge component $\gh_{\rm bulk}$ decouples from it, leaving
\begin{equation}\label{eq:Action2d}
    S_{4d}\left[ A[\gb, \gh_{\rm bulk} ]\right] = S_{2d}[\gb]\,.
\end{equation}
This equation may initially be seen as a definition of $S_{2d}[\gb]$ as a 4d integral over $\Sigma \times \CP$. However, since we have solved for the dependence of $L$ on $z$, the integration over $\CP$ can be explicitly performed, allowing $S_{2d}[\gb]$ to be written as a local 2d integral over $\Sigma$. We will not need its expression here, but refer to \cite{Liniado:2023uoo,CY3,unif,Benini,Lacroix:2020flf} where this has been done explicitly. See also Appendix~\ref{app:PCM} below, where we review the familiar example of the PCM.

Importantly, the functional $S_{2d}[\gb]$ governs the dynamics of $\gb$ on $\Sigma$. By construction, its equations of motion are equivalent to the flatness of the Lax connection $L_\pm[\gb]$, automatically ensuring classical integrability in the sense of existence of an infinite number of conserved quantities extracted from the connection's monodromy. Moreover, the Poisson-commutation of these quantities follows from the Hamiltonian analysis of~\cite{Vicedo}: in this context the function $\varphi(z)$ defining the 1-form $\omega=\varphi(z)\,\dd z$ controls the Maillet bracket of the Lax matrix and is called the twist function.

\paragraph{Diagonal gauge symmetry.} Recall that we reparametrised the gauge field $A$ in terms of $(\gh,L)$ in equation \eqref{eq:AgL}. It is important to note that this reparameterisation is in fact not unique. Indeed, it is invariant under the following transformation of $(\gh,L)$:
\begin{equation}\label{eq:gLv}
    \gh \longmapsto \gh\,v\,, \qquad L \longmapsto v^{-1}\,L\,v + v^{-1}\dd_{\Sigma} v \qquad \text{ with } \qquad v : \Sigma \to G\,,
\end{equation}
depending on a local $G$-valued parameter $v(x^+,x^-)$ on $\Sigma$, independent of $(z,\zb)$. Since this is simply a redundancy in the parametrisation of the gauge field, it should be interpreted as a gauge symmetry of the resulting 2d integrable model.

The field of the 2d model was defined as the projection $\gb$ of the jet $(\gh|_{z=p_r})_{r=1}^N$ onto the quotient $B \hspace{-1pt}\setminus\hspace{-1pt} D$. One easily checks that the transformation \eqref{eq:gLv} acts on this field by\footnote{One initially finds $\gj \mapsto \gj \, \Delta(v)$, acting on the $D$-valued jet $\gj$. This induces the transformation \eqref{eq:gv} on the $B \hspace{-1pt}\setminus\hspace{-1pt} D$-valued projection $\gb$ since the left $B$-action and the right $G_{\text{diag}}$-action on $D$ commute.}
\begin{equation}\label{eq:gv}
    \gb \longmapsto \gb \, \Delta(v)\,, \qquad \text{ with } \qquad \Delta( v) = (v,\dots,v) \; \in \; G_{\text{diag}}\,,
\end{equation}
where $G_{\text{diag}}$ is the diagonal subgroup of the defect group $D=G^N$.  Recall from the earlier that we can write the solution for the gauge field as a functional $A[\gb,\gh_{\rm bulk}]$. By construction, this functional should then be invariant under the redundancy \eqref{eq:gv} of $\gb$, and this can indeed be checked by a direct computation. In particular, this means that the 2d action \eqref{eq:Action2d} is also invariant:
\begin{equation}\label{eq:Symv}
    S_{2d}[\gb\,\Delta(v)]=S_{2d}[\gb]\,.
\end{equation}
This confirms our expectation that the transformation \eqref{eq:gv} is a gauge symmetry of the 2d integrable model. Recall from equation \eqref{eq:gQuot} that $\gb$ was valued in the quotient $B \hspace{-1pt}\setminus\hspace{-1pt} D$. The 2d model's true physical degree of freedom is thus a field valued in the double quotient $B \hspace{-1pt}\setminus\hspace{-1pt} D\,/\, G_{\text{diag}}$, which has dimension $M\,\dim \gf$ by equation \eqref{eq:dimK}. One can show that the 2d action is quadratic in this field's derivatives, and hence it defines a non-linear $\s$-model on the target space $B \hspace{-1pt}\setminus\hspace{-1pt} D\,/\, G_{\text{diag}}$.

Let us also recall that we wrote the 2d Lax connection $L_\pm[\gb]$ in terms of $\gb$. 
Coming back to its construction, one can check that
\begin{equation}\label{eq:L}
    L_\pm[\gb] \;\,\longmapsto\;\, L_\pm[\gb\,\Delta(v)] = v^{-1}\,L_\pm[\gb]\,v + v^{-1}\partial_\pm v\,,
\end{equation}
in agreement with equation \eqref{eq:gLv}. This is the well-known gauge transformation of the Lax connection, under which the 2d dynamics (the flatness ${\rm F}_{+-}[L]=0$) are clearly invariant.

Since it will be useful later, let us consider how the $\gf$-valued currents $\Jc_\pm^{(i,k)}[\gb]$ contained in the Lax connection (in its partial fraction decomposition \eqref{eq:Lz}) transform under \rf{eq:L}.
For $i\neq 0$, the currents correspond to poles of $L_\pm$ and simply transform covariantly under the $G_{\text{diag}}$--symmetry. In contrast, the constant term $\Jc^{(0,0)}_\pm[\gb]$ transforms as a connection:
\begin{equation}\label{eq:J0v}
    \Jc^{(0,0)}_\pm[\gb] \;\,\longmapsto\;\, \Jc^{(0,0)}_\pm[\gb\,\Delta(v)] = v^{-1}\Jc^{(0,0)}_\pm[\gb]v + v^{-1}\partial_\pm v\,.
\end{equation}
It thus behaves as a gauge field for the $G_{\text{diag}}$-symmetry, and on-shell the flatness of the Lax implies that $\Jc^{(0,0)}_\pm[\gb]$ is flat.

\section{RG flow of rational integrable \texorpdfstring{$\s$}{sigma}-models}\label{sec:rational}

In the previous section, we have reviewed the construction of a large class of rational integrable $\s$-models from the 4d Chern-Simons theory on $\Sigma \times \CP$. We will now show their 1-loop renormalisability in a systematic way and will prove the conjecture of~\cite{Delduc:2020vxy} on the RG flow of their twist 1-form $\omega$ (already proven in~\cite{Hassler} in some cases). We will begin by reviewing the `universal' treatment of 1-loop divergences \cite{Levine}, which leads to a simple path integral in terms of the zero-curvature equations. We will compute this path integral for the most general rational Lax connection and recast the resulting 1-loop divergences as a simple contour integral of the Lax connection. As an illustration, we also refer the reader to Appendix \ref{app:PCM}, where we give the explicit expressions of the universal counterterm and the flow of the twist 1-form in the simple example of the Principal Chiral Model.

\subsection{Review of `universal' 1-loop divergences}\label{subsec:univ1loop}
It was recently observed \cite{Levine} that the 1-loop divergences of integrable $\sigma$-models take a `universal' form in terms of the Lax connection. The basic idea is that the integrable structure heavily constrains the possible divergences---related to the fact that integrable models should be consistent truncations of the general $\sigma$-model RG flow \rf{ric}. Concretely, the proposal is that the 1-loop divergences can be built from the objects appearing in the classical Lax connection.

The argument of \cite{Levine} relies on the fact that 1-loop divergences are completely determined by the equations of motion. Since the latter take a canonical zero-curvature form ${\rm F}[L]=0$, the 1-loop divergences must take a canonical form in terms of the Lax connection. 
The bulk of that paper was making this argument precise at the level of the path integral, with a major subtlety being that some parts of the zero-curvature equations are not genuine equations of motion. Rather, they are Bianchi identities satisfied off-shell, as a consequence of the physical form $\mc J= \mc J[\gb]$ of the currents in the Lax connection in terms of the $\s$-model field $\gb$. In order to deal with this, it was necessary to assume the converse statement known as `\textit{Bianchi completeness}': that the Bianchi identities obeyed by the Lax currents $\mc J$ imply their physical form $\mc J=\mc J[\gb]$, at least to leading order when expanded around on-shell background fields. This condition seems to pick out a large class of $\sigma$-models and excludes, for example, the sine-Gordon model. In the context of 4d Chern-Simons theory with disorder defects, we expect that all of the resulting 2d theories satisfy the Bianchi completeness assumption\footnote{Indeed, the boundary conditions at the defects are equivalent to the currents in the Lax connection taking their form $\mathcal J=\mathcal J[{\gb}]$ in terms of the physical fields (up to $G_{\rm diag}$ gauge symmetry). At the same time, one can show that these boundary conditions trivialise a particular subset of the zero-curvature equations, which we identify as the `Bianchi identities'. Clearly, $\mathcal J=\mathcal J[{\gb}]$ is then a solution of these identities. Conversely, we expect at least to linear order in the background field expansion that these Bianchi identities imply the solution $\mathcal J=\mathcal J[{\gb}]$, giving Bianchi completeness. This is clearly reasonable since the number of Bianchi identities is equal to the difference between the number of currents components $\mc J_\pm$ and the number of scalar fields in $\gb$.}---although we will not prove it here.

The result is then as follows. Suppose the Lax connection has a given rational pole structure (with $z_i^+ \neq z_i^-)$, 
\be
L_\pm(z) =  \sum_{i=1}^{N_\pm} \sum_{k=0}^{m_i^\pm-1} \frac{1}{(z-z^\pm_i)^{k+1}} \, \Jc_\pm^{(i,k)} . \la{rela}
\ee
The zero-curvature equation of $L_\pm(z)$ is then equivalent to some list of `Lax' equations on the currents~$\Jc^{(i,k)}_\pm$,
\be
{\rm F}_{+-}[L] = 0\,, \ \ \forall z \qquad \Longleftrightarrow\qquad {\rm Lax}^A[\Jc] = 0 \ . \la{32}
\ee
For any Bianchi-complete theory realising such a Lax connection (whatever form the currents $\Jc$ take in terms of the physical fields), the 1-loop divergent terms in the effective action take the universal form: 
\be \la{up}
\widehat S^{(1)}[\bar{\Jc}] \sim -\frac{i}{2}  \log \int \mathcal D U_A \, \mathcal D j_\pm^{(i,k)} \ \exp i \int U_A\ {\rm Lax}^A [\bar{\Jc}+j] \ .
\ee
 where $\Jc=\bar{\Jc}+j$, with $\bar{\Jc}$ on-shell background fields and $j$ fluctuations, while the scalar fields $U_A$ are Lagrange multipliers enforcing the equations ${\rm Lax}^A[\bar{\Jc}+j]=0$ in the path integral. The symbol $\sim$ denotes that we are only keeping track of log-divergent terms.

Note that we assumed that there is no constant term (independent of $z$) in the Lax connection \rf{rela}. For the models arising from 4d Chern-Simons theory, such a term is generally present, corresponding to the current $\Jc^{(0,0)}$ in equation \rf{eq:Lz}. Crucially, however, this current can always be gauge-fixed on-shell to zero, meaning that we can work with the equations \eqref{rela}--\eqref{up} for these models as well. Let us explain how this works in detail. In their raw form, this class of theories have: (i) a diagonal $G$ gauge symmetry (see discussion at the end of Section \ref{subsec:int}) and (ii) a constant term $\Jc^{(0,0)}$ in the Lax connection. These two features are related, since  $\Jc^{(0,0)}$ transforms as a gauge connection. 
\textit{A priori}, the universal path integral will differ from \rf{up} in three ways. First, the path integral measure in \rf{up} should be divided by a gauge volume: $\mathcal D U_A \, \mathcal D j_\pm^{(i,k)} \to \frac{\mathcal D U_A \, \mathcal D j_\pm^{(i,k)}}{\text{vol(gauge)}}$. The zero-curvature equations will be modified, ${\rm Lax}^A[\Jc] \to \widetilde{\rm Lax}{}^A[\Jc]$, by the presence of $\Jc^{(0,0)}$. Moreover, there is an additional equation ${\rm F}_{+-}[\Jc^{(0,0)}]=0$. The honest path-integral computation then yields
\be \la{upg}
\widehat S^{(1)}[\bar{\Jc}] \sim -\frac{i}{2}  \log \int \frac{\mathcal D U \, \mathcal D U_A \, \mathcal D j_\pm^{(i,k)}}{\text{vol(gauge)}} \ \exp i \int \Big( U_A\ \widetilde{\rm Lax}{}^A [\bar{\Jc}+j] + \langle U, {\rm F}_{+-}[\Jc^{(0,0)}]\rangle\Big)\ ,
\ee
with the additional Lagrange multiplier $U$ valued in $\gf$. Integrating out $U$ imposes ${\rm F}_{+-}[\Jc^{(0,0)}]=0$ as a delta-function constraint, which may be solved by $\Jc^{(0,0)}_\pm = g^{-1} \del_\pm g$ (integrating over the field $g\in G$). The gauge symmetry now acts as $g \to g h$ where $h\in G$, so it may be fixed as $g=1$ (with trivial Faddeev-Popov determinant). In this gauge, the constant part of the Lax connection vanishes, $\Jc^{(0,0)}=0$, and the zero-curvature equations revert to $\widetilde{\rm Lax}{}^A[\Jc] \to {\rm Lax}^A[\Jc]$. As a result, we obtain the path-integral \rf{up} stated above, obtained by simply deleting the constant term in the Lax connection.

Proceeding with the universal path integral \rf{up}, we note that it essentially computes the volume of flat connections with a fixed pole structure, expanded around some background. Its divergent 1-loop part is readily evaluated: for example, for a Lax connection with only simple poles (with all $z^+_i\neq z^-_j$),
\be
L_+ = \sum_{i} \frac{1}{z-z^+_i} \, \Jc_+^{i}  \ , \qquad  L_- = \sum_{j} \frac{1}{z-z^-_j} \, \Jc_-^{j}  \ , 
\ee
the 1-loop divergent term in this path integral is found to equal \cite{Levine}
\be\label{eq:UnivSimpleZ}
\ddt \widehat S^{(1)}[\bar{\Jc }] = -  \frac{\cG}{2\pi} \, \sum_{i,j} \frac{1}{(z_i^+-z^-_j)^2} \int \ud^2 x \, \langle \bar{\Jc}_+^i , \bar{\Jc}_-^j \rangle \ . 
\ee
Here $\text{t}=\log\mu$ is the renormalisation group (RG) parameter (where $\mu$ is a renormalisation scale) and $\cG$ is the dual Coxeter number of $G$ (see footnote \ref{fn:Form}). The above expression is bilinear in the Lax currents $\Jc^i$, and depends rationally on the positions of the Lax connection's poles. It seems natural to reformulate it as a contour integral of the Lax connection itself, which is the sum of these currents multiplied by different poles in $z$. After some consideration, one observes that it is given by the following contour integral\footnote{NL would like to thank Konstantin Zarembo for a useful discussion of this point.}
\be
\ddt \widehat S^{(1)} = \frac{\cG}{2\pi} \, \int \ud^2 x \,  \oint_{{}_- \G_{+}} \frac{\ud z}{2\pi i} \, \langle L_+, \del_z L_-\rangle \ , \la{cl}
\ee
along a path ${}_- \G_{+}$ that encircles the poles of $L_+$ anticlockwise (with the poles of $L_-$ on the outside). In fact, we will show that this curious formula is correct in general.\footnote{We note that any constant term in the Lax connection (which can be gauge-fixed to zero on-shell as discussed above) does not contribute to the contour integral \rf{cl}: it is annihilated by the derivative $\del_z$, first in $L_-$, and then in $L_+$ after integrating by parts.}

To derive the result \rf{cl} for the most general rational Lax connection  \rf{rela}, we will need to use its zero-curvature equation, \textit{i.e.}\ the corresponding `Lax' equations in \rf{32}: 
\begin{subequations}\label{z}
\begin{align}
&\del_+ \mc \Jc_-^{(j,p)} + \sum_{m\geq p} \sum_{i,n} \frac{(-1)^{n+1}}{(z_i^+-z_j^-)^{m+n-p+1}} \,\begin{pmatrix}
    m+n-p \\
    n
\end{pmatrix} \, [\mc \Jc_+^{(i,n)},\mc \Jc_-^{(j,m)}] = 0 \ ,  \la{z1}\\
&\del_- \mc \Jc_+^{(i,q)} + \sum_{n\geq q} \sum_{j,m} \frac{(-1)^{m+1}}{(z_i^--z_j^+)^{m+n-q+1}} \,\begin{pmatrix}
    m+n-q \\
    m
\end{pmatrix}\, [\mc \Jc_-^{(j,m)},\mc \Jc_+^{(i,n)}] =0 \ . \la{z2}
\end{align}
\end{subequations}
These are obtained by simply writing down the zero curvature condition ${\rm F}_{+-}[L(z)]=0$ and performing a partial fraction decomposition.

\subsection{Computing the universal path integral}\label{subsec:PathInt}
Let us consider a more general class of equations,
\begin{subequations}\label{eq:OPeq}
\begin{align}
&\del_+ \mc J_-^I + \sum_{K,L}  [O^I_{KL}\cdot \mc J_+^K,\ \mc J_-^L] = 0 \ , \\
&\del_- \mc J_+^J + \sum_{K,L}  [P^J_{KL}\cdot \mc J_-^K, \ \mc J_+^L] = 0 \  .
\end{align}
\end{subequations}
where $O^I_{KL}$ and $P^J_{KL}$ are linear operators $\gf\to \gf$, the indices $I,J,K,L$ run over some sets (which are generically different for $\mc A_+$ and $\mc A_-$), and $\mc J^I\in \gf$ for each value of $I$. This general family clearly includes our equations \rf{z} as a simple example---and the elliptic case discussed in Section \ref{sec:elliptic} will be another example.

\begin{figure}[t]
\centering
\raisebox{-0.5\height}{\includegraphics[scale=0.14]{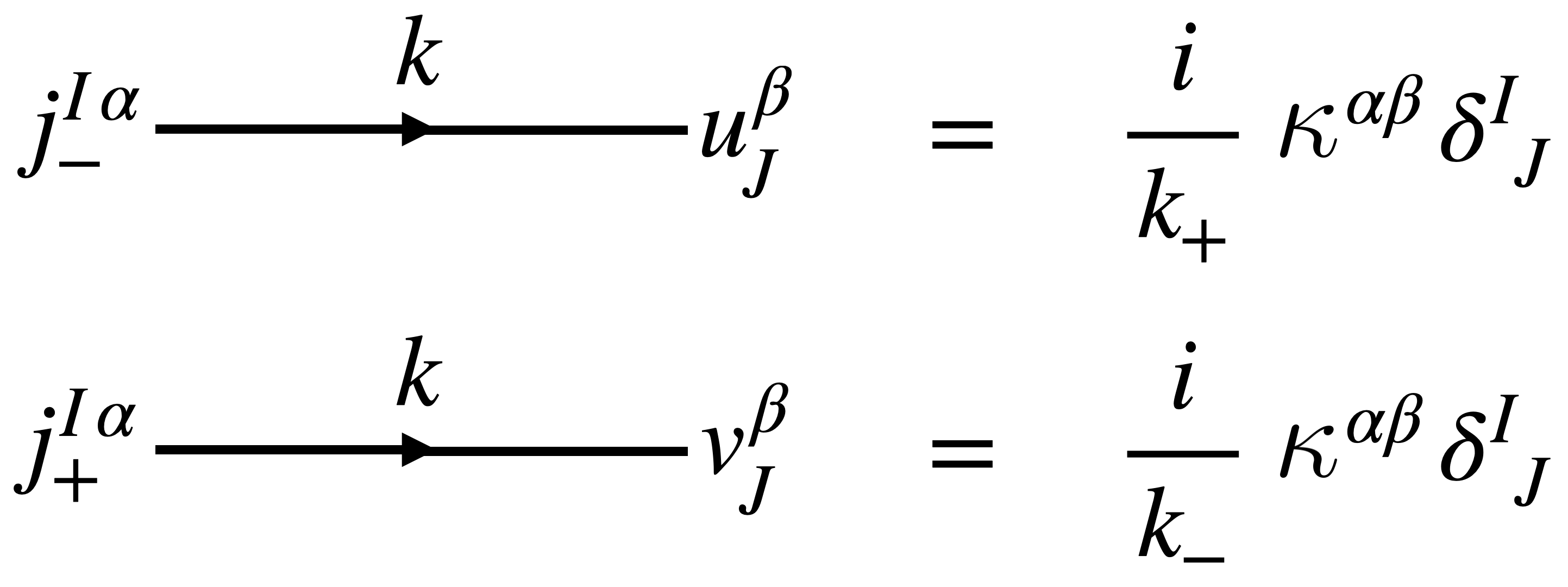}} 
\raisebox{-0.5\height}{\includegraphics[scale=0.14]{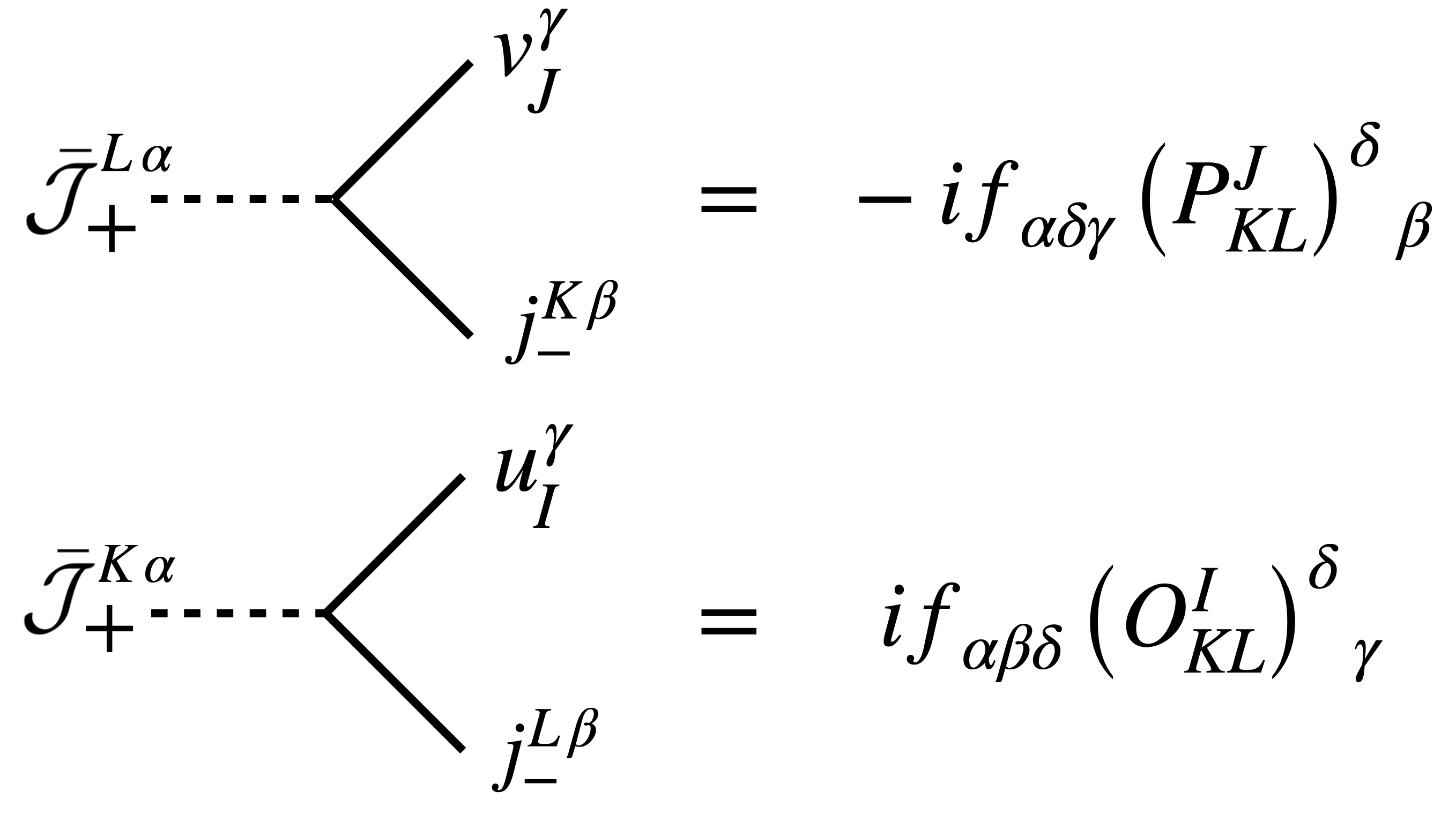}\hspace{1cm}\includegraphics[scale=0.14]{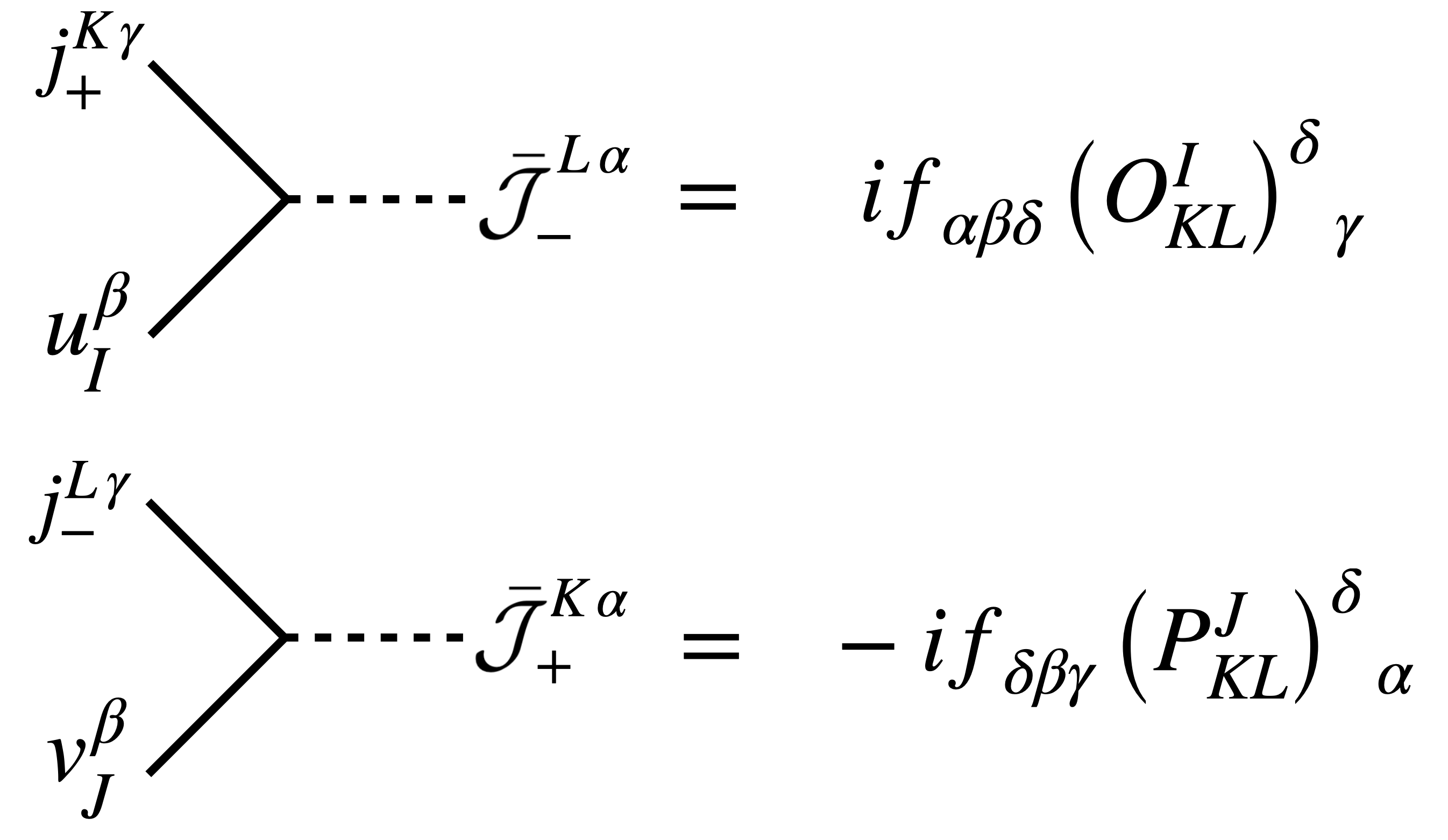}} 
\caption[]{\small The Feynman rules corresponding to the action in the exponent of the path integral \rf{PI}. Solid and dashed lines denote fluctuation and background fields, respectively. The fields are algebra-valued and decomposed as $u=u^\alpha\, T_\alpha$, where Greek indices run over a basis of algebra generators $\left\{T_\a\right\}_{\alpha=1}^{\text{dim}\mathfrak{g}}$ satisfying $[T_\b, T_\g] = f^\a{}_{\b\g} T_\a$ and are raised and lowered with $\kappa_{\a \b}\equiv \langle T_\a, T_\b \rangle$ and its inverse. The light-cone components of the 2d momenta are denoted by~$k_\pm$.
\label{FRules}}
\end{figure}
\begin{figure}[b]
\centering
\raisebox{-0.5\height}{\includegraphics[scale=0.19]{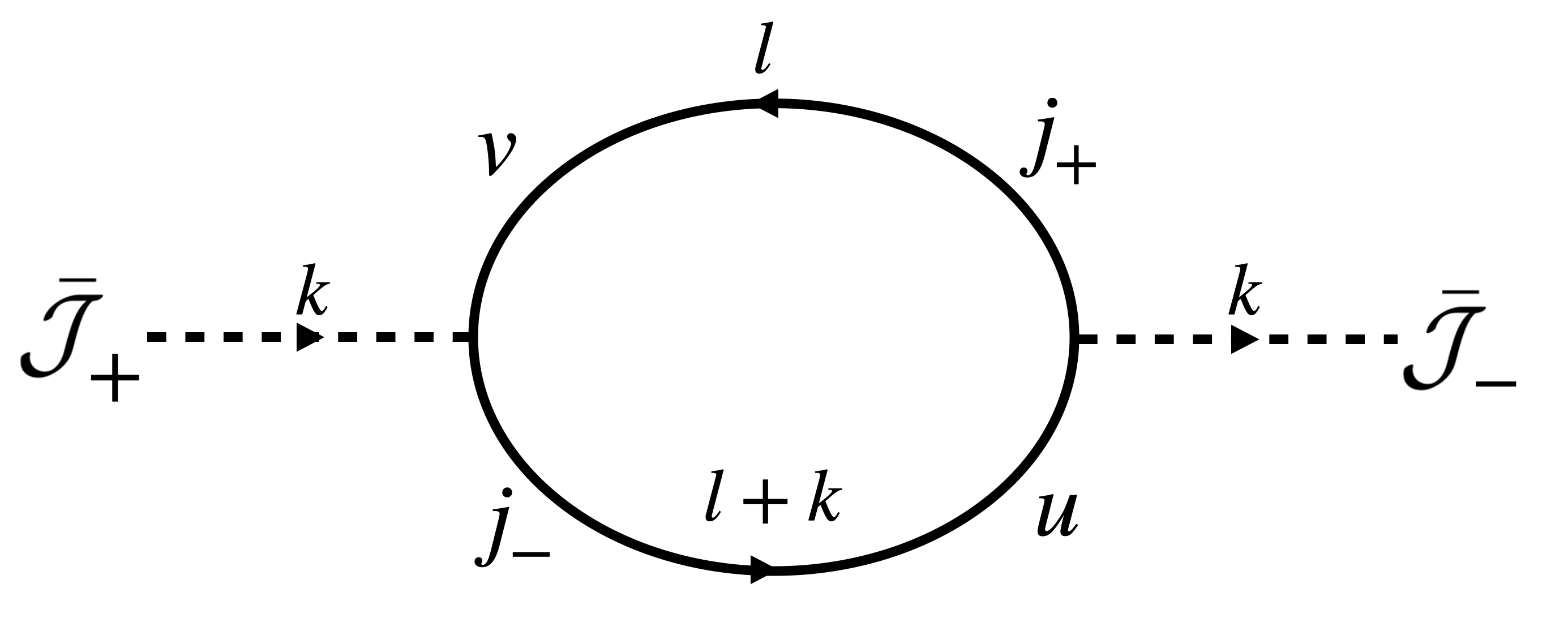}}
\caption[]{\small In the `universal' formulation, the divergent part of the 1-loop effective action \rf{ren} is given by a single Feynman diagram.\label{FDiag}} 
\end{figure}

We shall compute the universal path integral \rf{up} for these equations. Decomposing the Lagrange multipliers $U_A$ into $\gf$-valued scalars $u_I$ and $v_I$, it  takes the form (summing over repeated indices)
\be \begin{split} \label{PI}
&\widehat S^{(1)} \sim -\frac{i}{2}  \log \int \mathcal D u_I \ \mathcal D v_J  \ \mathcal D j_\pm \\
&\qquad \qquad \qquad \exp \Big[- i \int  \left\langle u_I, \,  \del_+  j_-^I +  [O^I_{KL}\cdot j_+^K,\ \bar{\mc J}_-^L] + [O^I_{KL}\cdot \bar{\mc J}_+^K, \ j_-^L]\right\rangle\\
&\qquad \qquad \qquad \qquad \quad \quad \,   + \left\langle v_J ,\,  \del_-  j_+^J +  [P^J_{KL}\cdot j_-^K,\ \bar{\mc J}_+^L] +[P^J_{KL}\cdot \bar{\mc J}_-^K,\  j_+^L]   \right\rangle\Big]\ .
\end{split} \ee
The action in the exponent is a simple theory of Lie-algebra valued 2d scalars $u_I$, $v_I$ and vectors $j_\pm^I$.  $\bar {\mc J}_\pm^I$ are background fields and $I$ is a flavour index. The theory's Feynman rules are shown in Figure~\ref{FRules}. There is a single divergent 1-loop diagram, shown in Figure~\ref{FDiag}, producing the following log-divergent contribution to the 1-loop effective action 
\be \la{ren}
\begin{split}
\widehat S^{(1)} &\sim -\frac 12 \int \ud^2 x \, \frac{\ud^2 k}{(2\pi)^2} e^{-i kx }\left[ \int \frac{\ud^2 l}{(2\pi)^2} \frac{1}{l_-(l+k)_+}\right] \ \Tr_\gf \big[ \ad_{\mathcal J_-^L}  \, O^J_{IL} \, \ad_{\mathcal J_+^K} \,  P^I_{JK}  \big]\\
&\sim\frac{1}{4\pi} \log{\frac\Lambda\mu}\int \ud^2 x \ \Tr_\gf \big[ \ad_{\mathcal J_-^L}  \, O^J_{IL} \, \ad_{\mathcal J_+^K} \,  P^I_{JK}  \big]\ ,
\end{split}
\ee
where $\Lambda$ is a UV-cutoff and $\mu$ is a renormalisation scale, and henceforth we are dropping the bar on background fields. Hence, denoting ${\rm t}=\log\mu$, the log-divergent terms in the effective action are given by
\be \la{renre}
\ddt \widehat S^{(1)} = - \frac{1}{4\pi } \int \ud^2 x \  \Tr_\gf \big[ \ad_{\mathcal J_-^L}  \, O^J_{IL} \, \ad_{\mathcal J_+^K} \,  P^I_{JK}  \big] \ .
\ee

\subsection{Universal divergences in the rational case\la{urat}}
For the Lax equations \rf{z} of a general rational Lax connection, we need to specialise to a case where these operators are proportional to the identity:
\begin{subequations}
\begin{align}
&O_{(\ell,q)\, (j,m)}^{(k,p)} = \delta_{k=j}\, \delta_{m\geq p}  \,  \frac{(-1)^{q+1}}{(z_\ell^+-z_j^-)^{m+q-p+1}} \,\begin{pmatrix}
   m+q-p \\
   q
\end{pmatrix}  \, \mathbb{I} \ , \\
&P_{(k,p)\, (i,n)}^{(\ell,q)} = \delta_{l=i}\, \delta_{n\geq q}  \,  \frac{(-1)^{p+1}}{(z_i^--z_k^+)^{p+n-q+1}} \,\begin{pmatrix}
   n+p-q \\
   p
\end{pmatrix} \, \mathbb{I} \ ,
\end{align}
\end{subequations}
where $\mathbb{I}:\gf\to\gf$ is the identity operator. Substituting these operators into \rf{renre} leads to (summing over $i,j$ and $m,n$)
\begin{align}
\hspace{-0.1cm}\no \ddt \widehat S^{(1)}  &= \frac{1}{4\pi} \sum_{\substack{i,j\\p\leq m,q\leq n}}\int \ud^2 x \  \Tr_\gf \big[ \ad_{\mathcal J_-^{(j,m)}}  \,  \ad_{\mathcal J_+^{(i,n)}}\big]  \frac{(-1)^{n}}{(z_i^+ - z_j^-)^{2+m+n}} \begin{pmatrix}
    m+q-p \\
    q
\end{pmatrix} \begin{pmatrix}
    n+p-q \\
    p
\end{pmatrix} \\
&= \frac{\cG}{2\pi} \sum_{{i,j,m,n}} \int \ud^2 x \ \frac{(-1)^{n+1}}{(z_i^+ - z_j^-)^{2+m+n}}\frac{(m+n+1)!}{m!\, n!} \, \langle \mc J_+^{(i,n)}, \mc J_-^{(j,m)} \rangle  \ , \la{uc}
\end{align} 
where in the second line we used the identity $\Tr_\gf[\ad_u \ad_v] = - 2\cG \langle u, v\rangle $ (see conventions in footnote \ref{fn:Form}) and standard identities for binomial coefficients to perform the sum.

Now, using the following identity 
\begin{align}
 \no \oint_{{}_- \G_{+}}  \frac{\ud z}{2\pi i} \, \langle L_+ , \del_z L_-\rangle &= 
\sum_{{i,j,m,n}}\oint_{z_i^+}\frac{\ud z}{2\pi i} \frac{1}{(z-z_i^+)^{n+1}}\del_z \Big[\frac{1}{(z-z_j^-)^{m+1}}\Big] \, \langle \mc J_+^{(i,n)}, \mc J_-^{(j,m)} \rangle  \\
&=\sum_{{i,j,m,n}} \frac{(-1)^{n+1}}{(z_i^+ - z_j^-)^{2+m+n}}\frac{(m+n+1)!}{m! \, n!} \, \langle \mc J_+^{(i,n)}, \mc J_-^{(j,m)} \rangle \ ,
\end{align}
we identify the quantity on the RHS of \rf{uc} as the expected contour integral. We have therefore proven \rf{cl} in general for the class of theories with rational Lax connections \eqref{rela}:
\be
\boxed{\ddt \widehat S^{(1)} = \frac{\cG}{2\pi} \, \int \ud^2 x \,  \oint_{{}_- \G_{+}} \frac{\ud z}{2\pi i} \, \langle  L_+, \del_z L_-\rangle \ .} \la{frr}
\ee

\subsection{Rewriting the RG flow in 4d Chern-Simons language}\label{subsec:4dFlow}

So far, our main result \eqref{frr} for the RG flow of rational integrable $\s$-models has been derived using only the analytic structure \eqref{rela} of the Lax connection $L_\pm(z)$, independently of the initial construction of these models. We now want to leverage their origin from 4d Chern-Simons theory, as reviewed in Section \ref{sec:review}. This will allow us to interpret the counterterm \eqref{frr} as a flow of the defining data of these models. To do so, we ought to rewrite it as a local 4d counterterm in terms of the Chern-Simons gauge field $A$.

\paragraph{A first attempt.} The attentive reader might worry that the formula \eqref{frr} appears to be asymmetric with respect to the $+$ and $-$ light-cone directions, and hence is not manifestly covariant. 
But in fact this quantity is symmetrical under swapping $(+\leftrightarrow -)$. Indeed, an integration by parts yields the integral of $ \langle L_-, \del_z L_+ \rangle$ over the contour with opposite orientation, which can be seen as a path ${}_+ \G_{-}$ enclosing the poles of $L_-$ (and not those of $L_+$). One can therefore rewrite \eqref{frr} in the more symmetrical form
\be
\ddt \widehat S^{(1)} = \frac{\cG}{8\pi^2 i} \, \int \ud^2 x \,  \left( \oint_{{}_- \G_{+}} \ud z \, \langle L_+, \del_z L_-\rangle + \oint_{{}_+ \G_{-}} \ud z \, \langle L_-, \del_z L_+ \rangle \right)\ . \la{frr2}
\ee
We now want to rewrite these contour integrals as a local 2d integral over $\CP$. This can be done using the identity\footnote{Here and below we use the normalisation $\ud^2 z = \ud z \wedge \ud \zb$.}
\be
\oint_{{}_g\G_f} {\ud z } \, f(z) \, g(z)  = -\int_{\CP} \ud^2 z\, \del_{\zb}f(z) \, g(z)  \ , \la{ContToSurfInt}
\ee
for $f$ and $g$ meromorphic functions with distinct sets of poles and ${}_g\G_f$ a contour wrapping the poles of $f$ but not those of $g$. Taking $f=L_+$ and $g=\del_z L_-$ for the first term and $f=L_-$ and $g=\del_z L_+$ for the second one, we can recast \eqref{frr2} as the 4d integral 
\be
\ddt \widehat S^{(1)} = -\frac{\cG}{8\pi^2 i} \int_{M}\ud^2 x\, \ud^2 z \Big(\langle\del_{\zb} L_+, \del_z L_-\rangle + \langle\del_{\zb} L_- , \del_z L_+\rangle \Big)\ .
\ee
Although this result looks promising, it is not manifestly possible to rewrite it covariantly in the language of differential forms, in terms of the exterior derivative $\ud$ and the connection $L$. This is mostly due to the presence of a plus sign rather than a minus sign between the two terms of the integrand. Solving this issue will require the introduction of a new ingredient, namely the function $\Psi(z)$.

\paragraph{Introducing the function $\bm{\Psi(z)}$.} The trick is to make the desired minus sign appear in our starting point \eqref{frr2} by considering:
\be
\ddt \widehat S^{(1)} = -\frac{1}{16\pi^2 i} \, \int \ud^2 x \,  \left( \oint_{{}_- \G_{+}} \ud z \, \Psi(z) \langle L_+, \del_z L_- \rangle - \oint_{{}_+ \G_{-}} \ud z \, \Psi(z) \langle  L_- , \del_z L_+ \rangle \right) \ .  \la{frr3}
\ee
The price to pay for this modification is the introduction in the integrand of a meromorphic function $\Psi(z)$ satisfying\footnote{We have included a global factor $-2\cG$ in the definition of $\Psi(z)$ for future convenience.}
\begin{equation}
\left\{\begin{matrix*}[l]
\Psi(z^\pm_i) = \mp 2\cG\\
\del_z^p\Psi(z^\pm_i)=0 \ , \quad p=1,\ldots,m_i^\pm-1 
\end{matrix*}\right. \,, \la{psi}
\end{equation}
where we recall that the points $z^\pm_i$ are the poles of $L_\pm(z)$ of orders $m_i^\pm \in \Z_{\geq 1}$ --- see equation \eqref{rela}. One easily checks that the condition \eqref{psi} leads to the same residues in equations \eqref{frr2} and \eqref{frr3}. However, it does not uniquely define $\Psi(z)$: the residual freedom in choosing $\Psi(z)$ is fixed by demanding its poles are the same as those of the twist 1-form $\omega=\vp(z) \dd z$ and it is bounded when $z\to\infty$. For now this is an arbitrary choice, but it will be helpful later when we wish to absorb the counterterm into a flow of $\omega$. Concretely, this means that $\Psi(z)$ takes the form
\begin{equation}\label{eq:psir}
    \Psi(z) = \sum_{r=1}^N \frac{\psi_r}{z-p_r} + \psi_\infty\,.
\end{equation}
The $N+1$ coefficients $\psi_r$ are then constrained by the conditions \eqref{psi}, which translate to a linear system of $N-2$ equations.\footnote{Explicitly, the number of equations is $\sum_{i=1}^{N_+} m_i^+ + \sum_{i=1}^{N_-} m_i^-$, which coincides with $2M=N-2$ using the identities \eqref{eq:M1} and \eqref{eq:M2}.} They can thus be expressed as rational functions of the parameters $(p_r,z_i^\pm)$ and 3 unconstrained variables (whose interpretation will be made clear in the next subsection).\footnote{For an example where $\Psi(z)$ can be worked out explicitly after fixing these variables, see equation \eqref{PsiPCM}. Note that appendix \ref{app:PCM} considers a 1-form $\omega$ with double poles, one of which is located at $z=\infty$, hence modifying the functional form of $\Psi(z)$.}

We now use the identity \eqref{ContToSurfInt} to recast the contour integrals of equation \eqref{frr3} as surface integrals over $\CP$ (taking $f=L_+$ and $g=\Psi(z)\del_{z}L_-$ for the first term and exchanging $+$ and $-$ for the second one). The result takes the form of a covariant 4d counterterm in terms of differential forms:
\be
\ddt \widehat S^{(1)} = -\frac{1}{16 \pi^2 i} \int_{M}  \Psi(z) \, \Tr[\ud L \wedge \ud L] \ .
\ee
Since $L$ only has legs in the $+$ and $-$ directions, this is the same as
\be
\ddt \widehat S^{(1)} = -\frac{1}{16 \pi^2 i} \int_{M}  \Psi(z) \, \Tr\left[ \text{F}[L] \wedge \text{F}[L]\right] = - \frac{1}{16 \pi^2 i} \int_{M}  \Psi(z) \, \ud {\rm CS}[L]\ . \la{4dc}
\ee

\paragraph{Counterterm in terms of $\bm A$.} We recall from equation \eqref{eq:AgL} that the 4d gauge field $A$ is related to $L$ by a \textit{formal} 4d gauge transformation $L=A^{\gh}$ (i.e.\ one not respecting the boundary conditions). The quantity $\ud {\rm CS}[A]$ is invariant under formal gauge transformations of $A$, since ${\rm CS}[A]$ transforms by a closed term. Hence we can replace $L$ by $A$ in the universal result \eqref{4dc}, yielding 
\be
\ddt \widehat S^{(1)} = -\frac{1}{16 \pi^2 i} \int_{M}  \Psi(z) \, \ud {\rm CS}[A] \ .  \la{combi}
\ee
This gives the interpretation of the counterterm as a local, gauge-invariant functional of the 4d gauge field.\footnote{As mentioned, the counterterm \rf{combi} is not only invariant under 4d gauge transformations, but under the strictly larger class of formal gauge transformations (this is what allowed us to pass from \eqref{4dc} to \eqref{combi}). The fact that we can write the counterterm as a functional of $A$ invariant under formal gauge transformations is equivalent to this counterterm being written in terms of the Lax connection $L$ only, independently of the pure formal gauge degree of freedom $\gh$. This is therefore a consequence of its `universal' form \cite{Levine}.} Geometrically, this is the integral of the so-called \textit{2$^{\text{nd}}$ Chern character} of the gauge connection $A$, weighted by the function $\Psi(z)$. It would be interesting to explore a geometric reformulation of the present work.

Let us also comment on the role of the gauge field component $A_z$ in the counterterm~\eqref{combi}. Recall that, in the classical 4d theory, $A_z$ completely decouples from the action, so it can be fixed to any value (see discussion below \eqref{eq:Action4d}). One may worry whether this is consistent with the counterterm \eqref{combi}, which superficially appears to depend on~$A_z$. However, simple manipulations show that $A_z$ decouples from it on-shell, since it only appears multiplied by the quantity $\text{F}_{+-}[A]$, which vanishes on-shell. Therefore, $A_z$ can be freely shifted to any given value also in this 1-loop counterterm, just as in the classical action.

\paragraph{Towards the flow of $\bm\omega$.} After integration by parts,  the 4d counterterm \eqref{combi} takes the form
\be
\ddt \widehat S^{(1)} = \frac{1}{16 \pi^2 i} \int_M \del_z \Psi(z) \, \ud z \wedge {\rm CS}[A] + \DT \ , \la{floweq}
\ee
where the second contribution is the `defect term'
\be
\DT = \frac{1}{16 \pi^2 i}  \int_M \del_{\bar z} \Psi(z) \, \ud \bar z \wedge {\rm CS}[A] \ , \la{dete}
\ee
which obviously localises along the poles of $\Psi$ (\textit{i.e.}\ that of $\omega$). Comparing with the classical 4d action \rf{eq:Action4d}, the first term in \rf{floweq} is clearly interpreted as the desired flow of the twist 1-form $\frac{\dd\;}{\dd\text{t}}\omega = \del_z \Psi(z)\ud z$, in agreement with the conjecture~\cite{Delduc:2020vxy} that we want to prove. In the next subsection, we will explain how the defect term \eqref{dete} originates from the flow of the gauge field's boundary conditions (including the location of the defects in $\CP$).

\subsection{Universal divergence as a flow of \texorpdfstring{$\omega$}{omega}}\label{subsec:ct}

In this subsection, we will reverse our point of view and start from the conjectured flow of the twist 1-form~\cite{Delduc:2020vxy,Derryberry:2021rne}, namely
\be
\boxed{\ \frac{\ud \omega}{\ud \text{t}}  = \del_z \Psi(z)\,  \ud z \ , \;} \la{fom}
\ee
with the other defining data  $\Zh^\pm$, $G$, $\kf$ being RG-invariant.
We will then show that the resulting variation of the classical action matches the universal counterterm \eqref{floweq}. In other words, we will prove that we can absorb that counterterm into a flow of the defining data, establishing the 1-loop renormalisability of this class of 2d theories.

As before, we only treat here the case where $\omega$ has simple real poles. For higher-order ones and more general reality conditions, the expression \eqref{floweq} of the universal counterterm is still valid. Similarly, the flow of $\omega$ takes the same form \eqref{fom}. However, we will generally find a non-trivial flow of the maximally isotropic subalgebra $\kf$ --- see Appendix \ref{app:GenPoles} for details.

\paragraph{Flow of $\bm\omega$.} To start with, we need to explain more precisely what we mean by the flow \eqref{fom} of the twist 1-form. Since we are focusing on the case of simple poles $p_r$, $\omega$ can be written as
\begin{equation}
    \omega =  \sum_{r=1}^N \frac{\ell_r}{z-p_r} \,\ud z\,,
\end{equation}
with residues $\ell_r$ as defined in equation \eqref{eq:Levels}. The parameters $(p_r,\ell_r)_{r=1}^N$ can be seen as encoding the choice of $\omega$. The flow of $\omega$ should then be understood as defining a flow of these parameters, according to
\begin{equation}\label{eq:OmFlow}
    \frac{\ud \omega}{\ud \text{t}}   = \left( \sum_{r=1}^N \frac{\ud \ell_r}{\dd\text{t}}\frac{1}{z-p_r} + \frac{\ud p_r}{\dd \text{t}}\frac{\ell_r}{(z-p_r)^2} \right) \dd z\,.
\end{equation}
Concretely, comparing to the right-hand side of \eqref{fom}, we find
\begin{equation}\label{eq:dlp}
    \frac{\dd \ell_r}{\dd \text{t}} = 0 \qquad \text{ and } \qquad \frac{\dd p_r}{\dd\text{t}} = -\frac{\psi_r}{\ell_r}\,,
\end{equation}
where $\psi_r$ are the coefficients defined in equation \eqref{eq:psir}.\footnote{As explained around equation \eqref{eq:psir}, $\psi_r$ is a function of the parameters $(p_r,\ell_r)$ and of 3 unconstrained variables. One shows~\cite{Delduc:2020vxy,LW2,Kotousov:2022azm} that these additional variables induce a $\text{t}$-dependent Möbius transformation of the poles $p_r$ along the flow, which can be undone by a change of spectral parameter. Therefore, they do not enter the RG flow of the $\s$-model itself.\label{foot:mobius}} In particular, the levels $\ell_r$ are RG-invariants: this is consistent with their interpretation as coefficients of Wess-Zumino terms in the 2d $\s$-model (see footnote \ref{foot:WZ}). The non-trivial RG flow is then contained in the flow of the poles $p_r$.

A subtle but technically important remark is that there are two different notions of $\text{t}$-derivatives of $\omega$ that we need to consider. The one $\frac{\ud \omega}{\dd \text{t}}$ defined in equation \eqref{eq:OmFlow} is a `formal derivative', mapping a rational 1-form to a rational 1-form. As explained above, this derivative is particularly useful to compactly encode the flow of the parameters $(p_r,\ell_r)$ defining $\omega$ and it is the one we use in equation \eqref{fom}. In contrast, it has to be distinguished from the derivative of $\omega$ seen as a `measure', which also includes distributions arising from the flow of the poles $p_r$. Let us explain this more precisely. In the context of 4d Chern-Simons, the 1-form $\omega$ is mostly used as a measure to integrate against. Namely, we are interested in integrals of the form $\int_M \omega \wedge \Lambda$, where $\Lambda$ is a 3-form on $M$, as for instance $\text{CS}[A]$ for the classical action \eqref{eq:Action4d}. We then define the `measure derivative' of $\omega$ through the variation of such integrals with respect to $\text{t}$:
\begin{equation}\label{eq:dtMeasure}
    \ddt \left( \int_M \omega \wedge \Lambda \right) = \int_M \left( \dtom \wedge \Lambda + \omega \wedge \frac{\dd \Lambda}{\dd \text{t}} \right)\,.
\end{equation}
With this characterisation, the two notions of derivatives are related in this case by
\begin{equation}\begin{split}
    \dtom &= \frac{\ud \omega}{\dd \text{t}} + \sum_{r=1}^N \ell_r\,\frac{\ud \overline{p_r}}{\dd \text{t}} \, \del_{\bar p_r} \Big( \frac{1}{z-p_r}\Big) \,\ud z\\
    &= \frac{\ud \omega}{\dd \text{t}} + 2\pi i \sum_{r=1}^N \ell_r\,\frac{\ud \overline{p_r}}{\dd \text{t}} \, \delta(z-p_r)\,\ud z
    \end{split}
\end{equation}
We refer to the appendix \ref{app:IntDer} for a more thorough explanation of this formula. Reinserting our starting point \eqref{fom} for the `formal derivative' $\frac{\ud \omega}{\dd \text{t}}$, we thus get
\begin{equation}\label{eq:dtDist}
    \dtom = \del_z \Psi(z)\,\ud z  + 2\pi i  \sum_{r=1}^N \ell_r\,\frac{\ud \overline{p_r}}{\dd \text{t}} \, \delta(z-p_r)\,\ud z\,.
\end{equation}
This is the object that will appear when taking RG derivatives of the Chern-Simons action.

\paragraph{Counterterm in the effective action.}  We now want to translate the conjectured beta-function \eqref{fom} for the `coupling' $\omega$ to a divergent counterterm in the 1-loop effective action~$\widehat S^{(1)}$. Taking the action to be evaluated on a classical background solution, there is no distinction between the 2d and 4d actions since $S_{2d}[\gb]=S_{4d}[A[\gb,\gh_{\rm bulk}]]$, where $A=A[\gb,\gh_{\rm bulk}]$ is a 4d classical solution and $\gb$ is a 2d one.\footnote{The identity $S_{2d}[\gb]=S_{4d}[A[\gb,\gh_{\rm bulk}]]$ is equation \eqref{eq:Action2d}, which served as a definition of the 2d action in the classical 4d-CS construction. Note that this also holds for an off-shell 2d field $\gb$, while the 4d gauge field $A[\gb,\gh_{\rm bulk}]$ is partially on-shell, having solved two of the three equations of motion to find the $z$-dependence of $A$. While treating one-loop divergences, we can treat $\gb$ as an on-shell background field, so that $A[\gb,\gh_{\rm bulk}]$ is also a fully on-shell 4d solution.} Thus, by taking the RG-derivative of the classical 4d action, the conjectured beta-function predicts the following counterterm:
\begin{align}\label{eq:flowS2d}
    \ddt \widehat S^{(1)}_{\rm conj} &= \frac{1}{16\pi^2 i} \int_M \dtom  \, \wedge {\rm CS}[A] + \Cvar \,, \no \\
   &= \frac{1}{16\pi^2 i} \int_M  \del_z \Psi(z) \, \ud z \wedge {\rm CS}[A]   + \Cdist + \Cvar  \ .
\end{align}
The `distributional' term $\Cdist$ is the additional contribution arising from the distributions in the measure derivative \eqref{eq:dtDist}. Meanwhile the `variational' term $\Cvar$ appears because the background field $A$ must run along the flow:
\be
\Cvar = \frac{\delta S_{4d}}{\delta A}\left[ \frac{\ud A}{\dd \text{t}}\right] \ . \la{vter}
\ee
Indeed, such a term is guaranteed since $A$ is subject to boundary conditions at positions in $\CP$ that are moving along the flow. Let us now evaluate these $\Cdist$ and $\Cvar$ explicitly.

\paragraph{The distributional term.} The distributional part of the measure derivative \eqref{eq:dtDist} contributes
\begin{equation}
\Cdist = \frac{1}{8\pi} \sum_{r=1}^N \ell_r\,\frac{\ud \overline{p_r}}{\dd \text{t}} \int_M  \, \delta(z-p_r)\,\ud z \wedge {\rm CS}[A]  \ .\label{eq:Cdist}
\end{equation}
To evaluate this term explicitly, note that, up to total derivatives and terms proportional to the equation of motion $\text{F}_{+-}[A]=0$, the 4-form $\ud z \wedge {\rm CS}[A]$ coincides with $-\ud z \wedge \ud \zb\,  \langle A , \del_{\zb} A \rangle $. Then, thanks to the delta functions, the integral over $z,\zb$ may be carried out to obtain
\begin{equation}\label{eq:Cdist2}
\Cdist = -\frac{1}{8\pi} \sum_{r=1}^N \ell_r\, \frac{\ud \overline{p_r}}{\dd \text{t}} \int_\Sigma  \langle A, \del_{\zb} A \rangle\Bigl|_{z=p_r}  \ .
 \end{equation}

\paragraph{The variational term.}
The variation of the action \rf{vter} is given by a bulk term, a boundary term and a surface defect term (see equation \rf{eq:varA}). Since the background field is on-shell along the flow, the bulk term vanishes. Moreover, the boundary term \eqref{eq:DeltaBoundary} still vanishes. However, as the positions of the pole defects are running, then the RG-derivative $\frac{\ud A}{\dd \text{t}}$ of the background field does not satisfy the boundary conditions at the poles, and the defect term \eqref{eq:DefectJet} contributes
\be
\Cvar = -\frac{1}{8\pi} \int_\Sigma \,\left\langle\hspace{-0.2cm}\left\langle  \left(\left. A \right|_{z=p_r} \right)_r,
\left( \left. \tfrac{\ud A}{\ud  t}\right|_{z=p_r}\right)_r\right\rangle\hspace{-0.2cm}\right\rangle_{\df}\,. \label{eq:Cvar}
\ee

The running of the background field $A$ along the flow is governed by the flow of $\omega$, and in particular of its poles $p_r$ that appear in the boundary condition
\be
(A|_{z=p_r})_{r} \;  \in \; \kf  \ .
\ee
Since $p_r$ are themselves running, differentiating this expression yields 
\be
 \ddt \Bigl( A\bigr|_{z=p_r} \Bigr)_r = \left(\frac{\ud A}{\dd \text{t}} \Bigr|_{z=p_r} + \frac{\ud p_r}{\dd \text{t}}\,\del_z A \bigr|_{z=p_r}  + \frac{\ud \overline{p_r}}{\dd \text{t}}\,\del_{\zb} A \bigr|_{z=p_r}  \right)_r \; \in \; \kf \ ,
\ee
where we are using the fact that the isotropic subalgebra $\kf \subset \df$ is taken to be RG-invariant. This statement, alongside the facts that  $\bigl(\left. A \right|_{z=p_r} \bigr)_r \in \kf$ and that $\kf$ is isotropic, allows us to replace $\bigl( \left. \tfrac{\ud A}{\ud \text{t}}\right|_{z=p_r}\bigr)_r$ by $- \bigl( \tfrac{\ud p_r}{\dd \text{t}}\,\del_z A \bigr|_{z=p_r}  + \tfrac{\ud \overline{p_r}}{\dd \text{t}}\,\del_{\zb} A \bigr|_{z=p_r}  \bigr)_r $ in \rf{eq:Cvar}:
\begin{align}
\Cvar &=  \frac{1}{8\pi} \int_\Sigma \,\left\langle\hspace{-0.2cm}\left\langle  \left(\left. A \right|_{z=p_r} \right)_r,
\left( \frac{\ud p_r}{\dd \text{t}}\,\del_z A \bigr|_{z=p_r}  + \frac{\ud \overline{p_r}}{\dd \text{t}}\,\del_{\zb} A \bigr|_{z=p_r}  \right)_r\right\rangle\hspace{-0.2cm}\right\rangle_{\df} \no \\
&= \frac{1}{8\pi} \sum_{r=1}^N \ell_r \int_\Sigma \,\left.\left\langle A, \frac{\ud p_r}{\dd \text{t}}\,\del_z A  + \frac{\ud \overline{p_r}}{\dd \text{t}}\,\del_{\zb} A \right\rangle \right|_{z=p_r} \,,\label{eq:Cvarfinal}
\end{align}
where we used the definition \rf{eq:FormDefectSimple} of the defect algebra's bilinear form $\dlangle \cdot, \cdot \drangle_{\df}$.

\paragraph{Matching the universal result.} Recall that our goal is to 
show that the counterterm \eqref{eq:flowS2d} corresponding to the conjectured flow of $\omega$ coincides with the universal result~\eqref{floweq}. Adding together the terms $\Cdist$ and $\Cvar$, we find that the `antiholomorphic' term cancels, leaving the simple expression
\begin{equation}\label{eq:Cdv}
   \Cdist+\Cvar =  \frac{1}{8\pi} \sum_{r=1}^N \ell_r  \frac{\ud p_r}{\dd \text{t}} \int_\Sigma \left.\left\langle A,\del_z A \right\rangle \right|_{z=p_r} \,.
\end{equation}
This quantity is precisely equal to the `defect' counterterm \rf{dete},
\be
\Cdist+\Cvar = \DT \ ,
\ee
evaluated in the frame where $A_z=0$ (recalling that the universal result is on-shell independent of $A_z$). This can be seen from the fact that with $A_z=0$ we have $\ud \bar z \wedge {\rm CS}[A] = \ud z \wedge \ud \zb \wedge \langle A, \del_z A \rangle$ and the explicit expression
\begin{equation}
    \del_{\zb}\Psi(z) = 2\pi i \sum_{r=1}^N \ell_r \frac{\ud p_r}{\dd \text{t}} \delta(z-p_r)\,,
\end{equation}
which is obtained from equations \eqref{eq:psir} and \eqref{eq:dlp}, with the delta-functions allowing the integral over $z,\zb$ to be carried out.

Altogether, we have proven that 
\be 
 \ddt \widehat S^{(1)}_{\rm conj} =  \frac{1}{16\pi^2 i} \int_M  \del_z \Psi(z) \, \ud z \wedge {\rm CS}[A]  + \DT = \ddt \widehat S^{(1)} \ ,
 \ee
as required. This shows that the universal counterterm $\widehat S^{(1)}$ can be absorbed as a flow of the defining parameters $(\omega,\Zh^\pm,G,\kf)$ of the models, proving their 1-loop renormalisability. As expected, the first term in equation \eqref{floweq} is directly tied to the running \eqref{fom} of the twist 1-form $\omega$.  On the other hand, the defect term was found to originate from the flow of the poles $p_r\in \CP$ and hence of the boundary conditions imposed on the gauge field at those locations. In particular, in the case where $\omega$ has simple poles, the isotropic subalgebra $\kf \subset \df$ encoding the boundary condition is RG-invariant. As mentioned earlier, this is no longer the case when $\omega$ has higher-order poles: rather, $\kf$ flows according to a 1-parameter family of automorphisms of $\df$. We refer to the appendix \ref{app:GenPoles} for details.

\section{RG flow of elliptic integrable \texorpdfstring{$\sigma$}{sigma}-models}\label{sec:elliptic}
In the analysis above, we assumed that the model belonged to the class of rational integrable field theories of the Belavin-Drinfel'd classification \cite{belavin_solutions_1982}. Namely, the spectral parameter $z$ was assumed to live in $\mathbb{CP}^1$, implying that the meromorphic Lax connection is a rational function of $z$, as in \eqref{rela}. However, an advantage of our approach is that it immediately generalises to other classes of models: here we will consider \textit{elliptic} integrable $\s$-models arising from disorder defects, examples of which were considered in \cite{CY3,Derryberry:2021rne,LW1,LW2}. The spectral parameter now lives on the complex torus
\begin{equation}
    \mathbb{T}\equiv \mathbb{C}/\Lambda\,,\qquad \Lambda=\left\{n_1\lambda_1+n_2\lambda_2 \, |\, n_1,n_2\in\mathbb{N}\right\}\,,\label{eq:defT}
\end{equation}
defined in terms of two basis vectors $\lambda_1,\lambda_2$ in $\mathbb{C}$. Their ratio $\l_2/\l_1\notin\mathbb{R}$ is called the \textit{modulus}. We will take $z$ throughout to be valued in the \textit{universal cover} $\td{\mathbb{T}}=\mathbb{C}$, in which case functions of $z$ should be doubly periodic with periods $\lambda_1$ and $\lambda_2$ in order to be well-defined on $\mathbb{T}$.
Note that a third class of models, the trigonometric integrable field theories, are a special case obtained by sending one of the $|\lambda_a|$ to infinity, while the rational integrable models are recovered by sending both to infinity. In particular, the RG flow of trigonometric models can be derived as a special case of what follows.

We will proceed in three steps. First, we will outline how elliptic integrable field theories can be constructed using 4d Chern-Simons theory, as recently highlighted in \cite{LW1}. Next, we will consider how the Lax connection depending on a torus-valued spectral parameter modifies the computation of 1-loop divergences: ultimately, we will still obtain the same contour integral \eqref{frr} in terms of the Lax connection. Finally, we will rewrite this result as a 4d counterterm and interpret it as an RG flow of the data defining the theory, in particular of the twist 1-form $\omega$. In the process, we will encounter a new term, coming from the fact that the periods of the torus generally flow under the RG, thus modifying the Riemann surface on which the 4d Chern-Simons theory is defined.

\subsection{Review of equivariant elliptic 4d Chern-Simons Theory}\label{subsec:elliptic4dcs}
Instead of reviewing 4d Chern-Simons theory from scratch, let us simply highlight the modifications to the theory constructed for the rational case in Section \ref{sec:review}. In the Belavin-Drinfel'd classification of integrable field theories, elliptic models only exist if the Lax connection takes its values in the algebra $\mathfrak{sl}_{\mathbb{C}}(N)$. As such, we naturally restrict our attention to gauge groups with complex algebra $\mathfrak{g}^{\mathbb{C}} = \mathfrak{sl}_{\mathbb{C}}(N)$ for the 4d Chern-Simons theory. 
The 4-dimensional manifold on which we construct the theory is $\Sigma\times\mathbb{T}$ and the action is
\begin{equation}
\label{eq:ellaction}    S_{4d}[A]=\frac{1}{16\pi^2i}\int_{\Sigma\times\mathbb{T}}\omega\wedge \text{CS}[A]\,,\qquad \omega=\varphi(z) \, \dd z\,.
\end{equation}
For $\omega$ to be a proper 1-form on the torus $\mathbb{T}$, $\varphi$ must satisfy
\begin{equation}
    \varphi(z+\lambda_a)=\varphi(z)\label{eq:lshift}\,.
\end{equation}
The zeros and poles of $\omega$ will again be denoted respectively by $z_i^\pm$ and $p_r$. For simplicity, we will once again require that all the poles $p_r$ are real and simple: complex and higher-order poles require a slight adaptation of the more general setup of Appendix \ref{app:GenPoles}. For the action to be real, one should take all $z_i^\pm$ to be real, and the periods $\lambda_1$ and $\lambda_2$ to be respectively real and imaginary (i.e.\ take the unit cell to be a rectangle)---but this will not affect the validity of what follows. Indeed, under these assumptions, it was argued in \cite{LW1} that the action \eqref{eq:ellaction} is real upon imposing the equivariance \eqref{eq:RealityA}, with the real form $\mathfrak{g}$ being $\mathfrak{sl}_\mathbb{R}(N)$.
As before, we will impose that the gauge field $A$ is smooth and regular, except at the $2M$ zeros of $\omega$ (counted with multiplicity), where we allow singularities of the type \eqref{eq:Disorder}. Similarly, at the $2M$ poles\footnote{Note that a 1-form on $\mathbb{T}$ has the same number of poles as zeros, as opposed to the relation \eqref{eq:M2} on $\CP$.} of $\omega$, we will again impose the boundary condition that the jet $\mathbb{A}$ is valued in a maximally isotropic subalgebra $\mathfrak{b}$ of the defect algebra $\mathfrak{d}$. The major difference between 4d Chern-Simons theory on $\Sigma\times\mathbb{T}$ and $\Sigma\times \mathbb{CP}^1$ is that we require $A$ to be multivalued on $\mathbb{T}$, such that $A(z+\lambda_a)\neq A(z)$. More precisely, we will enforce that $A$ transforms equivariantly under such shifts:
\begin{equation}
    A(z+\lambda_1)=\sigma_1\left[A(z)\right]\,,\qquad A(z+\lambda_2)=\sigma_2\left[A(z)\right]\,,\label{eq:Autdef}
\end{equation}
where $\sigma_{1,2}$ are two \textit{inner automorphisms} of $\mathfrak{sl}_\mathbb{C}(N)$, first considered in \cite{belavin_discrete_1981}. We refer to \cite{LW1} for more details and an explicit expressions of the $\sigma_a$. Although $A$ is multi-valued on $\mathbb{T}$,\footnote{Since $\left(\sigma_a\right)^N=\mathbb{I}$, then $A$ is in fact well-defined on a larger torus with periods $N\lambda_a$.} note that the Chern-Simons 3-form is single-valued, so the action \eqref{eq:ellaction} is well defined: this follows from \eqref{eq:Autdef} and the invariance of the bilinear form under automorphisms.

\paragraph{Dynamics.}
Having fixed the properties of the gauge field $A$, we can proceed with constructing the 4d Chern-Simons theory. As before, the bulk equations of motion obtained by varying \eqref{eq:ellaction} are
\begin{equation}
    \omega\wedge \text{F}[A]=0\,.\label{eq:elleom}
\end{equation}
 The variation of the action also contains the defect and boundary terms \eqref{eq:DeltaDefect} and \eqref{eq:DeltaBoundary}. The latter, which we recall here,
\begin{equation}
    \delta S^{\text{bdry}}_{4d}= \frac{1}{16\pi^2i}\int_M\dd \left(\omega\wedge\left<A,\delta A\right>\right)\,, \label{eq:DeltaBoundary2}
\end{equation}
was trivial in the rational case, but it will become important for the elliptic story since the boundary of $\mathbb{T}$ is non-empty when seen as a subset of $\mathbb{C}$. This boundary term still vanishes when its integrand is elliptic, as is the case at present due to \eqref{eq:lshift} and \eqref{eq:Autdef}; however, when we renormalise below, such a term will crop up with a non-elliptic integrand, giving a non-zero contribution.

\paragraph{$\bm{\gh}$ and the Lax connection.} As in the rational case, we reparametrise $A$ in terms of a connection $L=L_+\,\dd x^++L_-\,\dd x^-$ and a group-valued field $\gh$ in the form
\begin{equation}
    A\leftrightarrow (
\gh, L) \ , \qquad A=\gh L\gh^{-1}-\dd\gh\gh^{-1}\,.\label{eq:AgLE}
\end{equation}
The existence of such a $\gh$ is subtle \cite{CWY1,CWY2,Benini} but, as argued in \cite{LW1}, we expect it to exist thanks to the equivariance property \eqref{eq:Autdef} of $A$. Assuming the existence of $\gh$, it is uniquely\footnote{Up to multiplication by a matrix $N$-th root of unity.} specified by its equivariant transformation rule~\cite{LW1}
\begin{equation}
\gh(z+\lambda_a)=\sigma_a[\gh(z)]\,.\label{eq:gshift}
\end{equation}
In writing this down, we have for simplicity not distinguished between the automorphism acting on the algebra $\mathfrak{sl}_\mathbb{C}(N)$ as in \eqref{eq:Autdef} and on the group $\text{SL}_\mathbb{C}(N)$.

 With the parametrisation \eqref{eq:AgLE}, the equation of motion \eqref{eq:elleom} again implies that the connection $L=L_+\,\dd x^++L_-\,\dd x^-$ is flat and meromorphic in $z$. We will thus interpret it as the Lax connection of the resulting 2d theory. The allowed singularities of the gauge field imply that, around a zero $z_i^\pm$ of $\omega$ in the set $\hat{\mathbb{Z}}^\pm$, this Lax connection takes the form
\begin{equation}
    L_\pm(z)=\sum_{k=0}^{m_i^\pm-1}\frac{\mathcal{J}_\pm^{(i,k)}}{\left(z-z_i^\pm\right)^{k+1}}+O\left((z-z_i^\pm)^0\right)\,,\label{eq:LLocal}
\end{equation}
where $\mathcal{J}_\pm^{(i,k)}$ are a set of 2d currents on $\Sigma$. Moreover, these are the unique poles of $L_\pm(z)$ inside the unit cell $\mathbb{T}$. Secondly, the combination of \eqref{eq:Autdef} and \eqref{eq:gshift} requires that $L_\pm$ satisfies
\begin{equation}
    L_\pm(z+\lambda_a)=\sigma_a\left[L_\pm(z)\right]\,.\label{eq:Lshift}
\end{equation}

\paragraph{Extracting the 2d theory.} The 1-form $\omega$ has $2M$ poles $p_r$, at which 4d gauge transformations satisfy specific boundary conditions (see equation \eqref{gbc} and nearby). Fixing this 4d gauge symmetry, we once again conclude that the physical degrees of freedom of $\gh$ consist of a 2d field $\gb$ valued in the quotient $B\backslash D$ and built from the evaluations of $\gh$ at the poles $p_r$. Here, $D=G^{2M}$ is the defect group attached to these poles and $B$ is the maximally isotropic subgroup encoding the boundary conditions. Note that, when considering an elliptic integrable field theory, the 2d diagonal gauge symmetry \eqref{eq:gLv} of the rational model is no longer present, since it is incompatible with the equivariance \eqref{eq:gshift} of $\gh$: therefore, we will not need to further quotient by $G_\text{diag}$. The $B\backslash D$-valued field $\gb$ is thus completely physical and will serve as the fundamental field of the integrable 2d $\s$-model. In particular, the target space of this model has dimension $\dim (B\backslash D)=M\dim\gf$.

Let us now focus on the Lax connection $L_\pm(z)$, which satisfies the equivariance property \eqref{eq:Lshift} and has singularities \eqref{eq:LLocal} in terms of $\gf$-valued currents $\bigl\lbrace \mathcal{J}_\pm^{(i,k)} \bigr\rbrace_{i=1,\dots,N_\pm}^{k=0,\dots,m_i^\pm-1}$. As it turns out, this completely fixes $L_\pm(z)$ as a linear function of these currents (see the equation \eqref{eq:ellLR} below for the explicit expression). The light-cone components $L_\pm(z)$ are then each expressed in terms of $M\dim\gf$ degrees of freedom. Moreover, they are each subject to $\dim\kf=M\dim\gf$ boundary conditions at the poles $p_r$, relating these entries to the field $\gb$ and its $\del_\pm$-derivative. Solving these relations then allows one to express the Lax connection as a functional $L_\pm[\gb]$ of this 2d field.  Finally, similarly to the rational case (around \eqref{eq:Action2d}), the 2d action $S_{2d}[\gb]$ can be obtained by substituting this into the 4d action and integrating over the $\mathbb{T}$ directions. 

Contrary to the rational case \eqref{eq:Lz}, we note that the elliptic Lax connection $L_\pm(z)$ does not have a constant term $\mathcal{J}^{(0,0)}_\pm$, since it is forbidden by the equivariance property \eqref{eq:Lshift}. This aligns nicely with our previous observation that equivariance excludes $G_{\text{diag}}$ symmetry: indeed, as explained around equation \eqref{eq:J0v}, the current $\Jc^{(0,0)}_\pm$ played the role of a gauge connection for this symmetry in the rational case. These statements are further reflected in the counting of degrees of freedom of the theory. In the rational case, a twist 1-form with $2M$ zeros (with multiplicities) has $2(M+1)$ poles: there are then $(M+1)\dim\gf$ scalar components in the $B\backslash D$-valued field $\gb$ and $(M+1)\dim\gf$ current components in each of $L_+(z)$ and $L_-(z)$, all subject to the $G_{\text{diag}}$ gauge symmetry. In the elliptic case, a twist 1-form with the same number of zeros has 2 fewer poles, which precisely compensates for the absence of $\mathcal{J}^{(0,0)}_\pm$ and the $G_{\text{diag}}$ symmetry. Indeed, there are now $M\dim\gf$ scalar components in $\gb$ and $M\,\dim\gf$ current components in $L_\pm(z)$. Note that in both cases the physical target space of the $\sigma$-model has dimension $M\dim\gf$.

\paragraph{A digression on Belavin's $\mathcal{R}$-matrix.}
Having built field theories whose equations of motion can be expressed as the flatness of an elliptic Lax connection $L_\pm(z)$, we would like a more explicit expression for $L_\pm(z)$ in terms of the currents $\Jc^{(i,k)}_\pm$. This will allow us to repeat the steps of Section \ref{sec:rational} to renormalise the theory. Such an explicit form of  $L_\pm(z)$ was given in~\cite{LW1} in the case where it has simple poles. Its generalisation to arbitrary poles reads
\begin{equation}
    L_\pm(z)=\sum_{i=1}^{N_\pm}\sum_{k=0}^{m_i^\pm-1}\frac{(-1)^k}{k!}\left<\partial_z^k\,\mathcal{R}_{\text{Bel}}(z-z_i^\pm)_{\tn{12}},\mathcal{J}_{\pm,\tn{2}}^{(i,k)}\right>_{\tn{2}}\,.\label{eq:ellLR}
\end{equation}
To justify this expression, let us explain its ingredients and their properties. We use the notation $\tn{i}$ to denote the $i$'th copy of $\mathfrak{sl}_\mathbb{C}(N)$ in a tensor product and $\left<\cdot,\cdot\right>_{\tn{i}}$ to indicate that we trace only over the $i$'th tensor factor. $\mathcal{R}_{\text{Bel}}$ is \textit{Belavin's $\mathcal{R}$-matrix}, first constructed in \cite{belavin_solutions_1982}. This is a meromorphic map $\mathcal{R}_{\text{Bel}}:\mathbb{C}\to \mathfrak{sl}_{\mathbb{C}}(N)_{\tn{1}}\otimes\mathfrak{sl}_\mathbb{C}(N)_{\tn{2}}$, which satisfies
\begin{subequations}
\begin{equation}
    \mathcal{R}_{\text{Bel}}(z)_{\tn{12}}=\frac{C_{\tn{12}}}{z}+O(z)\,,\label{eq:Bel1}
    \end{equation}
    \begin{equation}
    \mathcal{R}_{\text{Bel}}(z+\lambda_a)_{\tn{12}}=\sigma_{a,\tn{1}}\left[\mathcal{R}_{\text{Bel}}(z)_{\tn{12}}\right]\,.\label{eq:Bel2}
    \end{equation}
\end{subequations}
In the first equation, $C_{\tn{12}}\in\mathfrak{sl}_\mathbb{C}(N)\otimes\mathfrak{sl}_\mathbb{C}(N)$ denotes the \textit{split Casimir} of $\mathfrak{sl}_\mathbb{C}(N)$, which itself satisfies the following completeness relation for any $X\in\mathfrak{sl}_\mathbb{C}(N)$:
\begin{equation}
    \left<C_{\tn{12}},X_{\tn{1}}\right>_{\tn{1}}=X_{\tn{2}}\label{eq:Casimir}\,.
\end{equation}
Let us now explain why the Lax connection takes the form \eqref{eq:ellLR}. From the 4d Chern-Simons construction, $L_\pm(z)$ is characterised by the following conditions: it is a meromorphic function of $z$, with expansion \eqref{eq:LLocal} around its poles $z_i^\pm$ and satisfying the equivariance property \eqref{eq:Lshift}. Using the two properties \eqref{eq:Bel1} and \eqref{eq:Bel2} of Belavin's $\mathcal{R}$-matrix, it is clear that the expression of $L_\pm(z)$ proposed in equation \eqref{eq:ellLR} satisfies these conditions. Conversely, any connection $L_\pm(z)$ satisfying these two conditions must take the form \eqref{eq:ellLR}.\footnote{Indeed, one easily shows that the difference between $L_\pm(z)$ and the right-hand side of equation \eqref{eq:ellLR} is a holomorphic function on the torus (with no poles). By Liouville's theorem, this difference is then a constant ($z$-independent) $\gf^\C$-valued current, which is further vanishing due to the equivariance property \eqref{eq:Lshift}. We therefore identified $L_\pm(z)$ with the right-hand side of \eqref{eq:ellLR}.}

\subsection{Universal result in terms of the  \texorpdfstring{$\mathcal{R}$}{R}-matrix}\label{sec:Rmat}
We will now generalise the treatment of the universal 1-loop divergences in Section \ref{subsec:univ1loop} to elliptic integrable field theories. In fact, our construction will work for any Lax connection that can be expressed in terms of an $\mathcal{R}$-matrix as:
\begin{equation}
    L_{\pm}(z)_{\tn{1}}=\sum_{i=1}^{N_\pm}\sum_{k=0}^{m_i^\pm-1}\frac{(-1)^k}{k!}\left<\partial^k_z\mathcal{R}(z-z_i^\pm)_{\tn{12}},\mathcal{J}_{\pm,\tn{2}}^{(i,k)}\right>_{\tn{2}}\label{eq:LR}\,.
\end{equation}
This clearly includes the elliptic Lax connection \eqref{eq:ellLR}, but also the rational one \eqref{eq:Lz}, after gauge fixing $G_\text{diag}$, if one takes $\mathcal{R}$ to be the \textit{Yangian $\mathcal{R}$-matrix}
\begin{equation}
    \mathcal{R}_\text{Yang}(z)_{\tn{12}}=\frac{C_{\tn{12}}}{z}\,,\label{Yang}
\end{equation}
as well as trigonometric integrable theories upon taking appropriate limits.
We will require that $\mathcal{R}$ satisfies the following three properties, which implies in particular that it belongs to the Belavin-Drinfel'd classification \cite{belavin_solutions_1982,Belavin:1983cYB} (see also the lectures~\cite{Torrielli:2016CI} for a review). First, it should be skew-symmetric, \textit{i.e.} $\mathcal{R}(z)_{\tn{12}}=-\mathcal{R}(-z)_{\tn{21}}$.
Secondly, to have an interpretation as a classical $\mathcal{R}$-matrix, it should satisfy the \textit{classical Yang-Baxter equation (cYBE)}
\begin{equation}
    \left[\mathcal{R}(z_{12})_{\tn{12}},\mathcal{R}(z_{13})_{\tn{13}}\right]+\left[\mathcal{R}(z_{12})_{\tn{12}},\mathcal{R}(z_{23})_{\tn{23}}\right]+\left[\mathcal{R}(z_{32})_{\tn{32}},\mathcal{R}(z_{13})_{\tn{13}}\right]=0\,,\label{eq:cYBE}
\end{equation}
with $z_{ij}=z_i-z_j$.
As we will see in the next paragraph, this relation will be crucial for us to rewrite the RG flow arising from the Lax connection \eqref{eq:LR} in a compact form. Lastly, around $z=0$ it should satisfy
\begin{equation}
    \mathcal{R}(z)_{\tn{12}}=\frac{C_{\tn{12}}}{z}+O(z)\label{eq:RAsym}\,,
\end{equation} 
which when inserted in \eqref{eq:LR} will allow the Lax connection's poles to be expressed in terms of the currents $\mc{J}^{(i,k)}_\pm$. We remark that this property is slightly stronger than the most general one in the Belavin-Drinfel'd classification, where the $O(z)$ correction may be replaced by $O(z^0)$. This technical difference, which is needed for some of the following computations, means that we are considering only a subset of $\mathcal{R}$-matrices from this classification. 

Note that, from our assumptions, it can furthermore be shown that any pole of $\mathcal{R}(z)$ at $z\neq 0$ will be related to the one at $z=0$ by equivariance \cite{belavin_solutions_1982}, in the following sense. The poles of $\mathcal{R}(z)$ form a real lattice $\Lambda\subset\C$ (which is 0-dimensional, 1-dimensional or 2-dimensional) with generators $\lambda_a$. Moreover, there exist automorphisms $\s_a$ of $\gf^\C$ such that $\mathcal{R}(z+\lambda_a)_{\tn{12}}=\sigma_{a,_{\tn{1}}}[\mathcal{R}(z)_{\tn{12}}]$. As a consequence, the poles of the Lax connection \rf{eq:LR} are the obvious ones at $z=z_i^\pm$ together with their equivariant images (with the residues related by the automorphisms $\sigma_a$).

\paragraph{Computing the 1-loop divergence.} 
Although the general form of the Lax \eqref{eq:LR} in terms of $\mc R$-matrices might seem abstract, as we will see below, the cYBE satisfied by the $\mc R$-matrix is exactly what is needed for the argument to go through. The computation is rather technical and we will relegate most of it to Appendix \ref{sec:RmatAppendix}, simply highlighting the main points here. 

First, we would like to extract the integrable field theory's equations of motion, which are the zero-curvature equations of the Lax connection \eqref{eq:LR}. In the rational case, this was done in equations \eqref{z} by decomposing the curvature of the Lax into partial fractions. As we show in Appendix \ref{subsec:EOM}, the cYBE \eqref{eq:cYBE} for general $\mathcal{R}$-matrices plays a precisely analogous role to this partial fraction decomposition, allowing us again to express the equations of motion of the currents in the following compact form: 
\begin{equation}
\begin{split}
    &\hspace{-0.3cm}\partial_+\mathcal{J}_{-,\tn{1}}^{(i,\ell)}+\sum_{j=1}^{N}\sum_{n=0}^{m_j^+-1}\sum_{m=\ell}^{m_i^--1}\frac{(-1)^n}{n!(m-\ell)!}\left[\left<\partial_{z_i^-}^{n+m-\ell}\mathcal{R}({z}_i^--{z}^+_j)_{\tn{12}},\mathcal{J}_{+,\tn{2}}^{(j,n)}\right>_{\tn{2}},\mathcal{J}_{-,\tn{1}}^{(i,m)}\right]=0\,,\\
    &\hspace{-0.3cm}\partial_-\mathcal{J}_{+,\tn{2}}^{(j,n)}+\sum_{i=1}^{N}\sum_{\ell=0}^{m_i^--1}\sum_{m=n}^{m_j^+-1}\frac{(-1)^\ell}{\ell!(m-n)!}\left[\left<\partial_{z_j^+}^{\ell+m-n}\mathcal{R}({z}_j^+-{z}^-_i)_{\tn{21}},\mathcal{J}_{-,\tn{1}}^{(i,\ell)}\right>_{\tn{1}},\mathcal{J}_{+,\tn{2}}^{(j,m)}\right]=0\,.
    \end{split}
\end{equation}
Since these equations of motion still match the abstract form assumed in \eqref{eq:OPeq},
the 1-loop calculation carried out in Section~\ref{subsec:PathInt} still applies. The general result \rf{renre} gives in this case (see Appendix \ref{subsec:1loop} for details) 
\begin{equation}
    \frac{\dd}{\dd\text{t}}\widehat S^{(1)}=\frac{\cG}{2\pi}\int\ud^2x\sum_{i,j=1}^N\sum_{\ell=0}^{m_i^--1}\sum_{n=0}^{m_j^+-1}\frac{(-1)^{\ell}}{\ell!n!}\left<\partial_{z^+_j}^{n+\ell+1}\mathcal{R}(z_j^+-z_i^-)_{\tn{12}},\mathcal{J}_{-,\tn{1}}^{(i,\ell)}\mathcal{J}_{+,\tn{2}}^{(j,n)}\right>_{\tn{12}} \ .
\label{eq:genflow}
\end{equation}
Remarkably, this can still be rewritten as the same contour integral formula as in the rational case:
\begin{align}
    \frac{\dd}{\dd\text{t}}\widehat{S}^{(1)}=\frac{c_{G}}{2
\pi}\int\ud^2x\oint_{{}_-\Gamma_+}\frac{\ud z}{2\pi i}\left<L_+(z),\partial_zL_-(z)\right>\,, \label{eq:contourell}
\end{align}
with ${}_-\Gamma_+$ encircling $\hat{\mathbb{Z}}^+$ but not $\hat{\mathbb{Z}}^-$.
To see that \eqref{eq:contourell} reproduces \eqref{eq:genflow}, we can express the contour integral as a sum of residues of $L_+$ at the poles $z=z_k^+$: 
\begin{align}
    &\frac{c_{G}}{2\pi}\oint_{{}_-\Gamma_+}\frac{\ud z}{2\pi i}\left<L_+(z),\partial_zL_-(z)\right> \label{eq:contell}\\
    =&\sum_{k=1}^N\text{res}\left\{\sum_{i,j=1}^N\sum_{\ell=0}^{m_i^--1}\sum_{n=0}^{m_j^+-1}\frac{(-1)^{n+\ell}}{n!\ell!}\left<\left<\partial_z^{\ell+1}\mathcal{R}(z-{z}_i^-)_{\tn{31}},\partial_z^{n}\mathcal{R}(z-z_j^+)_{\tn{32}}\right>_{\tn{3}},\mathcal{J}_{-,\tn{1}}^{(i,\ell)}\mathcal{J}^{(j,n)}_{+,\tn{2}}\right>_{\tn{12}}\right\}_{z=z_k^+}\,. \no
    \end{align}
Then expanding $\mathcal{R}(z-z^+_j)$ in the vicinity of such a pole using \eqref{eq:RAsym} allows us to identify
\begin{equation}
\begin{split}
    &\quad \ \text{res}\left\{\frac{(-1)^n}{n!}\left<\partial_z^{\ell+1}\mathcal{R}(z-{z}_i^-)_{\tn{31}},\partial_z^{n}\mathcal{R}(z-{z}^+_j)_{\tn{32}}\right>_{\tn{3}}\right\}_{z={z}^+_k}\\
    & =\text{res}\left\{\left<\partial_z^{\ell+1}\mathcal{R}(z-{z}^-_i)_{\tn{31}},C_{\tn{32}}\right>_{\tn{3}}\frac{\delta_{j=k}}{(z-{z}_k^+)^{n+1}}\right\}_{z={z}^+_k}=\frac{\delta_{j=k}}{n!} \, \partial_{z_j^+}^{n+\ell+1}\mathcal{R}({z}^+_j-{z}^-_i)_{\tn{21}}\,,
    \end{split}
\end{equation}
 where we used \eqref{eq:Casimir} in the last step. Substituting this into \eqref{eq:contell} then precisely reproduces \eqref{eq:genflow}.

\subsection{Interpretation as a flow of \texorpdfstring{$\omega$}{omega} and the torus periods}
We have seen that the features of the universal result from the rational case generalise rather seamlessly to the elliptic case. We rephrased the universal computation of the 1-loop divergences in terms of an $\mathcal{R}$-matrix appearing in the classical Lax connection. This formulation leads to the contour integral formula \eqref{eq:contourell} in general, and in particular in both the rational and elliptic cases. One therefore has good reason to hope that the further miracles of Section \ref{subsec:ct}---distributional and variational terms conspiring to allow the universal counterterm to be interpreted as a flow of $\omega$---should generalise to the elliptic case. We will show in this section that this is indeed the case. 

Even with all of the new complications associated with periodicity on the torus and the presence of its modulus, we will find that the universal result is interpretable as the \textbf{same flow} \eqref{fom} of the twist 1-form $\omega$ in terms of a suitably generalised function $\Psi$. From this, we will also extract the flow of the torus' periods. Our findings here are consistent with refs.\ \cite{Derryberry:2021rne,LW2}, which found the same flow of $\omega$ for elliptic models as for rational ones.

\paragraph{1-loop divergence as a 4d counterterm.}
Just as in the rational case, the contour integral formula \eqref{eq:contourell} is equivalent to the 4d counterterm \eqref{combi}, namely\footnote{In deriving this equality, we need to produce the elliptic equivalent of equation \eqref{frr2} from the counterterm \eqref{eq:contourell}. This is done by deforming the integration contour from ${}_-\Gamma_+$ (enclosing the points $\hat{\mathbb{Z}}^+$) to ${}_+\Gamma_-$ (enclosing $\hat{\mathbb{Z}}^-$) with opposite orientation. When working with contour integrals in $\C$, one might worry that this will produce contributions coming from the equivariant images of $\hat{\mathbb{Z}}^\pm$.
But, since the integrand $\langle L_+,\del_zL_-\rangle$ is elliptic, then this contour deformation really takes place on the compact torus where $\hat{\mathbb{Z}}^\pm$ are the only poles.}
\be
\ddt \widehat S^{(1)} = -\frac{1}{16\pi^2 i} \int_{\Sigma\times\mathbb{T}}  \Psi(z) \, \ud {\rm CS}[A] \,, \la{DSL}
\ee
where $\Psi:\mathbb{C}\rightarrow \mathbb{C}$ again satisfies \eqref{psi} at the zeros of $\omega$. To further fix this function, we still require that it has the same pole structure as the twist 1-form and additionally that it satisfies the following quasi-periodicity condition:
\begin{equation}
\exists\, \b_1,\beta_2\in\mathbb{C} \quad \text{such that} \quad \Psi(z+\lambda_a)=\Psi(z)-\beta_a\,\varphi(z)\, . \label{Psibeta}
\end{equation}
This requirement will be important later when we want to identify $\del_z\Psi$ as the flow of $\omega$. This fixes $\Psi$ up to two unconstrained variables, which will simply correspond to translations and dilations of $z$ along the RG-flow, similarly to the three unfixed parameters in the rational case discussed in footnote~\ref{foot:mobius}. (See also~\cite{LW2} for an explicit construction of $\Psi$ in the case of simple zeros).\footnote{Note that, up to the aforementioned dilation freedom, the $\beta_a$ are not independent of the parameters of the theory: just like the residues $\psi_r$ defined in \eqref{eq:psir}, they are specific functions of the data $(\mathbb{T},\omega,\Zh^\pm)$, obtained by solving the extrapolation problem for $\Psi$.}

\begin{figure}
	\centering
	\begin{tikzpicture}
		\filldraw[color=red,fill=red!30, thick] (9,0)--(10.5,0)--(12.67,6.5)--(2.17,6.5)--(2,6)--(11,6)--(9,0);
		\draw[line width= 0.5 mm,->] (0,0)--(4.4,0);
		\draw[line width= 0.5 mm,->] (4.4,0)--(4.6,0);
		\draw[line width= 0.5 mm,->] (11,6)--(10,3);
		\draw[line width= 0.5 mm,->] (2,6)--(1,3);
		\draw[line width= 0.5 mm,->](2,6)--(6.4,6);
		\draw[line width= 0.5 mm,->](6.4,6)--(6.6,6);
		\draw[line width= 0.5 mm] (0,0)--(9,0)--(11,6)--(2,6)--(0,0);
		
        \node[label=south: { $\mathcal{C}_2$}] at (4.5,0) {};
        \node[label=west: { $\mathcal{C}_1$}] at (1,3) {};
        \node[circle,fill=black,inner sep=0pt,minimum size=5pt,label=south west:{\scriptsize $z_0$}] at (0,0) {};
		\node[circle,fill=black,inner sep=0pt,minimum size=5pt,label=south west:{\scriptsize $z_0+\lambda_1$}] at (9,0) {}; 
		\node[circle,fill=black,inner sep=0pt,minimum size=5pt] at (11,6) {};
		\node[label=south west:{\scriptsize $z_0+\lambda_1+\lambda_2$}] at (10.8,6) {};
		\node[circle,fill=black,inner sep=0pt,minimum size=5pt,label=south east:{\scriptsize $z_0+\lambda_2$}] at (2,6) {};
		
		\node[circle,fill=black,inner sep=0pt,minimum size=5pt,label=south east:{\scriptsize $z_0+\lambda_1'$}] at (10.5,0) {}; 
		\node[circle,fill=black,inner sep=0pt,minimum size=5pt,label=north:{\scriptsize $z_0+\lambda'_1+\lambda_2'$}] at (12.67,6.5) {};
		\node[circle,fill=black,inner sep=0pt,minimum size=5pt,label=north east:{\scriptsize $z_0+\lambda_2'$}] at (2.17,6.5) {};
		
	\end{tikzpicture}
	
	\caption{The original unit cell representing $\mathbb{T}$ (marked in white) with basis vectors $\lambda_1$ and $\lambda_2$ and boundary marked in black. Under RG, the surface changes its shape and is defined in terms of new basis vectors $\lambda_1'$ and $\lambda_2'$: this introduces an additional area (marked in red) that should be integrated over in the effective action, which in turn results in $C_{\text{moduli}}$. $\mathcal{C}_1$ and $\mathcal{C}_2$ denote the contours along the left and bottom boundaries, respectively, with orientation indicated by the arrows.}
	\label{fig:torusflow}
\end{figure}
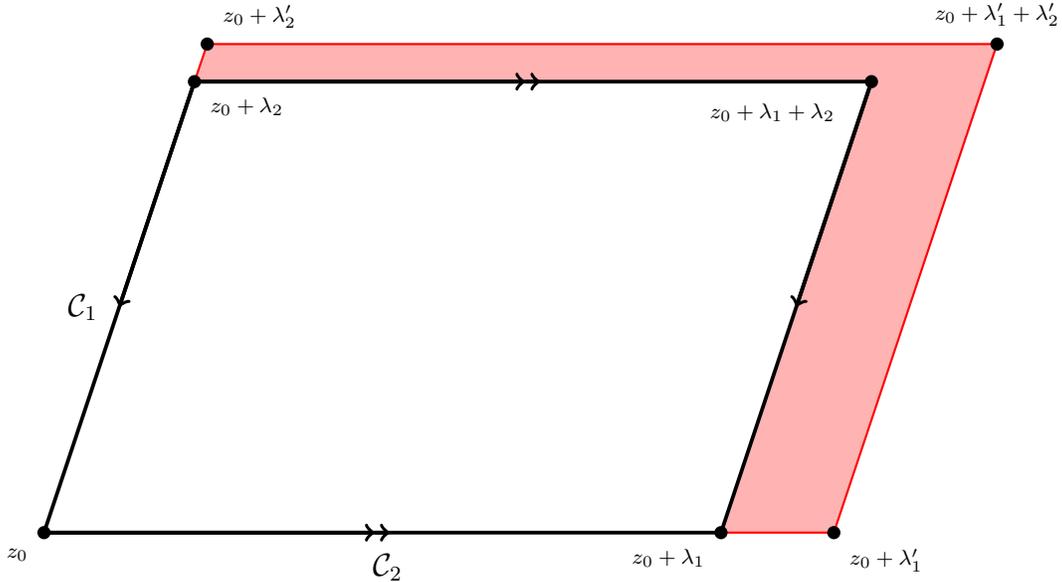

As in the rational case, we then use integration by parts in \rf{DSL} to move the derivative to $\Psi$. The result can be written as a bulk term plus lower-dimensional localised ones:
\be
\ddt \widehat S^{(1)} = \frac{1}{16\pi^2i} \int_{\Sigma\times\mathbb{T}}\del_z \Psi(z) \, \ud z \wedge {\rm CS}[A] + \DT +C_{\text{bdry}}\,, \label{ldl}
\ee
with
\be
\DT = \frac{1}{16\pi^2i}  \int_{\Sigma\times\mathbb{T}} \del_{\bar z} \Psi(z) \, \ud \bar z \wedge {\rm CS}[A] \,,\qquad C_{\text{bdry}}=-\frac{1}{16\pi^2i }\int_{\Sigma\times\partial \mathbb{T}}\Psi(z){\rm CS}[A]\,\label{eq:Cdefbound}
\ee
The first `defect' term is the same as in the rational case above, and localises to the poles of $\Psi$. 
The second `boundary' term, arising from integration by parts, is a new feature of the elliptic models, since the torus $\mathbb T$ when regarded as a unit cell inside its universal cover $\mathbb C$ has a boundary (as shown in Figure~\ref{fig:torusflow}).\footnote{Note that we assume that $A$ decays sufficiently fast towards the boundary of $\Sigma$ to discard any $\partial\Sigma$ term.} While the boundary term of any elliptic integrand (\textit{i.e.}\ one well-defined on the torus and thus doubly periodic) would cancel between parallel sides, the pseudo-ellipticity of $\Psi$ \eqref{Psibeta} leads to a non-vanishing boundary term
\begin{equation}\label{CBound}
	C_{\text{bdry}}=-\frac{1}{16\pi^2i}\sum_{a=1,2}\int_{\Sigma\times\mathcal{C}_a}\beta_a\,\varphi(z)\, \text{CS}[A]\,,
\end{equation}
with $\mathcal{C}_a$ the contours defined in Figure \ref{fig:torusflow}.
We will see shortly that similar boundary terms will appear naturally from the flow of $\omega$.

\paragraph{Flow of $\omega$.}
The first term in \eqref{ldl} seems to have a natural interpretation as the same flow of the twist 1-form as in the rational case:
\begin{equation}
	\frac{\ud \omega}{\ud \text{t}}=\partial_z\Psi(z) \, \dd z\,. \label{EF}
\end{equation}
The similarity in the RG flow between rational and elliptic theories was already conjectured in \cite{LW2,Derryberry:2021rne}. Recall that in the rational case, the flow of the poles and residues could be read off directly from that of $\omega$: the same equations \eqref{eq:dlp} remain true in the elliptic case, with $\psi_r$ identified with the residue of $\Psi(z)$ at $z=p_r$. Similarly, we find that the additional data in the elliptic case, namely the periods of the torus, also flow under RG and that this can be extracted from equation \eqref{EF} as well. Namely, requiring that the twist function remains elliptic \eqref{eq:lshift} under RG flow imposes
	\begin{equation}
		\frac{\dd}{\dd\text{t}}\left[\varphi(z+\lambda_a)-\varphi(z)\right] =
		\partial_z\Psi(z+\lambda_a) + \frac{\dd\lambda_a}{\text{dt}}\partial_z\varphi(z+\lambda_a) - \partial_z\Psi(z) \overset{!}{=} 0\,.\label{eq:readofdldt}
	\end{equation}
Using that $\varphi$ is elliptic and the pseudo-elliptic property \eqref{Psibeta} of $\Psi$ imposed earlier, we can then identify
\begin{equation}
    \frac{\dd\lambda_a}{\dd\text{t}}=\beta_a\label{lambdaflow}\,.
\end{equation}

We now want to prove the conjecture \eqref{EF}. As in the rational case, we will start from the conjectured flow of $\omega$ and show that the resulting flow of the 1-loop effective action matches the universal counterterm \rf{ldl}. A new feature is that the domain of integration (\textit{i.e.}\ the unit cell of the torus) itself flows, since we just found that the periods on the torus are running (see Figure \ref{fig:torusflow}). Leibniz' integration rule then tells us that the conjectured flow of the action takes the form
\begin{equation}
	\begin{split}
		&\frac{\dd}{\dd\text{t}}\widehat{S}^{(1)}_{\text{conj}}=\frac{1}{16\pi^2i}\int_{\Sigma\times\mathbb{T}}
		\frac{\dd\omega}{\text{dt}}\wedge\text{CS}[A]+C_{\text{dist}}+C_{\text{var}}+C_{\text{moduli}}\,,\\
        &\qquad \qquad C_{\text{moduli}}=\frac{1}{16\pi^2i}\int_{\Sigma\times\partial\mathbb{T}}\iota_{v}\left(\omega\wedge\text{CS}[A]\right)\,,
	\end{split}
\end{equation}
where the additional `moduli' term $C_{\text{moduli}}$ corresponds to the running of the modulus of the torus $\mathbb{T}$ with $v$ denoting the infinitesimal flow vector of the boundary, which will be discussed below. 
The quantities $C_{\text{dist}}$, coming from the difference between the flow of $\omega$ as a measure and as a 1-form, and $C_{\text{var}}$, coming from the fact that the background field $A$ depends on the parameters that are flowing, are broadly the same as in the rational case and we will explain their new features below.

\paragraph{The elliptic moduli term.} To evaluate the new moduli term, let us first determine the flow field $v=v^z\partial_z+v^{\bar{z}}\partial_{\bar{z}}$. Consider the expression of $z$ and $\bar{z}$ in terms of the periods in the following manner: 
\begin{equation}
	\begin{split}
		&z=z_0+s_1\lambda_1+s_2\lambda_2\,,\qquad\bar{z}=\bar{z}_0+s_1\bar{\lambda}_1+s_2\bar{\lambda}_2\,,\\
		&s_1=\frac{\bar{\lambda}_2(z-z_0)-\lambda_2 (\bar{z}-\bar{z}_0)}{V}\,,\qquad s_2=\frac{\lambda_1( \bar{z}-\bar{z}_0)-\bar{\lambda}_1(z-z_0)}{V}\,,
	\end{split}
\end{equation}
where the unit cell corresponds to the range $s_1,s_2\in[0,1]$ and we defined $V=\lambda_1\bar{\lambda}_2-\lambda_2\bar{\lambda}_1$ (which is proportional to the unit cell's area). We included an arbitrary constant shift $z_0$ in our choice of unit cell (see Figure \ref{fig:torusflow}); we will see that it is useful to choose it such that neither poles, zeros nor the lattice points $\Lambda$ in \eqref{eq:defT} are located on the boundary of the unit cell and the only lattice point in the interior is $z=0$. 
This implies that 
\begin{equation}
\begin{split}
	v^z(z,\bar{z})&=\frac{\dd z}{\text{dt}}=s_1\frac{\dd\lambda_1}{\text{dt}}+s_2\frac{\dd\lambda_2}{\text{dt}}
    \\
    &=\frac{1}{V}\left[\frac{\dd\lambda_1}{\text{dt}}\left(\bar{\lambda}_2(z-z_0)-\lambda_2(\bar{z}-\bar{z}_0)\right)+\frac{\dd\lambda_2}{\text{dt}}\left({\lambda}_1(\bar{z}-\bar{z}_0)-\bar{\lambda}_1(z-z_0)\right)\right]\,,
    \end{split}
\end{equation}
and similarly for $v^{\bar{z}}(z,\bar{z})=\overline{{v}^z(z,\bar{z})}$. We note the following (quasi-)elliptic properties of $v^    z$ and its derivatives:
\begin{equation}
	v^z(z+\lambda_a)=v^z(z)+\frac{\dd\lambda_a}{\text{dt}}\,,\qquad \partial_zv^z(z+\lambda_a)=\partial_zv^z(z)\,,\qquad\partial_{\bar{z}}v^z(z+\lambda_a)=\partial_{\bar{z}}v^z(z)\,,\label{eq:fshift}
\end{equation}
and similarly for $v^{\bar{z}}$.

With these properties of $v$ established, we can now return to the moduli term.
Using that $\iota$ is a derivation, we can write
\begin{equation}
	\iota_v\left(\omega\wedge\text{CS}[A]\right)=\iota_v(\omega)\text{CS}[A]-\omega\wedge\iota_v\left(\text{CS}[A]\right)=v^z(z)\varphi(z)\text{CS}[A]-\omega\wedge\iota_v\left(\text{CS}[A]\right)\,,
\end{equation}
where we used that $\omega$ only has legs along $\dd z$.
For the second term, we can use the expression \eqref{eq:CSA} for the Chern-Simons form to find
\begin{equation}
\begin{split}
	\iota_v\left(\text{CS}[A]\right){}&{}=    2\left<\iota_v[A],\text{F}[A]\right>-\left<\iota_v[A],\dd A\right>-\left<A,\iota_v(\dd A)\right>\,.
\end{split}\end{equation}
Focusing on the last term, we can use Cartan's magic formula to further rewrite
\begin{equation}
	\iota_v\left(\dd A\right)=
	-\dd (\iota_v[A])+A_\mu (\partial_\nu v^\mu)\dd x^\nu+v^\mu\partial_\mu A\,,
\end{equation}
where $\mu=(z,\bar{z})$. From this, we can gather that we have
\begin{align}
\int_{\partial\mathbb{T}\times\Sigma}\omega\wedge\iota_v\left(\text{CS}[A]\right) \no=	\int_{\partial\mathbb{T}\times\Sigma}\Big[&2\,\omega\wedge\left<\iota_v[A],\text{F}[A]\right>+\dd\left(\omega\wedge\left<A,\iota_v[A]\right>\right)-\dd\omega\wedge\left<A,\iota_v[A]\right>\\
	&-\omega\wedge\left<A,A_\mu\right>\partial_\nu v^\mu\dd x^{\nu}-\omega\wedge\left<A,\partial_\mu A\right>v^\mu\Big]\,.
	\end{align}
However, most of these terms actually vanish. The first term is killed by the on-shell equation of motion $\omega\wedge\text{F}[A]=0$; the second term is a total derivative and $\partial\mathbb{T}$ has no boundary (as it is itself a boundary); we chose $z_0$ such that $\omega$ was \textit{holomorphic} when restricted to $\partial\mathbb{T}$, so the third term vanishes; and we saw from \eqref{eq:fshift} that $\partial_\nu v^\mu$ is properly elliptic while we required in \eqref{eq:Autdef} that $A$ transforms equivariantly, which together kill the fourth term. However, the last term is non-vanishing precisely because $v$ is only quasi-elliptic \eqref{eq:fshift}: nevertheless, the partial cancellation means that the moduli term simplifies to
\begin{equation}
	C_{\text{moduli}}=-\sum_{a=1,2}\frac{1}{16\pi^2i}\int_{\Sigma\times\mathcal{C}_a}\frac{\dd\lambda_a}{\dd \text{t}}\, \varphi\,\text{CS}[A]+\frac{\dd\lambda_a}{\text{dt}}\, \omega\wedge\left<A,\partial_z A\right>+\frac{\dd\bar{\lambda}_a}{\dd\text{t}}\, \omega\wedge\left<A,\partial_{\bar{z}}A\right>\,,
\end{equation}
where we recall that the integration contours $\mathcal{C}_a$ are as defined in Figure~\ref{fig:torusflow}.
\paragraph{The elliptic distributional term.}
Like before, we can split $\frac{\dd \omega}{\text{dt}}\big|_{\rm meas}$ into the formal derivative $\frac{\dd \omega}{\text{dt}}$ and a distributional term:
\begin{equation}
	\frac{1}{16\pi^2i}\int_{\Sigma\times\mathbb{T}}
	\dtom\wedge\text{CS}[A]=\frac{1}{16\pi^2i}\int_{\Sigma\times\mathbb{T}}
	\frac{\dd\omega}{\text{dt}}\wedge\text{CS}[A]+C_{\text{dist}}\,. \la{bvt}
\end{equation}
Recall that the distributional contribution can heuristically be seen as arising from the $\bar{p}_r$ derivatives of $\omega$, as argued around \eqref{dpibar}. Since $\omega$ now also depends on $\lambda_a$, one might worry that we need to consider the $\bar{\lambda}_a$ derivatives of $\omega$ when computing the measure derivative. However, this is not the case: one can check that $\frac{\del \omega}{\del \overline{\lambda}}$ localises to the non-zero lattice points. Since we chose $z_0$ such that these were all located outside of the unit cell, such terms do not contribute to the measure derivative. Thus, we can once again conclude that
\begin{equation}
	C_{\text{dist}}=-\frac{1}{8\pi}\sum_{r=1}^N\ell_r\frac{\dd\bar{p}_r}{\dd\text{t}}\int_\Sigma\left<A,\partial_{\bar{z}}A\right>\Big|_{z=p_r}\,.
\end{equation}
\paragraph{The elliptic variational term.}
The situation is different for the variational term, where the elliptic case has some genuinely new features as compared to the rational case. The contribution from the defects still takes the form \eqref{eq:Cvarfinal}, since that computation was local and independent of the global topology of the Riemann surface. However, we now have to consider the additional boundary term \eqref{eq:DeltaBoundary2}, which leads to a variational contribution of the form
\begin{equation}
	\label{eq:flowdefect}\frac{1}{16\pi^2i}\int_{\Sigma\times\partial\mathbb{T}}\omega\wedge\left<A,\frac{\dd A}{\dd \text{t}}\right>\,.
\end{equation}
Similarly to how the quasi-elliptic property of $\Psi$ could be read off from requiring $\varphi$ to remain elliptic under RG \eqref{eq:readofdldt}, the corresponding behaviour of $\frac{\dd A}{\text{dt}}$ can be computed by requiring that the gauge field remains equivariant \eqref{eq:Autdef} under the flow, leading to
	\begin{equation}
		\frac{\dd}{\dd \text{t}}\left[A(z+\lambda_a)-\sigma_a(A(z))\right] =
		\frac{\dd A}{\dd \text{t}}(z+\lambda_a) + \frac{\dd\lambda_a}{\text{dt}}\partial_zA(z+\lambda_a)+\frac{\dd\bar{\lambda}_a}{\text{dt}}\partial_{\bar{z}}A(z+\lambda_a) - \frac{\dd \sigma_a[A]}{\dd \text{t}}(z) \overset{!}{=} 0.
	\end{equation}
Using that $\sigma_a$ is independent of $z$ and constant along the flow and reapplying \eqref{eq:Autdef}, we can then write
	\begin{equation}
		\frac{\dd A}{\text{dt}}(z+\lambda_a)=\sigma_a\left[\frac{\dd A}{\text{dt}}(z)\right]-\frac{\dd \lambda_a}{\text{dt}}\sigma_a[\partial_zA(z)]-\frac{\dd \bar{\lambda}_a}{\text{dt}}\sigma_a[\partial_{\bar{z}}A(z)]\,.\label{eq:dAshift}
	\end{equation}
Inputting the quasi-periodicity \eqref{eq:dAshift} in the boundary variational term \rf{bvt} and adding it to the defect variational term \eqref{eq:Cvarfinal} then yields the total result
\begin{align}
		C_{\text{var}}&=\frac{1}{8\pi} \sum_{r=1}^N \ell_r \int_\Sigma \,\left.\left\langle A, \frac{\ud p_r}{\dd \text{t}}\,\del_z A  + \frac{\ud \overline{p_r}}{\dd \text{t}}\,\del_{\zb} A \right\rangle \right|_{z=p_r}\,\\
		&\qquad +\frac{1}{16\pi^2i}\sum_{a=1,2}\int_{\Sigma\times\mathcal{C}_a}\frac{\dd\lambda_a}{\text{dt}}\,\omega\wedge\left<A,\partial_zA\right>+\frac{\dd\bar{\lambda}_a}{\text{dt}}\,\omega\wedge\left<A,\partial_{\bar{z}}A\right>\,. \no
	\end{align}
\paragraph{Matching the universal result.}
Combining these terms, we find as desired that
\begin{equation}C_{\text{moduli}}+C_{\text{dist}}+C_{\text{var}}=C_{\text{bdry}}+C_{\text{defect}}\,,
\end{equation}
where we used the identification \eqref{lambdaflow}.  We have thus shown that $\ddt \widehat S^{(1)} = \ddt \widehat S^{(1)}_{\rm conj}$. In other words,
the counterterm can again be interpreted as a flow of the defining parameters of the integrable theory. In particular, $\omega$ runs according to \rf{EF}, which now also implies that the periods of the torus run. The other defining data is RG-invariant in the case of simple poles. The case of higher-order poles (where the maximally isotropic subalgebra flows) follows straightforwardly from adapting Appendix \ref{app:GenPoles} to the elliptic setup.

\section{Discussion}\label{sec:disc}
Using the `universal' approach to computing 1-loop divergences, we have studied the renormalisation of the large class of classically integrable 2d $\sigma$-models (rational, trigonometric and elliptic) engineered with disorder surface defects in 4d Chern-Simons theory.
For this class, we have proven the folk theorem stated in the Introduction: these theories are renormalisable at the 1-loop order, with the data defining the theories running in a way that absorbs all of the 1-loop divergences. The twist 1-form $\omega$, which encodes among other things the positions of the defects, runs according to the flow \rf{tf}, as previously derived in particular cases and conjectured in general. The other data, encoding the boundary conditions at those defects, is invariant in the simplest case when the poles of $\omega$ are simple, and runs in general in a specific way found in Appendix \ref{app:GenPoles}. Finally, in the elliptic case, the flow of $\omega$ induces a particular flow of the periods of the spectral torus.

Along the way, we computed the `universal' path integral in generality. We encountered a first miracle: while intermediate steps took a rather complicated form, the result can be written as a very simple contour integral \rf{redi} over the spectral parameter. We then re-wrote this counterterm as a local, gauge-invariant 4d counterterm \rf{red2}. That term clearly suggests the desired flow of the twist 1-form $\omega$. However, making this precise required a second miracle, namely that certain defect terms arising in the universal result conspire to precisely match those required to make it interpretable as a flow of the twist 1-form. This miracle became even more non-trivial for the elliptic theories in Section~\ref{sec:elliptic}, where a similar precision matching happens with several additional terms, associated to the geometry of the torus and the flow of its periods.

These miracles appear, in this paper, as empirical observations, but it would be interesting in the future to understand how they come about---in other words, why these models conspire to be renormalisable. 
One may wonder if there is a deeper and simpler proof of this fact, relying on the structural features of the 4d Chern-Simons theory.
 In that language, the statement is that a certain class of integrable 2d defect theories in 4d Chern-Simons can be quantised in a consistent and renormalisable manner. We expect that a useful way to gain such an insight would be by perturbatively quantising the 4d Chern-Simons theory directly.
  Renormalising this non-Lorentz-invariant 4d theory in the presence of defects does not seem straightforward---particularly since the disorder defects naively correspond to places in spacetime where $\hbar$ diverges (see discussion in \cite{CWY1}).
Having said this, the 4d quantum effective action was argued in \cite{Levine4d} to localise at 1-loop order to the 2d one: so, at least to this order, the perturbative expansion may be expected to make sense.

The approach developed in this paper allows for a uniform treatment of the 1-loop renormalisation of a very large class of integrable $\sigma$-models. There are, however, some models not yet included in this approach. The simplest ones are the $\s$-models on $\mathbb{Z}_T$-cosets~\cite{Eichenherr:1979ci,Young:2005jv} and their integrable deformations~\cite{Delduc:2013fga,Hoare:2021dix,Osten:2021opf} and generalisations~\cite{Vicedo:2017cge,Arutyunov:2020sdo,Hamidi:2025sgg}. These theories are obtainable from affine Gaudin models or 4d Chern-Simons by imposing an additional equivariance property with respect to the cyclic group $\mathbb{Z}_T$, and we expect their 1-loop RG flow to take the same form \eqref{tf} as for the models considered here. Moreover, the universal divergence approach has been successfully applied to the simplest examples in this class \cite{Levine}. We thus expect the methods and results of our paper to extend to these theories, with only minor technical modifications. In particular, it would be interesting to show that their 1-loop divergences can also be written as the contour integral \eqref{redi} in terms of their Lax connection, using the techniques developed in Subsection \ref{sec:Rmat} based on $\mathcal{R}$-matrices. The $\mathcal{R}$-matrix underlying the $\mathbb{Z}_T$-equivariant models (see for instance~\cite{Vicedo:2017cge,Lacroix:2018njs}) presents two main differences compared to the ones considered in the present paper: it is not of difference form and satisfies a slightly weaker asymptotic condition than \eqref{eq:RAsym}. Adapting the method of Subsection \ref{sec:Rmat} to this larger class of $\mathcal{R}$-matrices is thus the main technical step required to include these theories in our approach.

A more complicated class of integrable $\s$-models that escape this approach are the higher-genus ones considered in~\cite[Section 15]{CY3} and~\cite{Derryberry:2021rne}. Their main difference from the models considered here is that their Lax connection is generally not a holomorphic function of the spectral parameter, although the generating function of conserved charges is. Moreover, they are only defined implicitly in terms of mathematical objects called Szeg\"o kernels, which are quite difficult to compute explicitly. However, Costello and Derryberry conjectured \cite{Derryberry:2021rne} that these models are also 1-loop renormalisable, with their RG flow taking the same form as the one considered here when written in terms of periods of $\omega$.\footnote{See~\cite{LW2} for the relation between that formulation of the RG flow and the one \eqref{tf} used in this paper.} It would thus be interesting to extend our approach to these theories. In particular, we hope that the methods of Subsection \ref{sec:Rmat} can be adapted to evaluate the universal divergences of these models, with the Szeg\"o kernel and its Fay identity playing a role akin to that of the $\mathcal{R}$-matrix and the cYBE. This is, however, quite speculative and a careful analysis of this question is a natural objective for future developments.

The last exceptions are the integrable $\s$-models on flag manifolds (see for instance~\cite{Bykov:2016rdv,Affleck:2021vzo}), which are of a quite different nature. In the 4d Chern-Simons language, they correspond to a very simple 1-form $\omega=\dd z$ on the plane, the cylinder or the torus. In particular, this 1-form has no zeros, meaning that there are no disorder defects.\footnote{In the Hamiltonian language, these theories are said to be ultralocal.} The currents in the Lax connection arise from another mechanism, associated with order defects (see \cite{CY3}), and take a different form than those considered here. In particular, we expect that this Lax connection does not satisfy the Bianchi completeness condition (see Subsection \ref{subsec:univ1loop}), so that the universal divergence approach does not obviously apply to these models. We refer to~\cite{Bykov:2020nal,Affleck:2021vzo} for some discussion of the renormalisation of these models using other methods.

\subsection*{Acknowledgements}
 We are grateful to K.~Costello, D.~Gaiotto, B.~Hoare, S.~Komatsu, A.~A.~Tseytlin, B.~Vicedo and K.~Zarembo for stimulating discussions. NL was supported by the European Research Council under the European Union's Seventh Framework Programme (FP7/2007-2013), ERC Grant agreement ADG 834878. NL is also grateful for the support of the Institut Philippe Meyer at the {\'E}cole Normale Sup{\'e}rieure during the first part of this project.

\addtocontents{toc}{\protect\setcounter{tocdepth}{1}}
\appendix 

\bigskip

\section{Arbitrary pole structure}
\label{app:GenPoles}
In this appendix, we generalise the presentation of 4d Chern-Simons theory on $\Sigma\times \CP$ to include higher order and complex poles in the twist 1-form. We will see that this requires us to change the definition of the defect algebra $\mathfrak{d}$ and its subalgebra $\mathfrak{b}$. This will force us to consider an RG flow of the subalgebra $\mathfrak{b}$, \textit{i.e.}\ of the boundary conditions imposed on the gauge field $A$ at the defects.

\subsection{Classical 4d Chern-Simons theory with arbitrary poles}

\paragraph{General twist 1-form.} Let $\omega=\vp(z)\,\dd z$ be a meromorphic 1-form on $\CP$ (with $\vp(z)$ a rational function of $z$). As in the main text we denote by $\Ph$ the set of poles of $\omega$ and we split its zeros into two subsets $\Zh^\pm$. We label these points as
\begin{equation}
    \Ph = \lbrace p_r \rbrace_{r=1}^N \qquad \text{ and } \qquad \Zh^\pm = \bigl\lbrace z_i^\pm \bigr\rbrace_{i=1}^{N_\pm}\,,
\end{equation}
and their respective multiplicities as $m_r,m_i^\pm \in \Z_{\geq 1}$ (the case treated in Section \ref{sec:review} is then $m_r=1$ for every $r$). We suppose that the number of zeros in $\Zh^+$ and in $\Zh^-$, counted with multiplicity, are the same and denote them by $M$. We then have
\begin{equation}\label{eq:M3}
    M = \sum_{i=1}^{N_+} m_i^+ = \sum_{i=1}^{N_-} m_i^- = \frac{1}{2} \sum_{r=1}^N m_r + 1\,,
\end{equation}
where the last equality follows from the Riemann-Hurwitz formula. Note that this forces the number of poles/zeros of $\omega$ (with multiplicities) to be even. We can write $\omega$ in the following factorised form, in terms of its poles, its zeros and an overall factor $K$:
\begin{equation}\label{eq:OmegaFacH}
    \omega = K \frac{\prod_{i=1}^{N_+} (z-z_i^+)^{m_i^+} \prod_{i=1}^{N_-} (z-z_i^-)^{m_i^-}}{\prod_{r=1}^{N} (z-p_r)^{m_r}}\, \dd z \, .
\end{equation}
It will also be useful to write its simple fraction decomposition
\begin{equation}\label{eq:OmegaPFD}
    \omega = \left( \sum_{r=1}^N \sum_{k=0}^{m_r-1} \frac{\ell_{r,k}}{(z-p_r)^{k+1}} \right) \dd z\,,
\end{equation}
introducing the \textit{levels} $\ell_{r,k}=\res_{z=p_r}(z-p_r)^k\omega$. The latter can be seen as rational functions of the parameters $(K,p_r,z_i^\pm)$ and generalise the residues \eqref{eq:Levels} considered for simple poles.

\paragraph{Reality conditions.} To ensure that the 4d Chern-Simons theory associated with $\omega$ is real, we need to impose certain reality conditions on this 1-form. Namely, we will ask that it is equivariant with respect to the complex conjugation on $\CP$, \textit{i.e.} that $\overline{\vp(z)} = \vp(\overline{z})$. In terms of the factorised parametrisation \eqref{eq:OmegaFacH}, this forces $K$ to be real while the poles and zeros are either real or come in pairs of complex conjugates with the same multiplicity. We will restrict to the case where all the zeros $z_i^\pm$ are real. In contrast, we will allow for complex poles. Up to a permutation, we can label them such that we have\vspace{-5pt}
\begin{itemize}\setlength\itemsep{1pt}
    \item $\Nr$ real poles $p_r\in \mathbb{R}$ with real levels $\ell_{r,k}\in\mathbb{R}$, for $r\in\lbrace 1,\dots,\Nr \rbrace$,
    \item $\Nc$ independent complex poles $p_r \in \C$ with levels $\ell_{r,k}\in\C$, for $r\in\lbrace \Nr+1,\dots,\Nr+\Nc \rbrace$,
    \item $\Nc$ conjugate poles $p_r = \overline{p}_{r-\Nc}$ with levels $\ell_{r,k} = \overline{\ell}_{r-\Nc,k}$, for $r\in\lbrace \Nr+\Nc+1,\dots,N \rbrace$.\vspace{-4pt}
\end{itemize}
The total number of poles is then $N=\Nr+2\Nc$. Taking into account Möbius transformations, the parameter space describing the choice of $\omega$ is of real dimension $N+N_++N_- - 2$. We further ask that the gauge field satisfies the same reality condition \eqref{eq:RealityA} as in the main text, in terms of the antilinear `complex conjugation' automorphism $\tau$ of $\gf^\C$ characterising the real form $\gf$. Together with the above restrictions on $\omega$, this ensures that the action \eqref{eq:Action4d} is real.

\paragraph{Defect term and jet.} The main differences in the case of arbitrary poles arise in the treatment of the \textit{defect term} in the variation of the action. This term still takes the form \eqref{eq:DeltaDefect}, involving the derivative $\dd\omega$. Rather than being a sum of Dirac distributions, like in the case of simple poles, $\dd\omega$ now also involves derivatives of Dirac distributions at higher-order poles (see appendix \ref{app:Int} and equation \eqref{eq:dOmDist} for more details). In terms of the levels $\ell_{r,k}$ defined in \eqref{eq:OmegaPFD}, the defect term then reads
\begin{equation}\label{eq:DeltaDefect3}
    \delta S^{\text{defect}}_{4d} = -\frac{1}{8\pi}\sum_{r=1}^N\; \sum_{k=0}^{m_r-1} \frac{\ell_{r,k}}{k!}\, \int_\Sigma \partial_z^k\,\langle A,\delta A \rangle \bigl|_{z=p_r}\,.
\end{equation}
This is the equivalent of equation \eqref{eq:DeltaDefect2} for simple poles.

In particular, the defect term now involves derivatives of the gauge field at the poles. This naturally leads us to consider the \textit{jet} of $A$ along the pole defects, defined as the collection
\begin{equation}\label{eq:JetHigher}
    \Ad_\pm = \Bigl( \frac{1}{k!}\, \partial_z^k A_\pm \bigl|_{z=p_r} \Bigr)_{r=1,\dots,N}^{k=0,\dots,m_r-1} \,.
\end{equation}
This is the higher-order generalisation of equation \eqref{eq:Jet}. Let us discuss the space in which this jet is valued, taking into account the reality conditions imposed on the poles and on $A$, \textit{i.e.} equation \eqref{eq:RealityA}. If $r\in\lbrace 1,\dots,\Nr\rbrace$, then $p_r$ is a real pole and $\partial^k_z A_\pm|_{z=p_r}$ is valued in the real form $\gf$. If $r\in\lbrace \Nr+1,\dots,\Nr+\Nc\rbrace$, then $(p_r,p_{r+\Nc})$ forms a pair of complex conjugate poles: the corresponding components of the jet are then valued in the complexified algebra $\gf^\C$ and related by complex conjugation, \textit{i.e.} $\partial^k_z A_\pm|_{z=p_r} = \tau\bigl( \partial^k_z A_\pm|_{z=p_{r+\Nc}} \bigr)$.  We can thus naturally see the jet $\Ad_\pm$ as an element of the vector space
\begin{equation*}
    \df = \left\lbrace \bigl( X_{r,k} \bigr)_{r=1,\dots,N}^{k=0,\dots,m_r-1}\;\left| \;  \begin{array}{ll}
       X_{r,k} \in \gf  & \text{ if } \;r\in\lbrace 1,\dots,\Nr\rbrace  \\[4pt]
       X_{r,k} = \tau\bigl( X_{r+\Nc,k} \bigr) \in \gf^\C  & \text{ if } \;r\in\lbrace \Nr+1,\dots,\Nr+\Nc\rbrace 
    \end{array} \right.\right\rbrace\,.
\end{equation*}
As the notation suggests, this is the \textit{defect Lie algebra}, although for the moment it is not yet defined as an algebra but only as a vector space. As such, it is isomorphic to the direct sum of $\sum_{r=1}^{\Nr} m_r$ copies of $\gf$ and $\sum_{r=\Nr+1}^{\Nr+\Nc} m_r$ copies of $\gf^\C$. In particular, it has real dimension
\begin{equation}
    \dim \df = 2(M+1)\dim\gf\,,
\end{equation}
where $M$ was defined through the equation \eqref{eq:M3}.

Using the generalised Leibniz rule for higher-order derivatives of products, one rewrites the defect term \eqref{eq:DeltaDefect3} in the compact form
\begin{equation}\label{eq:DefectJet2}
    \delta S^{\text{defect}}_{4d} = -\frac{1}{8\pi} \int_\Sigma \,\dlangle \Ad,\delta \Ad \drangle_{\df,\ellb}\,,
\end{equation}
in terms of the $\df$-valued one form $\Ad=\Ad_+\,\dd x^+ + \Ad_-\,\dd x^-$. In this equation, $\dlangle\cdot,\cdot\drangle_{\df,\ellb}$ denotes a natural symmetric bilinear form on $\df$ defined by
\begin{equation}\label{eq:FormDefect}
    \dlangle \Xd, \Yd \drangle_{\df,\ellb} \equiv \sum_{r=1}^N \sum_{k=0}^{m_r-1}  \sum_{l=0}^k \, \ell_{r,k}\; \langle X_{r,l}, Y_{r,k-l} \rangle\,,
\end{equation}
where $\Xd = ( X_{r,k} )_{r=1,\dots,N}^{k=0,\dots,m_r-1}$ and $\Yd = ( Y_{r,k} )_{r=1,\dots,N}^{k=0,\dots,m_r-1}$ are two generic elements of $\df$. For future convenience, and contrarily to the main text, we have added an index $\ellb$ in the notation $\dlangle\cdot,\cdot\drangle_{\df,\ellb}$ for this form, to indicate its dependence on the levels $\ellb = ( \ell_{r,k} )_{r=1,\dots,N}^{k=0,\dots,m_r-1}$. The equation \eqref{eq:DefectJet2} takes the exact same form as the one \eqref{eq:DefectJet} obtained in the main text, but with a generalised definition of the space $\df$, the jet $\Ad$ and the bilinear form $\dlangle\cdot,\cdot\drangle_{\df,\ellb}$. In the case of simple real poles, $\df$ becomes $\gf^N$ and the bilinear form \eqref{eq:FormDefect} reduces to the one \eqref{eq:FormDefectSimple} considered in the main text, as expected. One checks that this generalised pairing \eqref{eq:FormDefect} is also non-degenerate on $\df$. 

When considering an element $\Xd = ( X_{r,k} )_{r=1,\dots,N}^{k=0,\dots,m_r-1}$ of $\df$, one should recall that the last components $X_{r,k}$ with $r\in\lbrace \Nr+\Nc+1,\dots,N \rbrace$ are not independent from the previous ones, as they are related by the complex conjugation $X_{r,k} = \tau\bigl( X_{r-\Nc,k} \bigr)$. In particular, the data of the first components, up to $r=\Nr+\Nc$, is enough to fully characterise $\Xd\in\df$. For instance, the bilinear form \eqref{eq:FormDefect} can be rewritten in terms of these independent components as
\begin{equation*}
    \dlangle \Xd, \Yd \drangle_{\df,\ell} = \sum_{r=1}^{\Nr+\Nc} \sum_{k=0}^{m_r-1}  \sum_{l=0}^k \, \text{Re} \Bigl( \ell_{r,k}\;\langle X_{r,l}, Y_{r,k-l} \rangle \Bigr)\,,
\end{equation*}
where $\text{Re}(a)$ denotes the real part of $a\in\C$. In particular, this alternative form makes it apparent that $\dlangle\cdot,\cdot\drangle_{\df,\ellb}$ is real-valued. To obtain this expression, we have used the identity $\langle \tau(X),\tau(Y)\rangle = \overline{\langle X,Y \rangle}$, true for any $X,Y\in\gf^\C$.

\paragraph{Gauge symmetries and the defect Lie algebra structure.} So far, we have described $\df$ as a vector space equipped with a non-degenerate symmetric bilinear form. Let us now describe its Lie algebra structure. We will be guided by the behavior of the gauge transformations \eqref{eq:GaugeTransf} along the pole defects. We consider such a transformation $A\mapsto A^u$ and expand the gauge pararameter as $u=\text{Id} + V + O(V^2)$, where $V$ is an infinitesimal $\gf^\C$--valued function. This corresponds to taking the variation $\delta A = [A,V] + \dd V$ of the gauge field, which itself induces a certain variation $\delta \Ad_\pm \in \df$ of the jet \eqref{eq:JetHigher} along the pole defects. We will define the Lie bracket $[\cdot,\cdot]_{\df}$ on the defect algebra $\df$ so as to get
\begin{equation}\label{eq:InfGaugeJet}
    \delta\Ad_\pm =  [ \Ad_\pm, \mathbb{V} ]_{\df} + \del_\pm \mathbb{V}\,,
\end{equation}
where $\mathbb{V} = ( \frac{1}{k!} \partial_z^k V |_{z=p_r} )_{r=1,\dots,N}^{k=0,\dots,m_r-1}$ is the jet of $V$. This uniquely determines the bracket: after a few manipulations, we find
\begin{equation}\label{eq:LieDefect}
    \bigl[ \Xd, \Yd \bigr]_{\df} = \left( \sum_{l=0}^k \bigl[ X_{r,l}, Y_{r,k-l} \bigr] \right)_{r=1,\dots,N}^{k=0,\dots,m_r-1}\,,
\end{equation}
with $\Xd = ( X_{r,k} )_{r=1,\dots,N}^{k=0,\dots,m_r-1}$ and $\Yd = ( Y_{r,k} )_{r=1,\dots,N}^{k=0,\dots,m_r-1}$ in $\df$. That this bracket satisfies the Jacobi identity follows from the one of $\gf^\C$ and a little bit of algebra. One easily checks that in the case where all the poles $p_r$ are real and simple, this definition coincides with the one \eqref{eq:LieDefectSimple} and thus $\df$ becomes the direct product $\gf^N$, in agreement with the main text. The main generalisation introduced in equation \eqref{eq:LieDefect} is the Lie algebra structure involving the additional indices $k\in\lbrace 0,\dots,m_r-1\rbrace$ associated with higher-order poles, which are sometimes termed the \emph{Takiff modes}. Recall also that the last $\Nc$ components of a vector in $\df$ are related to the previous ones by the antilinear involution $\tau : \gf^\C \to \gf^\C$. The fact that $\tau$ is an automorphism, \textit{i.e.} it preserves the bracket $[\cdot,\cdot]$ on $\gf^\C$, ensures that the right-hand side of equation \eqref{eq:LieDefect} also satisfies these conditions. Therefore, it really defines a skew-symmetric bracket $[\cdot,\cdot]_{\df} : \df \times \df \to \df$. Finally, with a little bit of algebra, one shows that the bilinear form \eqref{eq:FormDefect} on $\df$ is ad-invariant with respect to the Lie bracket \eqref{eq:LieDefect}.

We note that each pole $p_r$ is naturally associated with a subalgebra of $\df$, composed by elements $\Xd = ( X_{r,k} )_{r=1,\dots,N}^{k=0,\dots,m_r-1}$ with $X_{s,k}=0$ for $s \neq r$. This subalgebra takes the form of a so-called \textit{Takiff Lie algebra}, which can be realised in terms of truncated loops. More precisely, a real pole of order $m$ is associated with the Takiff algebra $\Tg{m-1} \simeq \gf \otimes \mathbb{R}[\xi]/\mathbb{R}\xi^{m}$ and a pair of complex conjugate poles with its complexification $\Tgc{m-1} \simeq \gf \otimes \C[\xi]/\C\xi^{m}$. In particular, simple poles are attached to $\Tg{0} \simeq\gf$ or $\Tgc{0}\simeq\gf^\C$, \textit{i.e.} the gauge algebra or its complexification. In contrast, for higher-order poles, $\Tg{m-1}$ and $\Tgc{m-1}$ are more exotic Lie algebras, which in particular are not semi-simple and possess abelian ideals. In the end, the whole defect algebra can then be seen as the direct sum
\begin{equation}\label{eq:Takiff}
    \df = \left( \bigoplus_{r=1}^{\Nr} \Tg{m_r-1} \right) \oplus \left( \bigoplus_{r=\Nr+1}^{\Nr+\Nc} \Tgc{m_r-1} \right) \,.
\end{equation}

\paragraph{Lifts from algebras to groups.} Recall that $\gf$ is the Lie algebra of the connected gauge group $G$. Similarly, we lift the defect algebra \eqref{eq:Takiff} to a \textit{defect group}
\begin{equation}\label{eq:DefectGroup}
    D = \left( \prod_{r=1}^{\Nr} \TG{m_r-1} \right) \times \left( \prod_{r=\Nr+1}^{\Nr+\Nc} \TGc{m_r-1} \right)\,,
\end{equation}
with $\text{Lie}(D)=\df$. Here $\TG p$ (resp. $\TGc p$) denotes the real (resp. complexified) Takiff group of order $p$. $\TG p$ is often also called the $p^{\text{th}}$--jet group of $G$. Let us illustrate this concept with simple examples. The $0^{\text{th}}$--jet group is simply the group itself, \textit{i.e.} $\TG 0 \simeq G$. As a manifold, the $1^{\text{st}}$--jet group is identified with the tangent bundle $\TG{1} \simeq \TG\null$, whose elements are described by the choice of a point in $G$ and a tangent vector at this point. There exists a unique Lie group structure on this manifold which lifts the Lie algebra structure of $\Tg 1$: one then shows that $\TG 1$ is isomorphic to the semi-direct product $G \ltimes \gf$ (where the Lie algebra $\gf$ is seen as an abelian group with addition, on which $G$ acts by the adjoint action).

For brevity, we will not detail here the structure of the jet group $\TG p$  for general $p$ and refer to~\cite{Vizman:2013,Lacroix:2020flf} for a more thorough discussion. However, to give some intuition on this object, we include the following heuristic description. If $f: \mathbb{R} \to G$ is a $G$-valued function and $x_0$ is a given point in $\mathbb{R}$, the evaluation $\bigl( f(x_0), \partial_x f(x_0) \bigr)$ of $f$ and its derivative can naturally be seen as an element of the tangent space $\TG\null$, \textit{i.e.} the $1^{\text{st}}$--jet group. Similarly, the evaluation $\bigl( \partial_x^k f(x_0) \bigr)_{k=0,\dots,p}$ of $f$ and its first $p$ derivatives is naturally interpreted as an element of the $p^{\text{th}}$--jet group $\TG{p}$. As a manifold, the latter is isomorphic to the product of $G$ with $p$ copies of the vector space $\gf$. Moreover, the group structure on $\TG p$ is characterised by the property that $f \mapsto \bigl( \partial_x^k f(x) \bigr)_{k=0,\dots,p}$ is a homomorphism from the group of $G$-valued functions (equipped with pointwise multiplication) to the jet group $\TG p$.

After this Lie theoretic interlude, let us return to our 4d Chern-Simons theory. Earlier, the Lie algebra structure on $\df$ allowed us to describe the effect of infinitesimal gauge transformations on the jet $\Ad$ of the gauge field -- see equation \eqref{eq:InfGaugeJet}. As one may expect, the lift of this structure to the defect group $D$ is useful to discuss finite gauge transformations $A \mapsto A^u$. Indeed, in the same spirit as the heuristic discussion above, the jet $\uj = ( \partial_z^k u)_{r=1,\dots,N}^{k=0,\dots,m_r-1}$ of the gauge parameter can be naturally interpreted as a $D$-valued function on $\Sigma$. The induced gauge transformation on the jet then takes the form
\begin{equation}\label{eq:GaugeJetApp}
    \Ad_\pm \longmapsto \Ad_\pm^{\uj} = \uj^{-1}\Ad_\pm\uj + \uj^{-1}\partial_\pm \uj\,,
\end{equation}
where the adjoint action and Darboux derivatives are that of the defect group \eqref{eq:DefectGroup}.

\paragraph{Maximally isotropic subalgebra $\kf$.} Let us now return to the defect term \eqref{eq:DefectJet2}, written in terms of the jet $\Ad_\pm\in\df$ and the bilinear form $\dlangle\cdot,\cdot\drangle_{\df,\ellb}$. As in the main text, we ensure that this term vanishes by imposing the boundary condition \eqref{eq:BC}, \textit{i.e.} that $\Ad_\pm$ is valued in a maximally isotropic subalgebra $\kf$ of $\df$. Moreover, denoting by $B$ the unique closed subgroup of $D$ with $\text{Lie}(B)=\kf$, we restrict the local parameter $u$ of gauge transformations to have a jet $\uj = ( \partial_z^k u)_{r=1,\dots,N}^{k=0,\dots,m_r-1}$ valued in $B$, as in \eqref{gbc}. By equation \eqref{eq:GaugeJetApp}, this ensures that the boundary condition $\Ad_\pm\in\kf$ is preserved by these gauge transformations (this is the main reason behind the requirement for $\kf$ to be a subalgebra). Moreover, one can show~\cite[Theorem 4.2]{Benini} that the transformations satisfying this boundary condition leave the action \eqref{eq:Action4d} invariant and thus correspond to proper gauge symmetries. This is completely analogous to what was discussed in the main text around equation \eqref{eq:GaugeJet}, for the case of simple poles. We refer to this discussion for more details.

\paragraph{Summary.} In conclusion, we have defined the 4d Chern-Simons theory associated with the data $(\omega,\Zh^\pm,G,\kf)$, with $\omega$ now allowed to have an arbitrary pole structure. The key message of this appendix is the following: once the defect algebra $\df$ is appropriately generalised to the case of arbitrary poles, the rest of the construction follows exactly as in the case of real simple poles. In particular, following the construction of Subsection \ref{subsec:int} of the main text, one can extract a 2d integrable $\s$-model from this 4d setup. As in the case of simple real poles, the target space of this $\s$-model is the double coset $B \hspace{-1pt}\setminus\hspace{-1pt} D\,/\, G_{\text{diag}}$. The main difference here is hidden in the definition of the defect group $D$, which now takes the form \eqref{eq:DefectGroup} in terms of Takiff/jet groups, instead of $D=G^N$. To be complete, we still have to describe how the diagonal subgroup $G_{\text{diag}}$ of $D$ is defined in the case of higher-order poles. As mentioned above, a real pole $p_r$ is associated to the factor $\TG{m_r-1}$ in the defect group \eqref{eq:DefectGroup}. This factor naturally contains a subgroup isomorphic to $G$.\footnote{Recall the heuristic description of the group $\TG{p}$ as being composed by jets $\bigl(\partial_x^k f(x_0)\bigr)_{k=0,\dots,p}$ of functions $f 
: \mathbb{R} \to G$. In this context, the natural embedding of $G$ in $\TG{p}$ is obtained from constant functions $f$, whose jets take the form $\bigl( f,0,\dots,0)$.} Similarly, a pair of complex conjugate poles is associated with a jet factor $\TGc{m_r-1}$, which possesses a subgroup isomorphic to $G^\C$, which itself includes $G$ as a subgroup. We then define $G_{\text{diag}}$ as the diagonal embedding of $G$ as a subgroup in each of these jet factors.

\subsection{1-loop renormalisation for arbitrary pole structure}
\label{app:RGhigher}

\paragraph{Setup.} We now discuss the 1-loop renormalisation of the 2d integrable $\sigma$-model built in the previous subsection and associated with the data $(\omega,\Zh^\pm,G,\kf)$. The case where $\omega$ has simple real poles was treated in Section \ref{sec:rational} of the main text. Our main goal in this appendix is to generalise this result to the case of an arbitrary pole structure. Our starting point is the universal approach of~\cite{Levine}, which allows us to determine the 1-loop divergence of the model in terms of its Lax connection. The computation of this universal divergence in subsections \ref{subsec:univ1loop}--\ref{subsec:4dFlow} of the main text is in fact perfectly valid also for general poles. In particular, the equation \eqref{floweq} describing the RG dependence of the 1-loop effective action in terms of the 4d-CS gauge field $A$ also applies in the present case. For convenience, we recall it here:
\be
\ddt \widehat S^{(1)} = \frac{1}{16 \pi^2 i} \int_M \del_z \Psi(z) \, \ud z \wedge {\rm CS}[A] + \DT \ , \la{floweqApp}
\ee
where the second contribution is the `defect term'
\be
\DT = \frac{1}{16 \pi^2 i}  \int_M \del_{\bar z} \Psi(z) \, \ud \bar z \wedge {\rm CS}[A] \ . \la{deteApp}
\ee
In these equations, the function $\Psi(z)$ is defined as in subsection \ref{subsec:4dFlow}. More precisely, $\Psi(z)$ is characterised by the following properties: it has the same pole structure as $\vp(z)$ in the complex plane, is bounded when $z\to\infty$ and satisfies the property \eqref{psi} at the zeros $\Zh^\pm=\lbrace z_i^\pm \rbrace_{i=1}^{N_\pm}$ of $\omega$.

The main result left to generalise is that of subsection \ref{subsec:ct}, namely that we can absorb the flow \eqref{floweqApp} as a running of the defining parameters $(\omega,\Zh^\pm,G,\kf)$ of the model. We note that $G$ and the splitting of zeros into the sets $\Zh^\pm$ are essentially discrete data and thus cannot run. In contrast, the twist 1-form $\omega$ encodes an important part of the continuous parameters of the theory. As was the case for simple poles, these parameters are expected to run with the RG flow. More precisely, there exists a conjecture \eqref{fom} for the flow of $\omega$ regardless of its pole structure~\cite{Delduc:2020vxy}. We will recall it in the next paragraph and will discuss how its consequences differ in the case of higher-order poles. The behavior of the maximally isotropic subalgebra $\kf$ under renormalisation is a bit more subtle. It can depend on continuous parameters and could thus encode running couplings as well: its RG flow has not been thoroughly discussed so far in the literature. In the case of real simple poles, we showed in the main text that the universal divergence can be absorbed into a flow of only $\omega$, with $\kf$ remaining RG-invariant. As we will see in this appendix, this will no longer be true for general pole structures, where we will also encounter a non-trivial flow of $\kf$.

The structure of this subsection is the following. We will first describe the expected RG flow of $\omega$ and $\kf$, starting from the conjecture of~\cite{Delduc:2020vxy}. We will then show that the induced flow of the action matches the universal divergence computation \eqref{floweqApp}, thus proving that the model is 1-loop renormalisable.

\paragraph{Expected RG flow of $\bm\omega$.} We will work with the partial fraction decomposition \eqref{eq:OmegaPFD} of the twist 1-form $\omega=\vp(z)\,\dd z$, written in terms of the parameters $(p_r,\, \ell_{r,k})$. The latter are expected to flow with the RG-time $\text{t}$. This flow can be compactly encoded in the `formal derivative' of $\omega$ with respect to $\text{t}$, defined as
\begin{equation}\label{eq:FormalDerOm}
    \frac{\dd\omega}{\dd\text{t}} = \sum_{r=1}^N \sum_{k=0}^{m_r-1} \left( \frac{\dd \ell_{r,k}}{\dd\text{t}}\frac{1}{(z-p_r)^{k+1}} + \frac{\dd p_r}{\dd\text{t}} \frac{(k+1)\ell_{r,k}}{(z-p_r)^{k+2}} \right) \,\dd z \,.
\end{equation}
In terms of this object, the conjecture that we aim to prove~\cite{Delduc:2020vxy} (see also~\cite{Derryberry:2021rne,LW2}) is
\begin{equation}\label{eq:dPsiApp}
    \frac{\dd \omega}{\dd\text{t}} = \partial_z \Psi(z)\,\dd z\,,
\end{equation}
where $\Psi(z)$ is the function appearing in equation \eqref{floweqApp}. Recalling that $\Psi(z)$ has the same pole structure as $\vp(z)$, it will be useful to rewrite this function as
\begin{equation}
    \Psi(z) = - f(z)\,\vp(z)\,,
\end{equation}
where $f(z)$  has singularities at the zeros $\Zh^\pm=\lbrace z_i^\pm \rbrace_{i=1}^{N_\pm}$ of $\omega$, but is regular at its poles $\Ph=\lbrace p_r \rbrace_{r=1}^N$. Comparing equations \eqref{eq:FormalDerOm} and \eqref{eq:dPsiApp}, one can extract the flow of the poles $p_r$ and the levels $\ell_{r,k}$ predicted by the conjecture. We get
\begin{equation}\label{eq:FlowParam}
    \frac{\dd p_r}{\dd\text{t}} = f(p_r)\,, \qquad \frac{\dd \ell_{r,0}}{\dd\text{t}} = 0 \qquad \text{ and } \qquad \frac{\dd \ell_{r,k}}{\dd\text{t}} = \beta_{r,k} = k \sum_{l=k}^{m_r-1} \frac{f^{(l+1-k)}(p_r)}{(l+1-k)!}\, \ell_{r,l}\,.
\end{equation}
This generalises the equation \eqref{eq:dlp} found for simple poles.\footnote{The equation \eqref{eq:dlp} was written in terms of the residues $\psi_r$ of $\Psi(z)$. To compare with \eqref{eq:FlowParam}, we note that $\psi_r=-f(p_r)\,\ell_{r}$. The alternative formulation used here in terms of the function $f(z)$ will turn out to be quite convenient to treat general poles.} The main difference here is that the levels are not all RG-invariants. More precisely, the standard residues $\ell_{r,0}=\res_{z=p_r} \omega$ are still RG-invariants but the higher-order levels $\ell_{r,k}$ with $k>0$ are generally running.

\paragraph{RG flow of the bilinear form.} This non-trivial flow of the levels will be the main technical difference that we have to address in the case of higher-order poles. It means in particular that the bilinear form $\dlangle\cdot,\cdot\drangle_{\df,\ellb}$ does not stay constant along the RG flow, as its expression \eqref{eq:FormDefect} depends explicitly on the levels. More precisely, we get
\begin{equation}\label{eq:FlowForm}
    \frac{\dd\;}{\dd\text{t}} \dlangle \Xd, \Yd \drangle_{\df,\ellb} = \sum_{r=1}^N \sum_{k=1}^{m_r-1} \sum_{l=0}^{k} \beta_{r,k}\, \langle X_{r,l}, Y_{r,k-l} \rangle\,,
\end{equation}
for elements $\Xd = ( X_{r,k} )_{r=1,\dots,N}^{k=0,\dots,m_r-1}$ and $\Yd = ( Y_{r,k} )_{r=1,\dots,N}^{k=0,\dots,m_r-1}$ of $\df$ that do not depend on the RG-scale $\text{t}$. As a consequence, it follows that the maximally isotropic subalgebra $\kf$ will also have to run under the RG flow, since the notion of isotropy itself will evolve with the energy scale.

\paragraph{The derivation $\bm{\Delta}$.} Let us use the function $f(z)$ to define the following linear map on $\df$:
\begin{equation}\label{eq:DefDelta}
    \Delta : \;\; \displaystyle \Bigl( X_{r,k} \Bigr)_{r=1,\dots,N}^{k=0,\dots,m_r-1} \; \xmapsto{\hspace{1cm}} \; \displaystyle  \left(  \sum_{l=1}^k \frac{f^{(k+1-l)}(p_r)}{(k+1-l)!} \,l X_{r,l} \right)_{r=1,\dots,N}^{k=0,\dots,m_r-1}\,.
\end{equation}
Using the expression for $\beta_{r,k}$ in \eqref{eq:FlowParam}, one checks that the RG flow \eqref{eq:FlowForm} of the bilinear form $\dlangle \cdot,\cdot\drangle_{\df,\ellb}$ can be rewritten in terms of this map as
\begin{equation}\label{eq:FlowFormDelta}
   \frac{\dd\;}{\dd\text{t}} \dlangle \Xd, \Yd \drangle_{\df,\ellb} =  \dlangle\Delta\Xd,\Yd\drangle_{\df,\ellb} + \dlangle\Xd,\Delta\Yd\drangle_{\df,\ellb}\,.
\end{equation}
Moreover, $\Delta$ is a derivation of the defect Lie algebra $\df$, in the sense that
\begin{equation}\label{eq:DeltaDerivation}
    \Delta [\Xd,\Yd]_{\df} = \bigl[ \Delta\Xd, \Yd \bigr]_{\df} + \bigl[ \Xd, \Delta\Yd \bigr]_{\df}\,.
\end{equation}
This can be checked explicitly using the expression \eqref{eq:LieDefect} for the Lie bracket $[\cdot,\cdot]_{\df}$.

\paragraph{Exponentiating the derivation.} The map $\Delta : \df \to \df$ depends on the parameters encoded in $\omega$ and thus implicitly on the RG-scale $\text{t}$. We will `exponentiate' it by defining another $\text{t}$-dependent map $\rho : \df \to \df$ by the first order differential equation
\begin{equation}\label{eq:EdoSigma}
    \frac{\dd\;}{\dd\text{t}} \rho = \rho \circ \Delta\,, \qquad \text{ with } \qquad \rho|_{\text{t}=0} = \text{Id}_{\df}\,.
\end{equation}
Since we have specified its initial condition at $\text{t}=0$, this uniquely determines $\rho$. Formally, it can be seen as the $\text{t}$-ordered exponential of the integral of $\Delta$. We note that $\rho$ is then invertible on $\df$. Crucially, the fact that $\Delta$ is a derivation of $\df$ implies that $\rho$ is an automorphism, \textit{i.e.}
\begin{equation}\label{eq:SigmaAuto}
    \rho [\Xd,\Yd]_{\df} = \bigl[ \rho\Xd, \rho\Yd ]_{\df}\,.
\end{equation}

\begin{proof} Let us define a bilinear map $J: \df \times \df \to \df$ by
\begin{equation*}
    J(\Xd,\Yd) = \rho [\Xd,\Yd]_{\df} - \bigl[ \rho\Xd, \rho\Yd ]_{\df}\,.
\end{equation*}
We then want to prove that $J$ vanishes. Since $\rho|_{\text{t}=0} = \text{Id}_{\df}$, it is clear that $J|_{\text{t}=0}=0$ at the reference scale. Moreover, using the differential equation \eqref{eq:EdoSigma} obeyed by $\rho$, we get
\begin{equation*}
    \frac{\dd\;}{\dd\text{t}} J(\Xd,\Yd) = \rho \bigl( \Delta[\Xd,\Yd]_{\df} \bigr) - \bigl[ \rho (\Delta \Xd), \rho \Yd \bigr] -  \bigl[ \rho \Xd, \rho (\Delta \Yd) \bigr]\,.
\end{equation*}
We now use the derivation property \eqref{eq:DeltaDerivation} of $\Delta$ to get
\begin{equation*}
    \frac{\dd\;}{\dd\text{t}} J(\Xd,\Yd) = \rho  [\Delta\Xd,\Yd]_{\df} + \rho  [\Xd,\Delta\Yd]_{\df} - \bigl[ \rho (\Delta \Xd), \rho \Yd \bigr] -  \bigl[ \rho \Xd, \rho (\Delta \Yd) \bigr] = J(\Delta\Xd,\Yd) + J(\Xd,\Delta\Yd)\,.
\end{equation*}
Let us fix a basis $\lbrace \mathbb{T}_\alpha\rbrace$ of $\df$ and write $J(\mathbb{T}_\alpha,\mathbb{T}_\beta) = J_{\alpha\beta}\null^\gamma\,\mathbb{T}_\gamma$ and $\Delta(\mathbb{T}_\alpha) = \Delta_\alpha\null^\beta\,\mathbb{T}_\beta$, in `tensor form'. The above equation then translates to\vspace{-2pt}
\begin{equation*}
    \frac{\dd\;}{\dd\text{t}} J_{\alpha\beta}\null^\gamma = \Delta_{\alpha}\null^\eta\,J_{\eta\beta}\null^\gamma + \Delta_{\beta}\null^\eta\,J_{\alpha\eta}\null^\gamma\,.\vspace{-2pt}
\end{equation*}
The entries $J_{\alpha\beta}\null^\gamma$ thus obey a system of homogeneous first-order linear differential equations. Since their initial condition at $\text{t}=0$ vanishes, we deduce that $J_{\alpha\beta}\null^\gamma=0$ along the whole flow. This means that $J=0$, ending the proof of the automorphism property \eqref{eq:SigmaAuto}.
\end{proof}

\paragraph{Solution of the RG flow for the bilinear form.} We will now use the map $\rho$ to explicitly write the evolution of the bilinear form $\dlangle \Xd,\Yd \drangle_{\df,\ellb}$ under the RG flow. More precisely, we will prove that
\begin{equation}\label{eq:FormSigma}
    \dlangle \Xd,\Yd \drangle_{\df,\ellb} =  \dlangle \rho\Xd, \rho\Yd \drangle_{\df,\ellb_0}\,,
\end{equation}
where $\ellb_0 = \ellb|_{\text{t}=0}$ is the value of the levels at the reference scale $\text{t}=0$.

\begin{proof} Define a bilinear form $U : \df\times\df \to \R$ by
\begin{equation*}
    U(\Xd,\Yd) = \dlangle \mathbb \Xd,\Yd \drangle_{\df,\ellb} - \dlangle \rho\Xd, \rho\Yd \drangle_{\df,\ellb_0}\,.
\end{equation*}
We will prove that $U$ vanishes. Since $\ellb|_{\text{t}=0}=\ellb_0$ and $\rho|_{\text{t}=0}=\text{Id}_{\df}$, it is clear that $U |_{\text{t}=0}=0$ at the reference scale. Using the formula \eqref{eq:FlowFormDelta} for the flow of $\dlangle \cdot,\cdot \drangle_{\df,\ellb}$ and the differential equation \eqref{eq:EdoSigma} defining $\rho$, we get
\begin{equation*}
    \frac{\dd\;}{\dd\text{t}} U(\Xd,\Yd) = \dlangle\Delta\Xd,\Yd\drangle_{\df,\ellb} + \dlangle\Xd,\Delta\Yd\drangle_{\df,\ellb} - \dlangle \rho(\Delta\Xd), \rho\Yd \drangle_{\df,\ellb_0} - \dlangle \rho\Xd, \rho(\Delta\Yd) \drangle_{\df,\ellb_0}\,.
\end{equation*}
Therefore, we have
\begin{equation*}
    \frac{\dd\;}{\dd\text{t}} U(\Xd,\Yd) = U(\Delta\Xd,\Yd) + U(\Xd,\Delta\Yd)\,.
\end{equation*}
As for $J$ above, the entries of $U$ in a basis of $\df$ obey a system of homogeneous first-order linear differential equations. Since their initial condition at $\text{t}=0$ vanishes, they vanish for all $\text{t}$. This means that $U=0$, ending the proof of the identity \eqref{eq:FormSigma}.
\end{proof}

\paragraph{Expected flow of $\kf$.} As observed earlier, the choice of maximally isotropic subalgebra $\kf$ of $\df$ cannot stay the same along the RG flow since the bilinear form $\dlangle\cdot,\cdot\drangle_{\df,\ellb}$ defining the notion of isotropy is itself flowing. Let us denote by $\kf_0$ the choice made at the reference scale $\text{t}=0$, which is then isotropic with respect to the bilinear form $\dlangle\cdot,\cdot\drangle_{\df,\ellb_0}$. We note that the subspace $\rho^{-1}\kf_0 \subset \df$ is
\vspace{-7pt}
\begin{enumerate}\setlength\itemsep{1pt}
    \item a subalgebra of $\df$, since $\rho$ is an automorphism;
    \item isotropic with respect to $\dlangle\cdot,\cdot\drangle_{\df,\ellb}$, using equation \eqref{eq:FormSigma};
    \item of maximal dimension since $\dim(\rho^{-1}\kf_0)=\dim(\kf_0)=\frac{1}{2}\dim\df$.
\end{enumerate}
Therefore, the choice
\begin{equation}\label{eq:kt}
\kf = \rho^{-1}\kf_0    
\end{equation}
is a very natural candidate for the role of maximally isotropic subalgebra at RG-scale $\text{t}$. Equivalently, this can be reformulated as an infinitesimal RG flow of the subalgebra, \textit{i.e.}\ as a differential equation with respect to $\text{t}$: 
\be
\ddt \kf = - \Delta(\kf)\ , 
\ee
by which we simply mean that $\kf$ is the set of vectors $X_t$ evolved from $X_0\in \kf_0$ by the ODE $\ddt X_t = -\D(X_t)$.

\paragraph{RG dependence of the effective action.} We have arrived at the conjectured 1-loop flows \eqref{eq:dPsiApp} and \eqref{eq:kt} of $(\omega,\kf)$. They correspond to a particular dependence of the 1-loop effective action on the RG-scale $t=\log \mu$. To confirm this expectation, we need to show that this  $\rm t$-dependence matches the log-divergent universal counterterm \eqref{floweqApp} computed above. Thanks to the 4d Chern-Simons construction, the action of the $\s$-model can be written in terms of the 4d gauge field as \eqref{eq:Action2d}. The computation of the derivative of this action with respect to $\text{t}$ starts exactly as it did for simple poles in equation \eqref{eq:flowS2d} of the main text. In particular, there are two subtle points to take into account. First, when taking the derivative of the integral $\int_M \omega \wedge {\rm CS}[A]$, one needs to use the `measure derivative' $\dtom$ of $\omega$, which differs from the formal one \eqref{eq:FormalDerOm} by distributions arising from the flow of the poles $p_r$ -- see equation \eqref{eq:dtMeasure} and Appendix \ref{app:IntDer}. The second thing to account for is a variational term coming from the $\text{t}$-dependence of $A$ itself. Altogether, the conjectured counterterm in the effective action then takes the form
\begin{equation}\label{eq:flowS}
    \ddt \widehat S^{(1)}_{\rm conj} = \frac{1}{16i\pi^2} \int_M  \del_z \Psi(z) \, \ud z \wedge {\rm CS}[A]   + \Cdist + \Cvar  \ ,
\end{equation}
including a distributional term $\Cdist$ and a variational one $\Cvar$, which we compute below. 

\paragraph{Distributional term.} For simple poles, the measure derivative of $\omega$ was computed in equation \eqref{eq:dtDist}. In the present case, following Appendix \ref{app:IntDer}, this generalises to
\begin{equation}
    \dtom = \del_z \Psi(z)\,\ud z  + 2i\pi \sum_{r=1}^N \sum_{k=0}^{m_r-1} \frac{(-1)^k\ell_{r,k}}{k!}\frac{\dd\overline{p_r}}{\dd\text{t}} \,\del_z^k\delta(z-p_r)\,.
\end{equation}
The distributions in the second term contribute to the flow of the action by
\begin{equation}\label{eq:AppCdist}
    \Cdist = -\frac{1}{8\pi} \sum_{r=1}^N \sum_{k=0}^{m_r-1} \frac{\ell_{r,k}}{k!}  \overline{f(p_r)} \int_\Sigma  \del_z^k\langle A, \del_{\zb} A \rangle\Bigl|_{z=p_r} \,.
\end{equation}
To get this formula, we have carried out the integral over $(z,\zb)$, imposed the equation of motion ${\rm F}_{+-}[A]=0$, discarded some total derivatives and used the equation \eqref{eq:FlowParam} to express $\frac{\dd\overline{p_r}}{\dd\text{t}}=\overline{f(p_r)}$ (see \eqref{eq:Cdist2} and above for the analogous computation in the case of simple poles).

\paragraph{Variational term.} We now turn our attention to the variational term $\Cvar$ arising from the flow $\frac{\dd A}{\dd\text{t}}$ of the gauge field. Such a variation is computed using equation \eqref{eq:varA}. The bulk term in this equation vanishes on-shell due to the equation of motion $\text{F}[A]=0$, while the boundary term does not contribute since $A$ falls off at the boundary of $\Sigma$. The second term is a defect one and can be computed using the formula \eqref{eq:DefectJet2}. We find
\begin{equation}\label{eq:AppRGdef}
    \Cvar = -\frac 1{8\pi} \int_\Sigma \dlangle \Ad, \frac{\mathbbmss{}\mathbbm{d}\Ad}{\mathbbm{d}\text{t}} \drangle_{\mathfrak{d},\ellb}\,.
\end{equation}
In this equation, $\dfrac{\mathbbmss{}\mathbbm{d}\Ad}{\mathbbm{d}\text{t}}$ denotes the jet of $\dfrac{\dd A}{\dd\text{t}}$, defined as
\begin{equation}
    \frac{\mathbbmss{}\mathbbm{d}\Ad}{\mathbbm{d}\text{t}} = \left(\left. \frac{1}{k!} \, \partial_z^k \frac{\dd A}{\dd\text{t}} \right|_{z=p_r} \right)_{r=1,\dots,N}^{k=0,\dots,m_r-1}\; {\in \; \,\mathfrak{d}.}
\end{equation}
Crucially, this does not coincide with $\dfrac{\dd\Ad}{\dd\text{t}}$, \textit{i.e.} the $\text{t}$-derivative of the jet $\Ad = \bigl( \frac{1}{k!} \partial_z^k A|_{z=p_r} \bigr)_{r}^k$, since we are taking evaluations at points $p_r$ that are themselves $\text{t}$-dependent. Recalling from equation \eqref{eq:FlowParam} that $\dfrac{\dd p_r}{\dd\text{t}}=f(p_r)$, we get
\begin{equation}\label{eq:JetDtA}
     \frac{\mathbbmss{}\mathbbm{d}\Ad}{\mathbbm{d}\text{t}} = \frac{\dd \Ad}{\dd\text{t}} - \left( \frac{f(p_r)}{k!} \partial_z^{k+1} A \bigr|_{z=p_r}  +  \frac{\overline{f(p_r)}}{k!} \partial_{\zb} \partial_{z}^{k} A  \bigr|_{z=p_r} \right)_{r=1,\dots,N}^{k=0,\dots,m_r-1}\,.
\end{equation}

\paragraph{Flow of the boundary condition.} To compute the variational term \eqref{eq:AppRGdef}, we will need to take into account the boundary condition imposed on $\Ad$ and its behavior under the RG flow. This condition \eqref{eq:BC} is exactly where the maximally isotropic subalgebra $\kf$ enters the 4d Chern-Simons construction. In equation \eqref{eq:kt}, we have proposed a natural guess for the dependence of $\kf$ on the RG-scale, namely that $\kf = \rho^{-1}\kf_0$. In this expression, the dependence on $\text{t}$ is hidden in $\rho$, while $\kf_0$ is the initial condition for $\kf$ and thus is independent of $\text{t}$. Under this ansatz, the boundary condition \eqref{eq:BC} reads $\rho(\Ad) \in  \kf_0$. Taking the derivative with respect to $\text{t}$, using the definition \eqref{eq:EdoSigma} of $\rho$ and further acting by $\rho^{-1}$, we get
\begin{equation}
    \frac{\dd \Ad}{\dd\text{t}} + \Delta(\Ad) \; \in \; \kf\,.
\end{equation}
We now take the pairing with $\Ad$, which also belongs to the isotropic subspace $\kf$ by construction. We then find
\begin{equation}
    \dlangle \Ad, \frac{\dd \Ad}{\dd\text{t}} \drangle_{\df,\ellb} = -  \dlangle \Ad,\Delta(\Ad) \drangle_{\df,\ellb}\,.
\end{equation}
Together with equation \eqref{eq:JetDtA}, we rewrite \eqref{eq:AppRGdef} as
\begin{equation}\label{eq:AppRGdef2}
    \Cvar = \frac 1{8\pi} \int_\Sigma \dlangle \Ad, \Bd + \widetilde{\Bd}  \drangle_{\df,\ellb}\,,
\end{equation}
with
\begin{equation}\label{eq:B}
    \Bd = \Delta(\Ad) + \left( \frac{f(p_r)}{k!} \partial_z^{k+1} A \bigr|_{z=p_r}  \right)_{r}^{k} \qquad \text{ and } \qquad \widetilde{\Bd} = \biggl( \, \frac{\overline{f(p_r)}}{k!} \partial_{z}^{k} (\partial_{\zb}  A) \bigr|_{z=p_r}\biggr)_{r}^{k}\,\;.
\end{equation}

\paragraph{Computing the variational term.} Let us analyse in more detail the field $\Bd$ defined in equation \eqref{eq:B}. To do so, we use the explicit definition \eqref{eq:DefDelta} of the derivation $\Delta$ and apply it to the jet \eqref{eq:JetHigher}. The two terms entering $\Bd$ then combine to
\begin{equation}
    \Bd = \left( \sum_{l=0}^k \frac{f^{(k-l)}(p_r)}{l!(k-l)!} \partial_z^{l+1} A|_{z=p_r} \right)_{r=1,\dots,N}^{k=0,\dots,m_r-1} \,.
\end{equation}
From the generalised Leibniz rule, we recognise the quantity
\begin{equation}
    \Bd = \left( \frac{1}{k!}\, \partial_z^k \bigl( f(z)\, \partial_z A \bigr) \bigl|_{z=p_r} \right)_{r=1,\dots,N}^{k=0,\dots,m_r-1} \,.
\end{equation}
Substituting this expression into equation \eqref{eq:AppRGdef2} and using the explicit definition \eqref{eq:FormDefect} of $\dlangle\cdot,\cdot\drangle_{\df,\ellb}$, we obtain
\begin{align*}
    \Cvar = & \; \frac 1{8\pi} \sum_{r=1}^N \sum_{k=0}^{m_r-1} \ell_{r,k} \sum_{l=0}^k \frac{1}{l!(k-l)!} \int_\Sigma \Bigl\langle \partial^l_z A\bigl|_{z=p_r}, \partial^{k-l}_z\bigl( f(z)\, \partial_z A \bigr) \bigl|_{z=p_r} \Bigr\rangle\\
    & + \frac 1{8\pi}  \sum_{r=1}^N \sum_{k=0}^{m_r-1} \ell_{r,k} \sum_{l=0}^k \frac{\overline{f(p_r)}}{l!(k-l)!}  \int_\Sigma \Bigl\langle \partial^l_z A\bigl|_{z=p_r}, \partial^{k-l}_z\bigl( \partial_{\zb} A \bigr) \bigl|_{z=p_r} \Bigr\rangle\,.
\end{align*}
We then use the generalised Leibniz rule again to find
\begin{equation*}
    \Cvar = \frac 1{8\pi} \sum_{r=1}^N \sum_{k=0}^{m_r-1} \frac{\ell_{r,k}}{k!} \int_\Sigma \left.\left( \partial_z^k\Bigl( f(z)\,\langle A, \partial_z A \rangle )  \Bigr) + \overline{f(p_r)}\, \partial_z^k\langle A, \partial_{\zb} A \rangle \right) \right|_{z=p_r} \,.
\end{equation*}
Comparing with equation \eqref{eq:AppCdist}, we see that the second term coincides with the opposite of the distributional contribution $\Cdist$, and hence
\begin{equation}\label{eq:AppRGdef3}
    \Cdist + \Cvar = \frac 1{8\pi}  \sum_{r=1}^N \sum_{k=0}^{m_r-1} \frac{\ell_{r,k}}{k!} \int_\Sigma \partial_z^k\Bigl( f(z)\,\langle A, \partial_z A \rangle )  \Bigr) \Bigl|_{z=p_r} \,.
\end{equation}

\paragraph{Matching with the universal computation.} We now claim that the formula \eqref{eq:AppRGdef3} can be rewritten as the defect term \eqref{deteApp} appearing in the universal computation, \textit{i.e.} that
\begin{equation}\label{eq:dbarPsiCSA}
    \Cdist + \Cvar = \DT \, .
\end{equation} 
To prove this, we first recall that the component of $A$ along $\dd z$ decouples on-shell from the universal counterterm (see the discussion below equation \eqref{combi}) and can thus be set to zero in the computations. We then rewrite the defect term \eqref{deteApp} as
\begin{equation}\label{eq:dbarPsiAdA}
    \DT = \frac{1}{16i\pi^2} \int_M \del_{\zb} \Psi(z)\, \dd \zb \wedge {\rm CS}[A] = \frac{1}{16i\pi^2} \int_M \del_{\zb} \Psi(z)\, \dd z \wedge \dd \zb \wedge \langle A, \partial_z A \rangle\,.
\end{equation}
Recall that $\Psi(z)$ is a meromorphic function with poles $\Ph = \lbrace p_r\rbrace_{r=1}^N$. Therefore $\del_{\zb} \Psi(z)$ is a sum of derivatives of $\delta$-distributions located at these points and the above integral localises to $\Sigma\times\Ph$. We also recall that $\Psi(z)=-f(z)\vp(z)$, where $\Psi(z)$ and $\vp(z)$ share the same pole structure along $\Ph$, while the  poles of $f(z)$ coincide with the zeros $\Zh^\pm$ of $\vp(z)$ and thus disappear from the product $\vp(z)f(z)$. Writing $\del_{\zb} \Psi(z) = -f(z) \del_{\zb} \vp(z) - \del_{\zb} f(z)\vp(z)$, we see that the first term gives a contribution of the expected form, with distributions located along $\Ph$, while the second term would a priori give unwanted distributions along $\Zh^\pm$. We thus expect this second term to disappear: this is indeed the case since the distributions along $\Zh^\pm$ in $\del_{\zb} f(z)$ are multiplied by corresponding zeros in $\vp(z)$ and thus vanish in the distributional sense. We thus get $\del_{\zb} \Psi(z) \dd z \wedge \dd \zb = -f(z) \del_{\zb} \vp(z) \dd z \wedge \dd \zb = f(z)\dd\omega$, hence
\begin{equation}
    \DT  = \frac{1}{16i\pi^2} \int_M \dd \omega \wedge f(z)\,\langle A, \partial_z A \rangle\,.
\end{equation}
Inserting the partial fraction decomposition \eqref{eq:OmegaPFD}, the standard defect computation gives
\begin{equation*}
    \DT  = \frac 1{8\pi}  \sum_{r=1}^N \sum_{k=0}^{m_r-1} \frac{\ell_{r,k}}{k!} \int_\Sigma \partial_z^k\Bigl( f(z)\,\langle A, \partial_z A \rangle )  \Bigr) \Bigl|_{z=p_r} \,\,.
\end{equation*}
This expression matches \eqref{eq:AppRGdef3}, proving the claim \eqref{eq:dbarPsiCSA}.

To conclude, the identification \eqref{eq:dbarPsiCSA} implies that the universal counterterm \eqref{floweqApp} matches the conjectured flow \eqref{eq:flowS} of the effective action. We thus proved the renormalisability of the model, with the expected RG flow of $\omega$ and $\kf$.

\section{Integrals over \texorpdfstring{$\CP$}{CP1} and their derivatives}\label{app:Integrals}
\subsection{Integration against meromorphic 1-forms on \texorpdfstring{$\CP$}{CP1}}
\label{app:Int}

\paragraph{Generalities.} Let $\omega = \vp(z)\,\dd z$ be a \textit{meromorphic 1-form on $\CP$}. Our goal in this appendix is to define and study integrals of the form
\begin{equation}\label{eq:IntOmega}
    \int_{\CP} \omega \wedge \Lambda\,, \qquad \text{ where } \qquad \Lambda = f(z,\zb)\,\dd \zb
\end{equation}
is a \textit{smooth $(0,1)$-form on $\CP$}. Without loss of generality, we restricted ourselves to $\Lambda$ being a $(0,1)$-form, \textit{i.e.} having a component only along $\dd\zb$, since $\omega$ is proportional to $\dd z$. Note that $\Lambda=f(z,\zb)\,\dd \zb$ is not assumed to be an anti-holomorphic form, \textit{i.e.}\ it can depend on both $z$ and $\zb$, but is taken to be smooth on $\CP$. Integrals of the form \eqref{eq:IntOmega} are central in the 4d Chern-Simons theory, for instance in the definition of its action \eqref{eq:Action4d}.\footnote{Technically, the integrals considered in this appendix are slightly less general. Indeed, we supposed here that $\Lambda$ is smooth on $\CP$ while in the 4d-CS theory it would generally be allowed to have singularities, located at the zeros of $\omega$ so that these poles cancel in the integrand $\omega \wedge \Lambda$. We do not consider this more general setup for simplicity but expect most results of the appendix to extend to that case as well.} To be more precise, in that context, the integrals are taken over the direct product $\CP \times \Sigma$ and $\Lambda$ is replaced by a 3-form $f(z,\zb,x^+,x^-)\,\dd \zb\wedge \dd x^+ \wedge \dd x^-$, where $(x^+,x^-)$ are coordinates on $\Sigma$. This appendix can then be understood as dealing with the partial integration over $\CP$, working point-wise in $\Sigma$.

In the rest of this appendix, it will often be useful to write $\Lambda$ as a $\bar{\del}$-exact form, \textit{i.e.} as
\begin{equation}\label{eq:Lambdah}
    \Lambda = \bar{\del} h = \del_{\zb} h(z,\zb) \, \dd \zb\,,
\end{equation}
where $h$ is a smooth function on $\CP$. Such a rewriting always exists since the Dolbeaut cohomology group $H^{(0,1)}(\CP)$ is trivial. Moreover, the function $h$ is unique up to shifts by a constant term $h(z,\zb) \mapsto h(z,\zb) + a$ (indeed, the difference between two solutions of $\del_{\zb}h=f$ is by construction holomorphic on $\CP$ and thus constant by Liouville's theorem). We will call $h$ a $\bar{\del}$-primitive of $\Lambda$.

\paragraph{A simple example.} Before considering the most general case, let us investigate the simple example of the meromorphic 1-form
\begin{equation}\label{eq:Omegas}
    \omega_s = \ell\left(\frac{1}{z-p_1} - \frac{1}{z-p_2}\right) \dd z\,,
\end{equation}
with $p_1,p_2\in\C$. It has two simple poles at $z=p_1$ and $z=p_2$, with residues $+\ell$ and $-\ell$ respectively (note that the opposite signs ensure that $\omega$ is regular at infinity). Introducing polar coordinates $(\rho_r,\theta_r)$ by $z_r=p_r+\rho_r e^{i\theta_r}$ and using $\dd z\wedge \dd \zb = -2i\rho_r\, \dd \rho_r \wedge \dd \theta_r$, one sees that the 2-form $\omega_s \wedge \Lambda \sim -2ife^{-i\theta_r}\, \dd \rho_r \wedge \dd \theta_r$ is locally integrable around $z=p_r$ (with the singularity in $\omega$ being compensated by the factor $\rho_r$ in the measure). Therefore, the integral $\int_{\CP} \omega_s \wedge \Lambda$ converges, despite the presence of poles in $\omega_s$. Working in polar coordinates, the integration can be performed explicitly in terms of the $\bar{\del}$-primitive $h(z,\zb)$ introduced in equation \eqref{eq:Lambdah}, yielding
\begin{equation}\label{eq:IntOmegap}
    \int_{\CP} \omega_s \wedge \Lambda = 2i\pi\ell \bigl( h(p_1,\bar p_1) - h(p_2,\bar p_2) \bigr)\,.
\end{equation}
Recall that this $\bar{\del}$-primitive is not unique, as one can shift it by a constant, $h(z,\zb) \mapsto h(z,\zb) + a$. As a consistency check, we note that the right-hand side of equation \eqref{eq:IntOmegap} is invariant under such shifts, so is independent of the choice of primitive.

\paragraph{General case.} We now consider a general meromorphic 1-form $\omega$. We denote its poles by $p_r$, $r\in\lbrace 1,\dots,N \rbrace$, and their multiplicities by $m_r \in \Z_{\geq 1}$. For simplicity, we will work with a coordinate $z$ on $\CP$ such that each $p_r$ is in the finite complex plane. We further introduce the levels $\ell_{r,k}$ for $k\in\lbrace 0,\dots,m_r-1\rbrace$ through the partial decomposition
\begin{equation}\label{eq:OmegaApp}
    \omega = \sum_{r=1}^N \sum_{k=0}^{m_r-1} \frac{\ell_{r,k}}{(z-p_r)^{k+1}} \,\dd z \,.
\end{equation}
If $m_r>1$, then the 2-form $\omega \wedge \Lambda$ is not locally integrable around the pole $z=p_r$. To make sense of the integral \eqref{eq:IntOmega}, one needs to consider a regularised version of this 2-form, as explained in~\cite{Benini}. Essentially, it is obtained by adding a $\del$-exact form to $\omega \wedge \Lambda$:
\begin{equation}
    (\omega \wedge \Lambda)_{\text{reg}} = \omega \wedge \Lambda + \del\chi\,.
\end{equation}
The $(0,1)$-form $\chi$ is chosen in such a way that $(\omega \wedge \Lambda)_{\text{reg}} \sim \frac{a_r}{z-p_r} \dd z \wedge \dd \zb$ around each pole $p_r$, ensuring that $(\omega \wedge \Lambda)_{\text{reg}}$ is integrable on $\CP$. Such a $\chi$ always exists (see~\cite{Benini} for an explicit `minimal' construction). Moreover, different choices of $\chi$ would lead to regularised 2-forms $(\omega \wedge \Lambda)_{\text{reg}}$ that differ from the $\del$-derivative of a smooth form on $\CP$, whose integral vanishes. Therefore, this procedure uniquely defines a regularised integral $\int_{\CP} (\omega \wedge \Lambda)_{\text{reg}}$. In the rest of this appendix and the paper, we always assume that integrals $\int_{\CP} \omega \wedge \Lambda$ have been regularised in this way when $\omega$ has higher-order poles and omit the index `reg' to lighten the notation. Heuristically, this notion of regularisation amounts to discarding all integrals $\int_{\CP}\del\chi$ of $\del$-exact forms, even if $\chi$ posseses singularities.

With this definition, the integral $\int_{\CP} \omega \wedge \Lambda$ can be computed explicitly in terms of the pole structure of $\omega$ and the $\bar{\del}$-primitive $h$ of $\Lambda$. Namely, we find:
\begin{equation}\label{eq:IntOmegaGen}
    \int_{\CP} \omega \wedge \Lambda = 2i\pi \sum_{r=1}^N \sum_{k=0}^{m_r-1} \frac{\ell_{r,k}}{k!}\,\del^k_{z} h (p_r,\overline{p_r}) \,.
\end{equation}
This generalises the simple example \eqref{eq:IntOmegap} treated earlier, corresponding to two simple poles at $p_1$ and $p_2$. As one should expect, the right-hand side of equation \eqref{eq:IntOmegaGen} is independent of the choice of $\bar{\del}$-primitive of $\Lambda$. Indeed, a shift $h(z,\zb) \mapsto h(z,\zb)+a$ adds a term $2\pi i a\sum_{r=1}^N \ell_{r,0}$, which is proportional to the sum of all residues of $\omega$ and thus vanishes by the residue formula.

\paragraph{Distributions and integration by parts.} Note that $\omega \wedge \Lambda = \omega \wedge \bar{\del}h = \omega \wedge \dd h$, since $\omega$ is along $\dd z$. To be able to do concrete computations with integrals against meromorphic 1-forms, we want some standard rules such as integration by parts to hold, despite the presence of poles in $\omega$. In particular, we want to have 
\begin{equation}
    \int_{\CP} \omega \wedge \Lambda = \int_{\CP} \omega \wedge \dd h = \int_{\CP} \dd\omega \wedge h\,,
\end{equation}
without any boundary term since $\del\CP$ is empty. Comparing with the explicit result \eqref{eq:IntOmegaGen}, one sees that this requires treating $\dd\omega$ as a distribution. More precisely, we get
\begin{equation}\label{eq:dOmDist}
    \dd\omega = - \del_{\zb}\vp(z)\,\dd z\wedge \dd \zb = 2\pi i\,  \sum_{r=1}^N \sum_{k=0}^{m_r-1} \frac{(-1)^k\ell_{r,k}}{k!}\,\del_z^k\delta(z-p_r)\,\dd z\wedge \dd \zb\,.
\end{equation}
Here $\delta(z-p_r)$ is a 2d Dirac-distribution localised at $z=p_r$ (and normalised with respect to the measure $\dd z\wedge \dd \zb$) and $\del_z^k\delta(z-p_r)$ is its $k$-th derivative, satisfying
\begin{equation}
    \int_{\CP} \del_z^k\delta(z-p_r)\,h(z,\zb)\,\dd z\wedge \dd \zb = (-1)^k \del_z^kh(p_r,\overline{p_r})\,.
\end{equation}

\subsection{Derivatives of integrals}
\label{app:IntDer}

\paragraph{Formal and measure derivatives.} Let us consider the meromorphic 1-form $\omega$ in equation \eqref{eq:OmegaApp}, written in terms of complex parameters $\Pi_a = (p_r,\, \ell_{r,k})$. We now suppose that these parameters depend themselves on an external variable $\text{t}$. We then get a continuous family of $\text{t}$-dependent 1-forms $\omega$. We define its \textit{formal derivative} with respect to $\text{t}$ as
\begin{equation}\label{eq:dFormal}
    \frac{\dd\omega}{\dd\text{t}} = \sum_a \frac{\del \omega}{\del \Pi_a} \, \frac{\dd \Pi_a}{\dd\text{t}} = \sum_{r=1}^N \sum_{k=0}^{m_r-1} \left( \frac{\dd \ell_{r,k}}{\dd\text{t}}\frac{1}{(z-p_r)^{k+1}} + \frac{\dd p_r}{\dd\text{t}} \frac{(k+1)\ell_{r,k}}{(z-p_r)^{k+2}} \right) \,\dd z \,.
\end{equation}
This is still a meromorphic 1-form on $\CP$, whose data is equivalent to the knowledge of the flow of all parameters $\Pi_a = (p_r,\ell_{r,k})$.

The main goal of this subsection will be to study the behaviour of the integral \eqref{eq:IntOmega} when the external variable $\text{t}$ varies (also allowing $\Lambda$ to depend on $\text{t}$). Our main result will be that the $\text{t}$-derivative of this integral will take the form
\begin{equation}\label{eq:dMeas}
    \frac{\dd\;}{\dd\text{t}} \left( \int_{\CP} \omega \wedge \Lambda \right) = \int_{\CP} \left( \dtom \wedge \Lambda + \omega \wedge \dfrac{\dd\Lambda}{\dd\text{t}} \right) \,,
\end{equation}
where $\displaystyle\dtom$ is another object that we will call the \textit{measure derivative} of $\omega$. As we will see, the latter does not exactly coincides with the formal derivative $\dfrac{\dd\omega}{\dd\text{t}}$ and includes additional distributional terms, arising from the flow of the poles $p_r$.

\paragraph{Back to our simple example.} Before investigating the general case, it is instructive to consider the simple example treated in the previous subsection. There, we considered the 1-form $\omega_s$ given by equation \eqref{eq:Omegas}, possessing two simple poles and depending on 3 parameters $\Pi^a=(p_1,p_2,\ell)$. The integral $\int_{\CP} \omega_s \wedge \Lambda$ was computed explicitly in equation \eqref{eq:IntOmegap}, in terms of the $\bar{\del}$-primitive $h$. A direct derivation then yields
\begin{align}
    \frac{\dd}{\dd\text{t}} \left( \int_{\CP} \omega_s \wedge \Lambda \right) =& 2i\pi\frac{\dd\ell}{\dd\text{t}} \bigl( h(p_1,\bar p_1) - h(p_2,p_2) \bigr) + 2i\pi\ell \left( \frac{\dd p_1}{\dd\text{t}} \del_z h(p_1,\bar p_1) - \frac{\dd p_2}{\dd\text{t}} \del_z h(p_2,p_2) \right) \notag \\
     & \hspace{10pt} + 2i\pi\ell \left( \frac{\dd \overline{p_1}}{\dd\text{t}} f(p_1,\bar p_1) - \frac{\dd \overline{p_2}}{\dd\text{t}} f(p_2,p_2) \right) \notag \\
     & \hspace{10pt} + 2i\pi\ell \left( \frac{\dd h}{\dd\text{t}} (p_1,\bar p_1) - \frac{\dd h}{\dd\text{t}} (p_2,p_2) \right)  \,, \label{eq:dInts}
\end{align}
where we have used $\del_{\zb}h=f$ in the second line. The last line takes into account the explicit dependence of the function $h$ on $\text{t}$: namely, $\frac{\dd h}{\dd\text{t}} (z,\bar z)$ is defined as the $\text{t}$-derivative of $h(z,\bar z)$ keeping $(z,\bar z)$ constant. This derivative is related to the one of $f$ by $\frac{\dd f}{\dd\text{t}} (z,\zb) = \del_{\zb} \left( \frac{\dd h}{\dd\text{t}}(z,\zb) \right)$. Using the identity \eqref{eq:IntOmegap}, we can then rewrite the last line of equation \eqref{eq:dInts} as $\int_{\CP} \omega_s \wedge \frac{\dd\Lambda}{\dd\text{t}}$.

Comparing with equation \eqref{eq:dMeas}, the first and second lines of \eqref{eq:dInts} should therefore coincide with $\int_{\CP} \frac{\text{d}\omega_s}{\text{dt}}\bigr|_{\text{meas}} \wedge \Lambda$, serving as a definition of the measure derivative of $\omega_s$. On the other hand, its formal derivative reads
\begin{equation}\label{eq:dOs}
    \frac{\dd\omega_s}{\dd\text{t}} =  \frac{\dd \ell}{\dd\text{t}}\left( \frac{1}{z-p_1} - \frac{1}{z-p_2} \right) \dd z + \ell\left( \frac{\dd p_1}{\dd\text{t}} \frac{1}{(z-p_1)^2} - \frac{\dd p_2}{\dd\text{t}} \frac{1}{(z-p_2)^2} \right) \,\dd z\,.
\end{equation}
Using the identity \eqref{eq:IntOmegaGen} for the form $\frac{\dd\omega_s}{\dd\text{t}}$, we then find
\begin{equation*}
    \int_{\CP} \frac{\dd\omega_s}{\dd\text{t}} \wedge \Lambda =  2i\pi\frac{\dd\ell}{\dd\text{t}} \bigl( h(p_1,\bar p_1) - h(p_2,\bar p_2) \bigr) + 2i\pi\ell \left( \frac{\dd p_1}{\dd\text{t}} \del_z h(p_1,\bar p_1) - \frac{\dd p_2}{\dd\text{t}} \del_z h(p_2,\bar p_2) \right)\,.
\end{equation*}
This coincides with the first line of the right hand side of equation \eqref{eq:dInts}. We thus see that the formal derivative $\dfrac{\dd\omega_s}{\dd\text{t}}$ cannot coincide with the measure derivative $\dfrac{\text{d}\omega_s}{\text{dt}}\Bigr|_{\text{meas}}$. More precisely, they differ from a distributional term, whose presence accounts for the second line in equation \eqref{eq:dInts}. Explicitly, we find
\begin{equation}
    \frac{\text{d}\omega_s}{\text{dt}}\Bigr|_{\text{meas}} = \frac{\dd\omega_s}{\dd\text{t}} + 2i\pi\,\ell \left( \frac{\dd \overline{p_1}}{\dd\text{t}} \delta(z-p_1) - \frac{\dd \overline{p_2}}{\dd\text{t}} \delta(z-p_2) \right)\dd z\,.
\end{equation}

\paragraph{General case.} We can straightforwardly extend the analysis of the previous paragraph to the general meromorphic 1-form \eqref{eq:OmegaApp}. Namely, one can take the $\text{t}$-derivative of the integral \eqref{eq:IntOmegaGen} and compare it to the integral $\int_{\CP} \left( \frac{\dd\omega}{\dd\text{t}} \wedge \Lambda + \omega \wedge \frac{\dd\Lambda}{\dd\text{t}}\right) $, computed using the formal derivative \eqref{eq:dFormal}. One finds that these two quantities do not coincide and differ from a distributational term. Eventually, we extract the following relation for the measure derivative:
\begin{equation}\label{eq:dtMeasGen}
    \frac{\text{d}\omega}{\text{dt}}\Bigr|_{\text{meas}} = \frac{\dd\omega}{\dd\text{t}} + 2i\pi \sum_{r=1}^N \sum_{k=0}^{m_r-1} \frac{\dd\overline{p_r}}{\dd \text{t}}\,\frac{(-1)^k\ell_{r,k}}{k!} \,\del_z^k\delta(z-p_r)\,.
\end{equation}
We note that there is another (more symbolic) way of justifying this formula. Indeed, one can also interpret the measure derivative of $\omega$ as
\begin{equation}
    \frac{\text{d}\omega}{\text{dt}}\Bigr|_{\text{meas}} = \sum_a \left( \frac{\dd \Pi_a}{\dd\text{t}}\,\frac{\del \omega}{\del \Pi_a} + \frac{\dd \overline{\Pi_a}}{\dd\text{t}}\,\frac{\del \omega}{\del \overline{\Pi_a}}\right)\,,\label{dpibar}
\end{equation}
where we recall that $\Pi_a=(p_r,\ell_{r,k})$ are the complex parameters defining $\omega$. The first term is nothing but the formal derivative \eqref{eq:dFormal}, while the second term gives a distributional contribution using the formal rule
\begin{equation}
    \frac{\del}{\del \overline{p}} \left( \frac{1}{(z-p)^{k+1}} \right) = \frac{2i\pi}{k!}\, \del_p^k \delta(z-p) = \frac{2i\pi(-1)^k}{k!}\, \del_z^k \delta(z-p)\,.
\end{equation}

\paragraph{Distributional and variational terms.} We end this appendix by introducing some terminology that will be useful in the main text. Consider the expression \eqref{eq:dMeas} for the $\text{t}$-derivative of the integral $\int_{\CP} \omega \wedge \Lambda$. It will be useful to separate the different contributions in this equation by writing
\begin{equation}
    \frac{\dd\;}{\dd\text{t}} \left( \int_{\CP} \omega \wedge \Lambda \right) = \int_{\CP} \dfrac{\dd\omega}{\dd\text{t}} \wedge \Lambda + \Cdist + \Cvar \,.
\end{equation}
The first term involves only the formal derivative of $\omega$, as defined in equation \eqref{eq:dFormal}. The second one, called the \textit{distributional term}, is the contribution of the distributions in the measure derivative \eqref{eq:dtMeasGen}, which explicitly reads
\begin{equation}
    \Cdist = 2i\pi \sum_{r=1}^N \sum_{k=0}^{m_r-1} \frac{\dd\overline{p_r}}{\dd \text{t}}\,\frac{\ell_{r,k}}{k!} \,\del_z^k f(p_r,\bar p_r)\,.
\end{equation}
Finally, the last term arises from the explicit $\text{t}$-dependence of $\Lambda$ and is defined as
\begin{equation}
    \Cvar = \int_{\CP} \omega \wedge  \dfrac{\dd\Lambda}{\dd\text{t}}\,.
\end{equation}
In the main text, we will typically take $\int_{\CP} \omega \wedge \Lambda$ to be the 4d Chern-Simons action of a classical configuration of the gauge field $A$. In this context, $\Cvar$ then measures the variation of the action arising from the $\text{t}$-dependence of this field configuration: we will thus refer to it as the \textit{variational term}.

\section{1-loop RG flow for Lax in terms of \texorpdfstring{$\mathcal R$}{R}-matrix}\label{sec:RmatAppendix}
The purpose of this appendix is to carry out a general version of the universal computation of the 1-loop divergences of integrable $\sigma$-models by writing the Lax connection in terms of a classical $\mc{R}$-matrix: 
\begin{equation}
    L_{\pm}(z)_{\tn{1}}=\sum_{i=1}^{N_\pm}\sum_{k=0}^{m_i^\pm-1}\frac{(-1)^k}{k!}\left<\partial^k_z\mathcal{R}(z-z_i^\pm)_{\tn{12}},\mathcal{J}_{\pm,\tn{2}}^{(i,k)}\right>_{\tn{2}}\label{LRapp}
\end{equation}
This argument will generalise the one for rational integrable theories in Sections \ref{subsec:univ1loop}--\ref{subsec:PathInt}, replacing the role of partial fraction decomposition on $\mathbb {CP}^1$ with the classical Yang-Baxter equation (cYBE) \rf{eq:cYBE}. The argument will therefore apply as long as the Lax connection takes this form regardless of the spectral parameter's topology. In particular, it will allow our results to be extended to the elliptic case in Section \ref{sec:Rmat}. It should be noted that most of the computation does not rely on $\mathcal{R}:\mathfrak{g}\to\mathfrak{g}$ taking a difference form as we have written it in Section \ref{sec:elliptic} and in this appendix: however, the properties assumed in subsection \ref{sec:Rmat} actually enforce this as long as $c_G\neq 0$\cite{Belavin:1983cYB,Torrielli:2016CI}.

\subsection{Extracting equations of motion from zero curvature}\label{subsec:EOM}
Usefully for what follows, we will still be able to use the universal path integral computation from Section \ref{subsec:PathInt}, which is formulated abstractly in terms of equations of motion for the currents appearing in the Lax connection. Our first task is therefore to extract these equations of motion from the zero-curvature equation of the Lax connection,
\begin{equation}
    \partial_+L_-(z)-\partial_-L_+(z)+\left[L_+(z),L_-(z)\right]=0 \qquad \forall z \ .\label{eq:flat}
\end{equation}
Inserting the Lax connection \eqref{LRapp} expressed in terms of a general $\mc R$-matrix, we find the following equations of motion for the currents $\mathcal{J}_\pm^{(i,\ell)}$
\begin{align}
    0&=\sum_{i=1}^{N}\sum_{\ell=0}^{m_i^--1}\frac{(-1)^\ell}{\ell!}\left<\partial_z^\ell \mathcal{R}(z-{z}^-_i)_{\tn{31}},\partial_+\mathcal{J}^{(i,\ell)}_{-,\tn{1}}\right>_{\tn{1}}-\sum_{j=1}^{N}\sum_{n=0}^{m_i^+-1}\frac{(-1)^n}{n!}\left<\partial^{n}_z\mathcal{R}(z-z^+_j)_{\tn{32}},\partial_-\mathcal{J}^{(j,n)}_{+,\tn{2}}\right>_{\tn{2}}
    \no\\
    &\quad  +\sum_{i,j=1}^{N}\sum_{\ell=0}^{m_i^--1}\sum_{n=0}^{m_j^+-1}\frac{(-1)^{\ell+n}}{\ell!n!}\left<\left[\partial_z^{n}\mathcal{R}(z-{z}^+_j)_{\tn{32}},\partial_z^\ell \mathcal{R}(z-{z}^-_i)_{\tn{31}}\right],\mathcal{J}^{(i,\ell)}_{-,\tn{1}}\mathcal{J}^{(j,n)}_{+,\tn{2}} \right>_{\tn{12}}\,,
\label{eq:brute}
\end{align}
where the first line comes from $\partial_+L_-(z)-\partial_-L_+(z)$ and the second line from $[L_+(z),L_-(z)]$, and we have recalled that the commutator in \eqref{eq:flat} is only non-trivial in the third tensor factor of $\mathfrak{g}$. Using the cYBE \eqref{eq:cYBE} of $\mathcal{R}$ as well as the ad-invariance of the bilinear form $\left<\cdot,\cdot\right>$, we can re-express the trace in the second line as:
\begin{equation}
\begin{split}\label{eq:useLeib}
    (-1)^{\ell+n}&\bigg<\left[\partial^{n}_z\mathcal{R}(z-{z}^+_j)_{\tn{32}},\partial^\ell_z \mathcal{R}(z-{z}^-_i)_{\tn{31}}\right],\mathcal{J}^{(i,\ell)}_{-,\tn{1}}\mathcal{J}^{(j,n)}_{+,\tn{2}} \bigg>_{\tn{12}}\\ \qquad \qquad &=+(-1)^{n}\partial_{z^-_i}^{\ell}\bigg<\mathcal{R}(z-{z}^-_i)_{\tn{31}},\left[\left<\partial^{n}_{z_i^-}\mathcal{R}({z}^-_i-{z}^+_j)_{\tn{12}},\mathcal{J}_{+,\tn{2}}^{(j,n)}\right>_{\tn{2}},\mathcal{J}^{(i,\ell)}_{-,\tn{1}}\right]\bigg>_{\tn{1}}\\
    \qquad \qquad & \  \quad -(-1)^{\ell}\partial^{n}_{{z}_j^+}\bigg<\mathcal{R}(z-{z}^+_j)_{\tn{32}},\left[\left<\partial^\ell_{z^+_j} \mathcal{R}({z}^+_j-{z}^-_i)_{\tn{21}},\mathcal{J}_{-,\tn{1}}^{(i,\ell)}\right>_{\tn{1}},\mathcal{J}^{(j,n)}_{+,\tn{2}}\right]\bigg>_{\tn{2}}\,.
    \end{split}
\end{equation}
This rewriting collects all the $z$-dependence into a single $\mathcal{R}$-matrix, which makes the following analysis easier. Note, however, that this comes at the cost of having one of the two derivatives act on both $\mathcal{R}$-matrices. Distributing these derivatives using Leibniz rule and re-shuffling some terms, we are left with the commutator $[L_+,L_-]$ being written as
\begin{equation}\label{eq:RafterLeib}
\begin{split}
   \hspace{-1cm}&\sum_{i=1}^{N}\sum_{\ell=0}^{m_i^--1}\frac{(-1)^\ell}{\ell!}\left<\partial_z^\ell \mathcal{R}(z-{z}_i^-)_{\tn{31}},\sum_{j=1}^{N}\sum_{n=0}^{m_j^+-1}\sum_{m=\ell}^{m_i^--1}\frac{(-1)^n}{n!(m-\ell)!}\left[\left<\partial^{n+m-\ell}_{z_i^-}\mathcal{R}({z}_i^--{z}^+_j)_{\tn{12}},\mathcal{J}_{+,\tn{2}}^{(j,n)}\right>_{\tn{2}},\mathcal{J}_{-,\tn{1}}^{(i,m)}\right]\right>_{\tn{1}}
    \\
    \hspace{-1cm}-&\sum_{j=1}^{N}\sum_{n=0}^{m_j^+-1}\frac{(-1)^n}{n!}\left<\partial^{n}_z\mathcal{R}(z-{z}_j^+)_{\tn{32}},\sum_{i=1}^{N}\sum_{\ell=0}^{m_i^--1}\sum_{m=n}^{m_j^+-1}\frac{(-1)^\ell}{\ell!(m-n)!}\left[\left<\partial^{\ell+m-n}_{z_j^+}\mathcal{R}({z}_j^+-{z}^-_i)_{\tn{21}},\mathcal{J}_{-,\tn{1}}^{(i,\ell)}\right>_{\tn{1}},\mathcal{J}_{+,\tn{2}}^{(j,m]}\right]\right>_{\tn{2}}\,.
    \end{split}
\end{equation}
We then notice the similarity between \eqref{eq:RafterLeib} and the first line of \eqref{eq:brute}. Requiring these two lines to cancel for \textit{any} value of $z$, we can read off the equations of motion for the currents:
\begin{subequations}
\label{eq:ellpm}
\begin{align}
    &\partial_+\mathcal{J}_{-,\tn{1}}^{(i,\ell)}+\sum_{j=1}^{N}\sum_{n=0}^{m_j^+-1}\sum_{m=\ell}^{m_i^--1}\frac{(-1)^n}{n!(m-\ell)!}\bigg[\Big<\partial^{n+m-\ell}_{z^-_i}\mathcal{R}({z}_i^--{z}^+_j)_{\tn{12}},\mathcal{J}_{+,\tn{2}}^{(j,n)}\Big>_{\tn{2}},\mathcal{J}_{-,\tn{1}}^{(i,m)}\bigg]=0\,,
\\
\label{eq:ellmp}
    &\partial_-\mathcal{J}_{+,\tn{2}}^{(j,n)}+\sum_{i=1}^{N}\sum_{\ell=0}^{m_i^--1}\sum_{m=n}^{m_j^+-1}\frac{(-1)^\ell}{\ell!(m-n)!}\bigg[\Big<\partial_{z^+_j}^{\ell+m-n}\mathcal{R}({z}_j^+-{z}^-_i)_{\tn{21}},\mathcal{J}_{-,\tn{1}}^{(i,\ell)}\Big>_{\tn{1}},\mathcal{J}_{+,\tn{2}}^{(j,m)}\bigg]=0\,.
\end{align}
\end{subequations}
	These are the generalisations of \eqref{z1} and \eqref{z2} to also include the elliptic integrable field theories.
    
\subsection{1-loop divergence}\label{subsec:1loop}
The equations of motion \eqref{eq:ellpm}, match the general form in equation \eqref{eq:OPeq} of the main text, with the operators $O$ and $P$ identified as
	\begin{subequations}
    \begin{align}
		&O^{(i',k)}_{(j',m)\,(i,\ell)}\left[X\right]_{\tn{1}}=\delta_{i'=i}\delta_{\ell\geq k}\frac{(-1)^m}{m!(\ell-k)!}\bigg<\partial^{m+\ell-k}_{z_i^-}\mathcal{R}({z}_i^--{z}^+_{j'})_{\tn{12}},X_{\tn{2}}\bigg>_{\tn{2}}\,,\\
		&P^{(j',m)}_{(i',k)\,(j,n)}\left[X\right]_{\tn{1}}=\delta_{j'=j}\delta_{n\geq m}\frac{(-1)^k}{k!(n-m)!}\bigg<\partial^{n+k-m}_{z_j^+}\mathcal{R}(z_j^+-z^-_{i'})_{\tn{12}},X_{\tn{2}}\bigg>_{\tn{2}}\,,
        \end{align}
	\end{subequations}
    for $X\in\mathfrak{g}$.
    Thus, the general computation of the universal path integral in Section \ref{subsec:PathInt} applies, and the general result \eqref{renre} gives in this case
	\begin{equation}
		\frac{\dd}{\dd\text{t}}
        \widehat S^{(1)}=- \frac{1}{4\pi } \int \ud^2 x \, \Tr_{\mathfrak{g}}\left[\text{ad}_{\mathcal{J}_-^{(i,\ell)}}O^{(i',k)}_{(j',m)\,(i,\ell)}\text{ad}_{\mathcal{J}_+^{(j,n)}}P^{(j',m)}_{(i',k)\,(j,n)}\right]\,,\label{eq:renell}
	\end{equation}
	where summation over all indices is implied. Note that, given a basis $\{T^\alpha\}_{\alpha=1}^{\text{dim}\,\mathfrak{g}}$ of the Lie algebra, for an operator $Q:\mathfrak{g}\mapsto\mathfrak{g}$ the trace can be written as $\Tr_\mathfrak{g}[Q]=\left<T^\alpha,Q[T_\alpha]\right>$, where the split Casimir $C$ is defined by $C_{\tn{12}}=T^\alpha_{\tn{1}}\,T_{\alpha,\tn{2}}$. We would then like to unpack \eqref{eq:renell} by essentially reversing the steps that led us from \eqref{eq:brute} to \eqref{eq:RafterLeib}. We use Leibniz theorem and ad-invariance to write it in the form 
	\begin{align}\label{eq:reverse}
		&\Tr_{\mathfrak{g}}\left[\text{ad}_{\mathcal{J}_-^{(i,\ell)}}O^{(i',k)}_{(j',m)\,(i,\ell)}\text{ad}_{\mathcal{J}_+^{(j,n)}}P^{(j',m)}_{(i',k)\,(j,n)}\right]
  \\
  &\ \ \ =\sum_{i,j=1}^{N}\sum_{\ell=0}^{m_i^--1}\sum_{n=0}^{m_j^+-1}\frac{1}{\ell!n!}\, \partial^{\ell}_{z_i^-}\partial^{n}_{z_j^+}\left<\left[\mathcal{R}(z_i^--z_j^+)_{\tn{12}},\mathcal{R}(z_j^+-z_i^-)_{\tn{23}}\right],\left[\mathcal{J}_{-,\tn{1}}^{(i,\ell)},C_{\tn{13}}\right]\mathcal{J}_{+,\tn{2}}^{(j,n)}\right>_{\tn{123}}\,. \no
  \end{align}
	It would then be natural to use the cYBE to simplify the commutator. However, we have to be careful: we will seemingly generate terms where the two arguments of the $\mathcal{R}$-matrix are the same, but this is precisely where we assumed in \eqref{eq:RAsym} that $\mathcal{R}$ has a pole. Using the asymptotic behaviour of $\mathcal{R}(z)\sim C/z$ alongside the property $\left[C_{\tn{12}},X_{\tn{1}}\right]=-\left[C_{\tn{12}},X_{\tn{2}}\right]$ and the skew symmetry of $\mathcal{R}$, we find 
	\begin{equation}
		\left[\mathcal{R}(-z)_{\tn{12}},\mathcal{R}(z)_{\tn{23}}\right]=-\partial_z\left[\mathcal{R}(z)_{\tn{21}},C_{\tn{13}}\right]\,.
	\end{equation}
We can insert this in \eqref{eq:reverse}, to obtain
\begin{equation}
\begin{split}
    &\partial^{\ell}_{z_i^-}\partial^{n}_{z_j^+}\left<\left[\mathcal{R}(z_i^--z_j^+)_{\tn{12}},\mathcal{R}(z_j^+-z_i^-)_{\tn{23}}\right],\left[\mathcal{J}_{-,\tn{1}}^{(i,\ell)},C_{\tn{13}}\right]\mathcal{J}_{+,\tn{2}}^{(j,n)}\right>_{\tn{123}}
    \\
    =&(-1)^{\ell+1}\partial_{z_j^+}^{n+\ell+1}\bigg<\Big<\left[\mathcal{R}(z_j^+-z_i^-)_{\tn{21}},C_{\tn{13}}\right],\left[\mathcal{J}^{(i,\ell)}_{-,\tn{1}},C_{\tn{13}}\right]\Big>_{\tn{13}}\mathcal{J}^{(j,n)}_{+,\tn{2}}\bigg>_{\tn{2}}\,.
\end{split}
\end{equation}
We can then use the identity $\left<[X_{\tn{1}},C_{\tn{13}}],[Y_{\tn{1}},C_{\tn{13}}]\right>_{\tn{13}}=2c_G\left<X_{\tn{1}},Y_{\tn{1}}\right>_{\tn{1}}$ to compute the trace over $\tn{3}$, which leaves us with a simple expression for the log-divergent terms in the 1-loop effective action:
\begin{equation}
    \frac{\dd}{\dd\text{t}}\widehat S^{(1)}=\frac{\cG}{2\pi}\int\ud^2x\sum_{i,j=1}^N\sum_{\ell=0}^{m_i^--1}\sum_{n=0}^{m_j^+-1}\frac{(-1)^{\ell}}{\ell!n!}\left<\partial_{z^+_j}^{n+\ell+1}\mathcal{R}(z_j^+-z_i^-)_{\tn{12}},\mathcal{J}_{-,\tn{1}}^{(i,\ell)}\mathcal{J}_{+,\tn{2}}^{(j,n)}\right>_{\tn{12}} \ .
\end{equation}
\section{Example: the Principal Chiral Model}\label{app:PCM}
This appendix reviews the simplest model obtained from 4d Chern-Simons, the Principal Chiral Model (PCM): a $\sigma$-model with a Lie group as its target space. We will follow refs.~\cite{CY3,unif}, and refer the reader to~\cite{Lacroix:2021iit} for a more introductory disccussion.

\paragraph{The 4d setup.} Let us consider the twist 1-form
\begin{equation}\label{eq:OmegaPCM}
    \omega = h\,\frac{1-z^2}{z^2}\,\dd z\,,
\end{equation}
depending on a real parameter $h$. It has simple zeros at $z_1^\pm = \pm 1$ and double poles at $p_1=0$ and $p_2=\infty$. The latter have vanishing residues, \textit{i.e.}  $\ell_{r,0}=\res_{z=p_r} \, \omega = 0$, which will correspond to the absence of Wess-Zumino terms in the resulting 2d action. We note that we have moved away slightly from the conventions of the main text and Appendix \ref{app:GenPoles}, by allowing the twist 1-form to have a double pole at infinity. While this choice makes most formulae in this appendix simpler, it is inconsequential: all the results of the main text apply to this case, sometimes with minor modifications (see also the related discussion below equation \eqref{eq:ghPCM}).

Since $\omega$ has poles at $p_1=0$ and $p_2=\infty$, we need to impose boundary conditions on the gauge field at these points. To obtain the PCM, we choose the simple boundary conditions\footnote{These boundary conditions can be reformulated in the language used in the main text and Appendix \ref{app:GenPoles}, in terms of defect algebras. More precisely, the defect algebra is $\df = (\gf \ltimes \gf_{\text{ab}})^2$, where $\gf_{\text{ab}}$ is the vector space $\gf$ equipped with an abelian Lie algebra structure, on which $\gf$ acts by the adjoint action. The boundary condition \eqref{eq:BCpcm} can then be translated as the statement that the jet of the gauge field belongs to the isotropic subalgebra $\kf = (\lbrace 0 \rbrace \ltimes \gf_{\text{ab}})^2$ of $\df$.}
\begin{equation}\label{eq:BCpcm}
    A_\pm \bigl|_{z=0} =  A_\pm \bigl|_{z=\infty} = 0\,.
\end{equation}

\paragraph{The 2d field.} Let us reparameterise the gauge field $A$ in terms of $(\gh,L)$, as in equation \eqref{eq:AgL}. The group-valued field $\gh$ is subject to the 4d gauge symmetry $\gh \mapsto u\gh$, where the gauge parameter is a smooth function on $\CP\times\Sigma$ required to satisfy $u|_{z=0} = u|_{z=\infty} = \text{Id}$ to preserve the boundary condition \eqref{eq:BCpcm}. The only degrees of freedom left in $\gh$ are thus the 2d group-valued fields $g=\gh |_{z=0}$ and $\widetilde{g}=\gh |_{z=\infty}$. The model is further invariant under the redundancy $\gh \mapsto \gh v$, where $v:\Sigma \to G$ is independent of $(z,\zb)$ -- see Section \ref{subsec:int}. This results in a local 2d $G_{\text{diag}}$-symmetry acting as $(g,\widetilde{g}) \mapsto (gv,\tilde{g}v)$. For simplicity, we will fix this gauge symmetry by setting $\widetilde{g}$ to the identity, and we are left with only $g$. To summarise, the only physical degree of freedom is a 2d field $g : \Sigma \to G$, and $\gh$ satisfies
\begin{equation}\label{eq:ghPCM}
    \gh|_{z=0} = g \qquad \text{ and } \qquad \gh|_{z=\infty} = \text{Id}\,.
\end{equation}
Note that we have used the $G_{\text{diag}}$ redundancy to fix the value of $\gh$ at $z=\infty$. In practice, this makes the model behave quite similarly to one where $\omega$ has no pole at infinity. This actually means that most formulae used in the main text and Appendices \ref{app:GenPoles} and \ref{app:Int} are still valid in the present case without modifications.

\paragraph{Lax connection.} Following equation \eqref{eq:Lz}, the Lax connection can be written as
\begin{equation}\label{eq:LPCMJ}
    L_\pm(z) = \frac{\Jc^{(1)}_\pm}{z \mp 1} + \Jc^{(0)}_\pm\,.
\end{equation}
Recall the relation $A_\pm = \gh L_\pm \gh^{-1} - (\del_\pm \gh)\gh^{-1}$ between this connection and the gauge field. Using the equation \eqref{eq:ghPCM}, the boundary conditions \eqref{eq:BCpcm} translate to the constraints $L_\pm(0)=g^{-1}\del_\pm g$ and $L_\pm(\infty)=0$. This fixes $\Jc^{(0)}_\pm=0$ and $\Jc^{(1)}_\pm = \mp g^{-1}\del_\pm g$: as a result, we obtain the Lax connection 
\begin{equation}\label{eq:LPCM}
    L_\pm(z) = \frac{g^{-1}\del_\pm g}{1 \mp z}\,.
\end{equation}
Note that the vanishing of the constant term $\Jc^{(0)}_\pm$ in $L_\pm(z)$ is a consequence of the gauge-fixing $\gh|_{z=\infty} = \text{Id}$ of the local $G_{\text{diag}}$-symmetry. This is a choice of gauge similar to the one done in Section \ref{sec:rational} of the main text---see the discussion above \eqref{upg}.

\paragraph{Action.} Recall that $A=L^{\gh^{-1}}$. The standard identity for the gauge transformation of the Chern-Simons 3-form gives
\begin{equation}
    {\rm CS}[A] = {\rm CS}\bigl[L^{\gh^{-1}}\bigr] = \langle L , \dd L \rangle + \dd\, \langle \gh^{-1} \dd \gh, L \rangle +  \frac{1}{3} \langle \gh^{-1}\dd \gh, \gh^{-1}\dd \gh \wedge \gh^{-1}\dd \gh \rangle\,.
\end{equation}
The first term does not contribute to the action since $\omega\wedge \del_{\zb} L = 0$. We will further work in the so-called archipelago gauge~\cite{unif}, where $\gh$ is the identity everywhere except in a neighbourhood of $p_1=0$, where it interpolates between a plateau at the value $g$ and the identity (in agreement with the condition \eqref{eq:ghPCM}). Following~\cite{CY3,unif}, the third term in the above equation yields a WZ term in the action proportional to the residue $\res_{z=0} \, \omega$. In the present case, the latter vanishes, so that we are only left with the contribution of the second term. Discarding integrals of spatial derivatives, we get 
\begin{equation}
    S = \frac{1}{16i\pi^2} \int_{\CP\times\Sigma} \omega \wedge \overline{\del}\, \langle \gh^{\,-1} \dd_\Sigma \gh, L \rangle\,.
\end{equation}
This can be computed using the rule \eqref{eq:IntOmegaGen}. Due to our choice of gauge for $\gh$, we only get contributions from the double pole $p_1=0$, which has levels $\ell_{1,0}=0$ and $\ell_{1,1}=h$. This gives
\begin{equation}
    S = \frac{h}{8\pi} \int_\Sigma \langle g^{-1} \dd_\Sigma g, \del_z L\bigr|_{z=0} \rangle
\end{equation}
or, in light-cone coordinates,
\begin{equation}
    S = \frac{h}{8\pi} \int_\Sigma \dd x^+ \wedge \dd x^-\, \left( \langle g^{-1} \del_+ g, \del_z L_-\bigr|_{z=0} \rangle - \langle g^{-1} \del_- g, \del_z L_+\bigr|_{z=0} \rangle \right)\,.
\end{equation}
Equation \eqref{eq:LPCM} provides the quantities $\del_z L_\pm \bigr|_{z=0} = \pm g^{-1} \del_\pm g$. Then taking into account $\dd^2 x = -2\,\dd x^+ \wedge \dd x^-$, we end up with the 2d action
\begin{equation}\label{eq:PCM}
    S = \frac{h}{8\pi} \int_\Sigma \dd^2 x\, \langle g^{-1} \del_+ g, g^{-1} \del_- g \rangle\, ,
\end{equation}
which is precisely the standard action of the Principal Chiral Model.

\paragraph{RG flow.} The 1-loop renormalisation of the PCM \eqref{eq:PCM} is a standard computation using the formulation as a Ricci flow \eqref{ric} (without B-field). One then finds that the model is renormalisable, with the coupling $h$ running according to
\begin{equation}\label{eq:FlowPCM}
    \ddt h = c_G\,.
\end{equation}
This RG flow translates to the following log-divergent term in the 1-loop effective action:
\begin{equation}
    \ddt \widehat S^{(1)} = \frac{\cG}{8\pi} \int_\Sigma \dd^2x   \,  \bigl\langle g^{-1}\del_+ g,g^{-1}\del_- g \bigr\rangle \, ,
\end{equation}
by simply taking the ${\rm t}$-derivative of the classical action and using the expression \rf{eq:FlowPCM} for $\ddt h$. In terms of the currents $\Jc^{(1)}_\pm = \mp g^{-1}\del_\pm g$ appearing in the Lax connection \eqref{eq:LPCMJ}, this expression can be written as
\begin{equation}
    \ddt \widehat S^{(1)} = - \frac{\cG}{2\pi} \frac{1}{(z^+_1 - z^-_1)^2} \int_\Sigma \bigl\langle \Jc^{(1)}_+, \Jc^{(1)}_- \bigr\rangle \, \dd^2x \,,
\end{equation}
which precisely agrees with the universal result \eqref{eq:UnivSimpleZ} above.\\

Let us now turn to the RG flow of the twist 1-form $\omega$. According to the general formula \eqref{fom}, this flow should be controlled  by a specific meromorphic function $\Psi(z)$ (see subsections \ref{subsec:4dFlow} and \ref{app:RGhigher}). In the present case, it takes the form
\begin{equation}
  \label{PsiPCM}  \Psi(z) = -\cG \left( \frac{1}{z} + z \right)\,.
\end{equation}
As required, it shares the same poles as $\omega$ and satisfies $\Psi(z_1^\pm)=\Psi(\pm 1) = \mp 2\cG$.\footnote{Contrarily to the setup of subsections \ref{subsec:4dFlow} and \ref{app:RGhigher}, $\Psi(z)$ is not bounded at infinity due to the linear term $z$. This is because in this appendix we allowed $\omega$ to have a double pole at infinity.} The flow \eqref{fom} then becomes
\begin{equation}
    \ddt \omega = \del\Psi = \cG \left( \frac{1}{z^2} - 1 \right)\dd z\,.
\end{equation}
Comparing this with the form \eqref{eq:OmegaPCM} of $\omega$, one can extract from this formula the same flow \eqref{eq:FlowPCM} of the coupling $h$.

\bigskip

\bibliography{bib}
\bibliographystyle{JHEP}

	\end{document}